%% file: sample631.tex
 \documentclass[twocolumn]{aastex631}
\usepackage{graphicx}	
\usepackage{amsmath}	

\input{macros}

\begin{document}

\title{Unveiling the Cosmic Chemistry: Revisiting the Mass-Metallicity Relation \\ with JWST/NIRSpec at 4 $ < z < $ 10}

\author[0000-0002-5222-1337]{Arnab Sarkar}
\affiliation{Kavli Institute for Astrophysics and Space Research,
Massachusetts Institute of Technology, 70 Vassar St, Cambridge, MA 02139}
\email{arnabsar@mit.edu}

\author[0000-0002-4469-2518]{Priyanka Chakraborty}
\affiliation{Center for Astrophysics $\vert$ Harvard \& Smithsonian, 60 Garden St, Cambridge, MA 02138}

\author[0000-0001-8593-7692]{Mark Vogelsberger}
\affiliation{Kavli Institute for Astrophysics and Space Research,
Massachusetts Institute of Technology, 70 Vassar St, Cambridge, MA 02139}

\author{Michael McDonald}
\affiliation{Kavli Institute for Astrophysics and Space Research,
Massachusetts Institute of Technology, 70 Vassar St, Cambridge, MA 02139}

\author[0000-0002-5653-0786]{Paul Torrey}
\affiliation{University of Virginia, Virginia, USA}

\author[0000-0002-8111-9884]{Alex M. Garcia}
\affiliation{University of Virginia, Virginia, USA}

\author{Gourav Khullar}
\affiliation{University of Pittsburgh, Pittsburgh, PA, 15260}

\author{Gary J. Ferland}
\affiliation{University of Kentucky, Lexington, KY}

\author{William Forman}
\affiliation{Center for Astrophysics $\vert$ Harvard \& Smithsonian, 60 Garden St, Cambridge, MA 02138}

\author{Scott Wolk}
\affiliation{Center for Astrophysics $\vert$ Harvard \& Smithsonian, 60 Garden St, Cambridge, MA 02138}

\author{Benjamin Schneider}
\affiliation{Kavli Institute for Astrophysics and Space Research,
Massachusetts Institute of Technology, 70 Vassar St, Cambridge, MA 02139}

\author[0000-0002-1379-4482]{Mark Bautz}
\affiliation{Kavli Institute for Astrophysics and Space Research,
Massachusetts Institute of Technology, 70 Vassar St, Cambridge, MA 02139}

\author[0000-0002-3031-2326]{Eric Miller}
\affiliation{Kavli Institute for Astrophysics and Space Research,
Massachusetts Institute of Technology, 70 Vassar St, Cambridge, MA 02139}

\author[0000-0002-4737-1373]{Catherine Grant}
\affiliation{Kavli Institute for Astrophysics and Space Research,
Massachusetts Institute of Technology, 70 Vassar St, Cambridge, MA 02139}

\author{John ZuHone}
\affiliation{Center for Astrophysics $\vert$ Harvard \& Smithsonian, 60 Garden St, Cambridge, MA 02138}

\date{Accepted for publication in ApJ}

\begin{abstract}
We present star formation rates (SFR), the mass-metallicity relation (MZR), and the SFR-dependent MZR across redshifts 4 to 10 using 81 star-forming galaxies observed by the JWST NIRSpec employing both low-resolution PRISM and medium-resolution gratings, including galaxies from the JADES GOODS-N and GOODS-S fields, the JWST-PRIMAL Legacy Survey, and additional galaxies from the literature in Abell 2744, SMACS-0723, RXJ2129, BDF, COSMOS, and MACS1149 fields. These galaxies span a 3 dex stellar mass range of $10^7 < M_{\ast}/M_{\odot} < 10^{10}$, with an average SFR of $7.2 \pm 1.2 M_{\odot} {\rm yr}^{-1}$ and an average metallicity of $12+{\rm log(O/H)} = 7.91 \pm 0.08$. Our findings align with previous observations up to $z=8$ for the MZR and indicate no deviation from local universe FMR up to this redshift. Beyond $z=8$, we observe a significant deviation $\sim 0.27$ dex) in FMR, consistent with recent JWST findings. We also integrate CEERS (135 galaxies) and JADES (47 galaxies) samples with our data to study metallicity evolution with redshift in a combined sample of 263 galaxies, revealing a decreasing metallicity trend with a slope of $0.067 \pm 0.013$, consistent with IllustrisTNG and EAGLE, but contradicts with FIRE simulations. We introduce an empirical mass-metallicity-redshift (MZ--$z$ relation): $12+{\rm log(O/H)}=6.29 + 0.237 \times{\rm log}(M_{\ast}/M_{\odot}) - 0.06 \times (1+z)$, which accurately reproduces the observed trends in metallicity with both redshift and stellar mass. This trend underscores the ``Grand Challenge'' in understanding the factors driving high-redshift galactic metallicity trends, such as inflow, outflow, and AGN/stellar feedback--and emphasizes the need for further investigations with larger samples and enhanced simulations.

\end{abstract}

\keywords{James Webb Space Telescope--High redshift Galaxies--Chemical abundances}



\section{Introduction}
Stellar mass and 
gas-phase metallicity are
two of the most fundamental physical 
properties of galaxies, serving as key 
indicators of galaxy evolution.
Star formation primarily enhances the metal 
content of a galaxy, but this enrichment is 
temporarily offset by the inflow of 
cosmological gas and large-scale galactic 
winds,
with inflowing gas fueling long-term star 
formation and outflows enriching the 
interstellar medium (ISM)
\citep{2011MNRAS.415...11D, 2013ApJ...772..119L}.
The exchange of baryons in and out of 
galaxies affects their stellar masses 
($M_\star$), metallicities ($Z$), and star 
formation rates (SFRs), impacting the mass–
metallicity relation (MZR) and the 
fundamental metallicity relation (FMR; 
$M_\star$–SFR–$Z$). 
Therefore, understanding how these 
quantities evolve with cosmic time and in 
relation to one another
is vital for deciphering the processes 
that control star formation in 
galaxies and drive galactic evolution.

The initial evidence of a 
mass-metallicity relation (MZR) was 
demonstrated in 
\citet{1979A&A....80..155L}, who found a 
relationship between total mass and 
metallicity in irregular and blue compact 
galaxies.
Due to challenges in obtaining reliable 
galaxy masses, subsequent studies have used 
optical luminosity as a proxy, showing a 
clear correlation between blue luminosity 
and metallicity, with more 
luminous galaxies exhibiting higher 
metallicities 
\citep[e.g.,][]{1987ApJ...317...82G, 1994ApJ...420...87Z}.
The development of reliable stellar 
population synthesis models 
\citep{2003MNRAS.344.1000B} has since 
enabled more accurate stellar mass 
measurements from spectral energy 
distributions (SEDs). 
In the local Universe, 
\citet{2004ApJ...613..898T} found a tight 
correlation between galaxy stellar mass and 
gas-phase oxygen abundance in star-forming 
galaxies, 
based on data from $\sim$ 53,000 
galaxies in the Sloan Digital Sky Survey
(SDSS) early data release 
\citep{2000AJ....120.1579Y, 2003AJ....126.2081A}. 
Numerous studies have since identified 
the mass-metallicity relation in 
galaxies up to $z \sim$ 2.3 
\citep{2006ApJ...644..813E, 2006ApJ...646..107E, 2011ApJ...730..137Z, 2013ApJ...765..140A},
demonstrating that the MZR evolves with 
redshift, showing higher metallicities at 
lower redshifts for a given stellar mass 
\citep{2013ApJ...771L..19Z}. 
\citet{2008A&A...488..463M} extended the 
mass-metallicity relation up to $z \sim$ 
3.5, finding that its evolution is much 
stronger than observed at lower redshifts.

\citet{2010MNRAS.408.2115M} identified 
an anti-correlation between metallicity and 
SFR in a large sample of SDSS galaxies, 
indicating that galaxies of the same 
stellar mass with higher SFRs tend
to have lower gas-phase metallicity,
a finding later extended to low-mass 
galaxies by \citet{2011MNRAS.414.1263M}.
The observed anti-correlation is 
attributed to the interplay between the 
infall of pristine gas, which fuels star 
formation, and the outflow of enriched 
material \citep{2011MNRAS.416.1354D,  2013MNRAS.430.2891D, 2015MNRAS.452..486D, 2017MNRAS.469.2121S, 2018NatAs...2..179C}. 
This relationship remains largely unevolved 
between $z \sim$ 0 and $z \sim$ 3 
\citep[e.g.,][]{2010MNRAS.408.2115M, 2013ApJ...765..140A, 2014ApJ...797..126S,  2019A&A...627A..42C, 2020MNRAS.491..944C,2021ApJ...914...19S}, 
and is therefore referred to as the 
Fundamental Metallicity Relation
(or FMR)
\citep[e.g.,][]{2008ApJ...672L.107E,2010A&A...521L..53L}.

In the pre James Webb Space Telescope
(JWST) era,  accurate measurements of galaxy 
chemical abundances were limited to
redshift $z=3.3$ 
\citep{2021ApJ...914...19S}, 
thereby restricting the study of  MZR 
and FMR for galaxies at higher redshifts.  
With its wide range of spectroscopic 
capabilities and unparalleled sensitivity
in the near- and mid-infrared
band, the JWST
and its near-infrared spectrograph NIRSpec 
\citep{2022A&A...661A..81F, 2022A&A...661A..80J}  have
revolutionized our ability to explore
and analyze galaxies from the earliest 
epochs of the universe. 
Recently,
the use of nebular emission lines and line 
ratios in star-forming galaxies has proven 
to be a crucial tool for probing their gas 
properties, and providing detailed 
measurements of their metallicities, stellar 
masses, and star formation rates for 
galaxies with redshifts up to $z \sim$ 10 
\citep{2023ApJ...950L...1S, 2023ApJS..269...33N, 2024ApJ...962...24S,  2024A&A...684A..75C}
detected through observational campaigns 
such as the JWST Early Release Observations 
\citep[JWST-ERO,][]{2022ApJ...936L..14P}, 
JWST Advanced Deep Extragalactic Survey
\citep[JADES,][]{2023arXiv230602465E}, 
Through the Looking GLASS  \citep[GLASS-
JWST,][]{2022ApJ...935..110T}, and the 
Cosmic Evolution Early Release Science  
\citep[CEERS,][]{2023ApJ...946L..13F} 
programs.

The exceptional capabilities of 
JWST have allowed us, for the first time, 
to test whether high-redshift galaxies 
adhere to the same mass-metallicity-SFR 
relation as the extensively studied galaxies 
with redshifts up to $z \sim 3$. 
Using JWST/NIRSpec observations from the 
Abell 2744 and RXJ-2129 regions, 
as well as the CEERS survey, 
\citet{2023NatAs...7.1517H} studied
the mass-metallicity relation of galaxies
at redshifts $z = 7 - 10$, 
finding that these high-redshift galaxies 
have lower gas-phase metallicities
compared to local star-forming galaxies 
at $z \approx 0$.
Similar findings were reported by 
\citet{2023ApJ...957...39L},
who analyzed a sample of 11 galaxies
within the redshift range $7.2 < z < 9.5$ 
and provided a quantitative statistical 
inference of the mass-metallicity relation 
at $z \approx 8$, concluding that galaxies 
at this epoch are less metal-enriched than 
those in the local universe.

Using observations from the public 
spectroscopy programs -
ERO, GLASS, and CEERS observed with 
JWST/NIRSpec, \citet{2023ApJS..269...33N} 
investigated the evolution of the
mass-metallicity relation across 
redshifts $z = 4 - 10$.
By analyzing a sample of 135 galaxies,
their study revealed that the MZR
exhibits only small evolution 
towards lower metallicity when
compared to the well-established 
relation at $z \sim 2 - 3$. 
\citet{2024A&A...684A..75C} investigated 
the metallicity properties of low-mass 
galaxies within the redshift range 
$3<z<10$ using deep NIRSpec 
spectroscopic data from the JADES program, 
and also found a mild evolution of mass 
metallicity relation at $z>3$ 
indicating a trend of slightly 
decreasing metallicity.
Similar findings were reported by 
\citet{2023ApJ...950...67M}, 
who studied the mass-metallicity 
relation in galaxies at  $5 < z < 7$ 
using the first deep 
JWST/NIRCam wide-field slitless 
spectroscopic observations, and by 
\citet{2023ApJ...950L...1S},
who examined galaxies at $2.7 < z < 6.5$ 
from the CEERS survey, finding no 
significant evolution in the mass-
metallicity relation within that redshift 
range.

Earlier studies have suggested that 
the local FMR may not be applicable at 
redshifts \( z \gtrsim 3.5 \) 
\citep{2014A&A...563A..58T, 2016ApJ...822...42O}.
However, many of these studies, based on 
small sample sizes and predominantly 
featuring starburst galaxies, do not 
accurately represent the typical galaxy 
population at these high redshifts.  
Interestingly, recent JWST observations have 
revealed divergence from the local
FMR at $z > 5$ \citep{2023NatAs...7.1517H, 2023ApJS..269...33N, 2023arXiv230706336L, 2024A&A...684A..75C}. 
\citet{2023NatAs...7.1517H} observed 
a clear offset in the FMR from 
that of local galaxies using 
the same sample of $7.2<z<9.5$ galaxies.
\citet{2023ApJS..269...33N} found that the 
FMR remains largely unchanged from
\( z = 0 \) to \( z = 4 \text{--} 8 \), 
but exhibits a significant decrease in 
metallicity at \( z > 8 \).
\citet{2024A&A...684A..75C} also reported 
a deviation in the FMR, finding that high-
redshift galaxies exhibit a substantial 
metal deficiency compared to local galaxies 
with similar stellar mass  and star 
formation rate.
The origins of these offsets remain 
ambiguous, potentially indicating either 
authentic deviations from the FMR, 
systematic inaccuracies in metallicity 
determinations, or limitations inherent to 
current observational methodologies. 
The limited sample size used within 
individual studies, particularly in 
the \( z = 8 \text{--} 10 \) range, 
undoubtedly contributes to this ambiguity.

\begin{figure}
\centering
\includegraphics[width=0.5\textwidth]{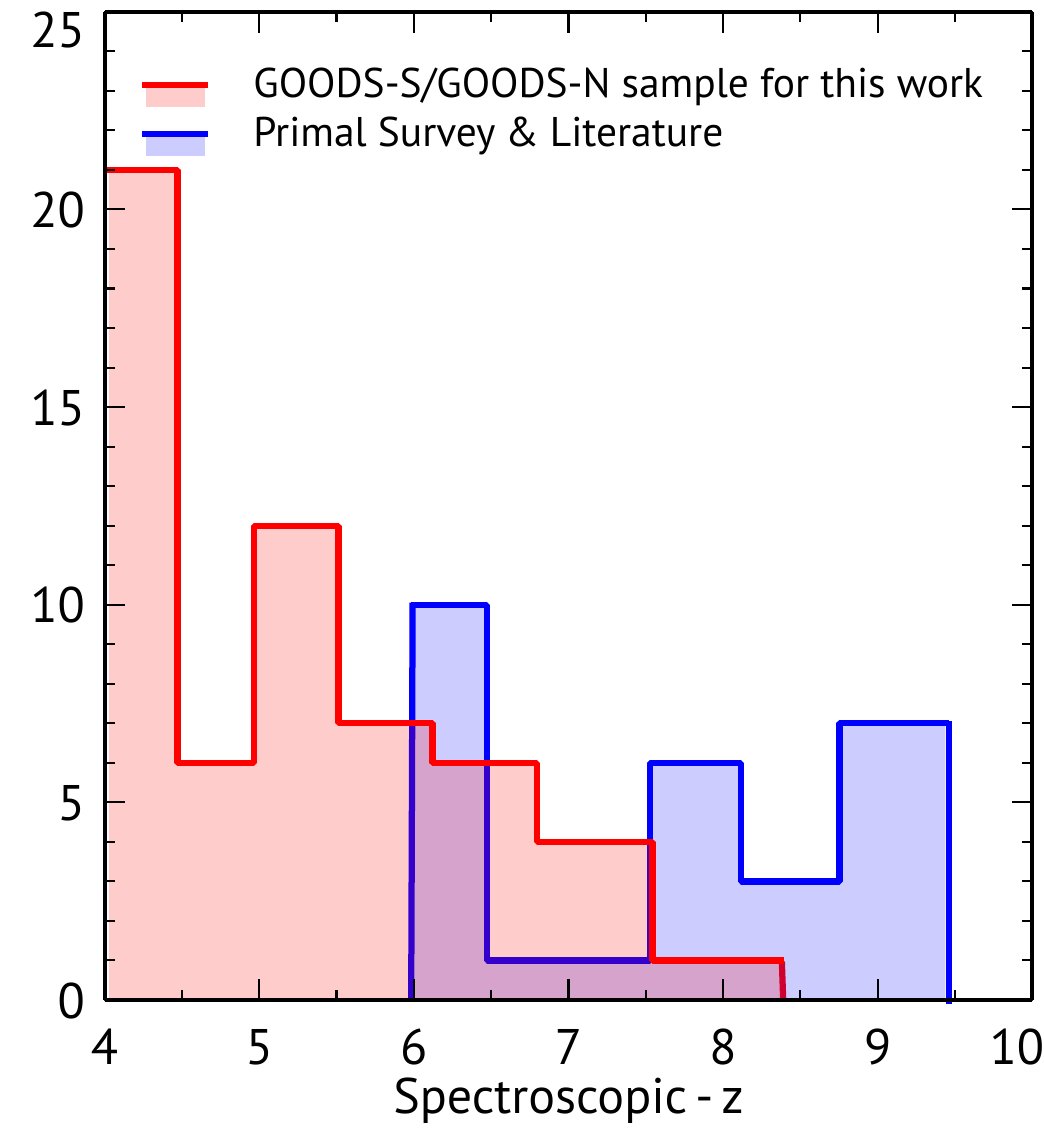}
    \caption{ Redshift distribution of the galaxies selected for the analysis of stellar mass-metallicity-SFR correlations. The red histogram shows the redshift distribution of galaxies from the GOODS-N and GOODS-S Data Release 3
    \citep{2024arXiv240406531D}, whereas the blue histogram represents the distribution of galaxies from the PRIMAL survey
    \citep{2024arXiv240402211H}, supplemented with additional samples from the literature
    \citep[e.g.,][]{2024arXiv240303977V,2024ApJ...967L..42W,2024arXiv240708616M}.}
    \label{fig:hist}
\end{figure}

In this paper, we investigate 
the mass-metallicity-SFR relations and 
their evolution across redshifts
for a sample of 81 star-forming 
galaxies ranging from redshifts 
$4 < z < 10$, utilizing observations 
from the NIRSpec instrument,
employing both its low-resolution 
{\it PRISM} and medium-resolution 
grating capabilities.
Our dataset includes:
\begin{itemize}
    \item 54 galaxies from the 
    GOODS-S and GOODS-N fields, 
    part of the JADES public data 
    release 3 \citep{2024arXiv240406531D}.

    \item 22 galaxies as reported in  
    JWST-PRIMAL Legacy Survey 
    \citep{2024arXiv240402211H}.

    \item 5 galaxies selected
    from various fields
    as reported in literature: Abell 2744, BDF, COSMOS 
    \citep{2024arXiv240303977V}, RXJ2129 
    \citep{2024ApJ...967L..42W}, and 
    MACS1149 
    \citet{2024arXiv240708616M}.
\end{itemize}
This sample of galaxies, previously
never utilized for studying the 
SFR-MZR, provides a novel and independent 
dataset for investigating the SFR-MZR at
high redshifts. 
This allows us to not only compare 
and contrast our findings with previous 
studies and simulations but also to 
offer independent constraints on MZR and
FMR 
\citep[e.g.,][]{2014Natur.509..177V,2020NatRP...2...42V,2018ApJS..238...33D,2019MNRAS.487.1529D}.
Details on these 
galaxies are 
provided in 
Table \ref{tab:goods}, 
\ref{tab:goodss}, \ref{tab:primal},
and \ref{tab:literat}
in Appendix,
their distribution across redshifts 
is depicted in Figure \ref{fig:hist}.
The organisation of the paper 
is as follows. 
In Section  \ref{data}, 
we discussed data analysis and
methods involved in fitting
{\it PRISM}
and grating spectrum. 
In Section \ref{results},
we present the physical properties of 
our galaxy sample, including  emission 
line-flux ratios, 
AGN contamination removal, 
metallicity measurements, and  
mass-metallicity relation. 
In Section \ref{sec:MZR} 
and \ref{FMR}, 
we discuss the mass-metallicity
relation and the
fundamental
metallicity relation and their 
evolution.
In Section \ref{sec:simulation}, 
we probe the evolution of
metallicity with redshift and
MZ--$z$ relation.
In Section \ref{summary}, 
we summarize our findings. 

Throughout this paper, we adopt
the AB magnitude system  
\citep{1983ApJ...266..713O} and
cosmological parameters reported by 
\citet{2020A&A...641A...6P}: 
the Hubble constant \( H_0 = 67.4 \, 
\text{km s}^{-1} \text{Mpc}^{-1} \), matter 
density parameter \( \Omega_M = 0.315 \), 
and dark energy density
\( \Omega_\Lambda = 
0.685 \).

\section{Observations and Methods}\label{data}
For this work, we assemble a 
sample of
81 galaxies in redshift range
4 $< z <$ 10 
primarily derived from
publicly available programs or
data releases, such as JADES (data release 3;
e.g.,
\citealt{2024arXiv240406531D,2022ARA&A..60..121R,2023arXiv230602465E,2023arXiv230602467B,2023NatAs...7..622C}), 
JWST-PRIMAL Legacy Survey
\citep{2024arXiv240402211H}, and
from literature
\citep[e.g,][]{2024arXiv240303977V,2024ApJ...967L..42W,2024arXiv240708616M}. 
{ We select
these galaxies based on 
below criteria:
\begin{itemize}
    \item 
Spectra 
are obtained through 
the application of
the NIRSpec
low-resolution configuration 
{\it PRISM}/CLEAR,
covering the spectral range of 
0.6--5.3$\mu$m,
and medium resolution gratings 
(R$\sim$1000), including G140M/F070LP 
(0.7--1.27$\mu$m), G235M/F170LP
(1.66--3.07$\mu$m), and G395M/F290LP 
(2.87--5.10$\mu$m).

\item The minimum 
signal-to-noise ratios
should be $\geq 3$ for
faint lines 
(such as H$\gamma$, [$\nii$]$\lambda$6584, and 
[$\sii$]$\lambda\lambda$6716,31) and 
$\geq 5$
for bright lines (such as
[$\oiii$]$\lambda\lambda$5007,4959, H$\alpha$, and $H\beta$).

\item For each 
galaxy, 
[$\oiii$]$\lambda\lambda$5007
and 
$H\beta$ should be present
with
required signal-to-noise ratio,
as well as one of
below emission line or pairs 
should be present
[$\oii$]$\lambda\lambda$3727,29, [$\nii$]$\lambda$6584 \& 
H$\alpha$, and 
[$\sii$]$\lambda\lambda$6716,31
\& H$\alpha$.

\item These galaxies
should be star-forming
galaxies. 
Mass-Excitation 
diagnostic has been
utlized to identify 
star-forming galaxies 
from AGN, as discussed in
Section \ref{sec:agn_contam}.

\end{itemize}

}

If medium-resolution grating spectra
are unavailable or do not cover
the spectral range needed to
capture three key emission lines-
$H\beta$, [$\oii$]$\lambda$3727,29,
[$\oiii$]$\lambda\lambda$5007,4959-
we instead utilize low-resolution 
{\it PRISM} spectra. 
Out of 81 galaxies in our study, 
67 have spectra observed with both 
{\it PRISM} and 
medium-resolution gratings. 
However, 9 galaxies 
lack medium-resolution grating spectra.
For most of the galaxies,
we utilize multi-object 
spectroscopy observations using the 
micro-shutter assembly (MSA) of
NIRSpec on-board JWST
\citep{2022A&A...661A..81F}.
During these observations, three 
micro-shutters were activated for each
target.
An exposure protocol was employed
consisting of a three-point nodding 
sequence along the slit,
ensuring comprehensive coverage
and improved data quality for each 
target.
For each JADES GOODS-S and GOODS-N
targets, 
the flux-calibrated 1D 
and 2D spectra were produced by the 
JADES 
team using a custom pipeline developed by
the ESA NIRSpec Science Operations Team 
(SOT) and Guaranteed Time Observations
(GTO) teams. 
For detailed descriptions of the data
reduction steps and methods,
we refer readers to
\citet{2023arXiv230602467B}, 
\citet{2024A&A...684A..75C}, 
\citet{2024arXiv240406531D} 
and references therein.
For this paper, we adopted the reduced 
and flux-calibrated 
{\it medium}-tier 1D and 2D spectra of
hundreds of targets , which were
released publicly as part of 
JADES Data Release 3\footnote{\url{https://jades-survey.github.io/scientists/data.html}}
\citep{2024arXiv240406531D}.

For targets selected from the
JWST-PRIMAL Legacy Survey,
we utilized DAWN JWST Archive (DJA), 
containing 
reduced images, photometric catalogs, 
and spectroscopic data for public 
JWST data products 
\footnote{\url{https://dawn-cph.github.io/dja}}
\footnote{\url{https://s3.amazonaws.com/msaexp-nirspec/ extractions/nirspec_graded_v2.html}}.
The DJA spectroscopic archive 
(DJA-Spec) includes observations 
from major programs
such as CEERS 
\citep{2022ApJ...940L..55F}, GLASS-DDT 
\citep{2022ApJ...935..110T},
JADES \citep{2023arXiv230602467B}, and 
UNCOVER 
\citep{2022arXiv221204026B}. 
For detailed data reduction
processes we refer readers to 
\citet{2024arXiv240402211H}.

Additionally, we selected 
5 galaxies from the literature, 
including three galaxies located
in the Abell 2744, BDF, and COSMOS 
fields with redshifts of 7.89, 7.11,
and 6.36, respectively 
\citep{2024arXiv240303977V}. 
The other two galaxies, at 
redshifts 8.16 and 9.11, 
were observed in the RXJ2129 
and MACS1149 galaxy cluster fields, 
as reported by 
\citet{2024ApJ...967L..42W} and 
\citet{2024arXiv240708616M}, 
respectively.
For these 5 galaxies, 
we utilized the redshift,
stellar mass, $H\beta$ flux,
and metallicity measurements
reported in their respective studies.

\section{Spectral fitting}
We discuss the detailed spectral fitting
processes below.

\subsection{{\it PRISM/CLEAR} spectra}\label{sec:prism_spectra_fitting}
To measure the stellar mass and spectral
energy distribution (SED)-based redshift
of each targets, we fitted the PRISM 
spectra by using the SED fitting code
$\bagpipes$ \citep{2018MNRAS.480.4379C}.
$\bagpipes$ creates detailed model
galaxy spectra and fits them to
photometric and spectroscopic observations
\citep{2008MNRAS.384..449F}.
This method produces posterior 
distributions of galaxy properties 
for each source in the sample. 
$\bagpipes$ is versatile, capable
of modeling galaxies with various  
star formation histories (SFHs),
including delayed-$\tau$, constant, 
and bursts 
\citep[e.g.,][]{2020ApJ...904...33L,2024arXiv240605306C}.

{
For this study, 
we used a constant star-formation model with the minimum and maximum
star formation ages were 
allowed to vary between 0 and 2 Gyr.} We adopted stellar
population synthesis models based on
the 2016 version of the BC03 models 
\citep{2003MNRAS.344.1000B}. 
These models assume the initial mass 
function (IMF) from 
\citet{2002Sci...295...82K} and include 
nebular line and continuum emissions 
using $\Cloudy$ \citep{2023RMxAA..59..327C}.
The SED-fitting was conducted over a broad 
range of parameters, with stellar mass 
log$(M_{\ast}/M_{\odot})$ varying between
4 and 13, and 
stellar metallicities log$(Z/Z_{\odot})$
ranging from 0.005 to 2.5. 
$\bagpipes$ assumes 
the solar abundances from 
\citet{1989GeCoA..53..197A} and 
incorporates ISM depletion factors 
and He and N scaling relations
from \citet{2000ApJ...542..224D}.

The ionization parameter for nebular 
line and continuum emissions was varied 
between $-4$ and $-1$. 
We adopted the Calzetti dust attenuation
curve \citep{2000ApJ...533..682C} with
an extinction parameter $A_{V}$ ranging 
from 0 to 4.
Additionally, to address birth-cloud 
dust attenuation, we introduced a
multiplicative factor ($1 < \eta < 2$) 
to the dust model.
This accounts for the increased dust
attenuation typically observed around 
$\hii$ regions, 
which is usually double that of the
general ISM within the galaxy's first 
10 Myr \citep{2023arXiv230602467B}. 
To model this effect, we set the maximum 
age of the birth-cloud to 0.01 Gyr 
\citep{2023arXiv230602467B}.

We note that the spectral resolution 
of the {\it PRISM} spectra varies 
significantly with wavelength, ranging
from R $\sim$ 30 at 1.2$\mu$m to a peak 
of R $\sim$ 300 at the cutoff wavelength
of 5 $\mu$m.
Since we only use the {\it PRISM}
spectra to estimate the stellar mass 
and SED-based redshift, we chose not
to fit these spectra with variable
resolution settings in $\bagpipes$. 
This simplifies our analysis 
and avoids the systematics that 
come with adjusting the resolution 
settings for different wavelengths.
Instead, we focused on extracting 
reliable stellar mass and
SED-based redshift estimates.
{
The uncertainties on best-fit
parameters are derived from
the posterior distribution of
each parameter.
The resulting best-fit
stellar masses
for our sample 
of galaxies are 
listed in
Table \ref{tab:goods}, 
\ref{tab:goodss}, \ref{tab:primal},
and \ref{tab:literat}
in Appendix.
}

\subsection{Emission
line flux measurement}
We conduct measurements of
emission-line fluxes for each
target using publicly released 
1D medium-resolution grating spectra 
obtained through G140M/F070LP,
G235M/F170LP, and G395M/F290LP 
dispersers/filters. 
For 9 galaxies, 
medium-resolution grating spectra 
are unavailable; 
therefore, we use {\it PRISM} 
spectra to measure emission-line
fluxes.
Our analysis focused on extracting 
crucial emission lines,
including hydrogen Balmer lines,
[$\oii$]$\lambda\lambda$3727,29,
[$\oiii$]$\lambda\lambda$5007,4959,
[$\nii$]$\lambda$6584, and 
[$\sii$]$\lambda\lambda$6716,31.
For each emission line, we applied
Gaussian profile fitting to
accurately determine their fluxes. 
The errors associated with these
flux measurements were computed by 
combining noise levels from spectral
bins within the 
Full Width at Half Maximum (FWHM) 
centered on the Gaussian peak, 
providing robust estimates of
measurement uncertainties.
In low-resolution {\it PRISM} 
spectra, the doublet
lines are typically blended, such as
[$\oii$]$\lambda\lambda$3727,29 and 
[$\oiii$]$\lambda\lambda$5007,4959.
Consequently, we use a single 
Gaussian profile to represent
each of the [$\oii$] and [$\oiii$]
blend.
To estimate the 
[$\oiii$]$\lambda$5007 line-flux,
we assume a theoretical 
flux ratio of 2.98 for the
[$\oiii$]$\lambda\lambda$5007,4959 
doublet
\citep{2000MNRAS.312..813S}.

To evaluate the quality of our measurements, 
we calculate the signal-to-noise ratios 
(S/N) for each emission line. 
We establish a minimum S/N criterion
of $\geq$ 3$\sigma$ for including
a given emission line in our 
subsequent metallicity calculations. 
Specifically, our sample selection
criteria required galaxies to 
exhibit detectable H$\beta$ and [$\oiii$], in 
addition to having at least
one emission line—[$\oii$]$\lambda\lambda$3727,29,
[$\nii$]$\lambda$6584, or 
[$\sii$]$\lambda\lambda$6716,31—measured at 
or above the 3$\sigma$ confidence level.
This methodical approach enable us 
to robustly measure and validate 
emission-line fluxes across our sample, 
ensuring that only reliable data 
points are utilized in deriving 
the gas-phase metallicities of
galaxies at high redshifts.

\section{Determining the Physical Properties of Galaxies}\label{results}

\subsection{Dust corrections and Line ratios}
Our primary objective in this study is to 
measure SFR and gas-phase  
metallicity of galaxies at high redshifts,
which are significantly impacted by the 
dust reddening. 
To achieve accurate measurements, 
we carefully accounted for the 
corrections due to
dust reddening on the key emission 
lines before using their fluxes
for further analysis. 
Based on the empirical extinction 
relationship established by 
\citet{1994ApJ...429..582C},
the intrinsic luminosities (dust-corrected), 
$L_{\rm int}$ of the emission lines,
can be estimated using
\begin{equation}
L_{\text{int}}(\lambda) = L_{\text{obs}}(\lambda) 10^{0.4 k(\lambda) E(B-V)},
\end{equation}
where, $L_{\rm obs}(\lambda)$ represents
the observed luminosities,
$k_{\lambda}$ denotes the 
extinction coefficient at wavelength 
$\lambda$, and the 
specific reddening curve $k_{\lambda}$ 
was adopted from 
\citet{2000ApJ...533..682C}.

We use three different approaches to 
determine the dust-corrected flux: 
\begin{itemize}
    \item For galaxies at 
    $z < 6.75$,
    where both H$\alpha$ and H$\beta$ are 
    detected with a S/N $\geq$ 3,
    we estimate $E(B - V)$ using the Balmer 
    decrement method. 
    We assume an intrinsic flux ratio
    of H$\alpha$/H$\beta$ = 2.86 
    \citep{2006agna.book.....O} and
    apply the dust extinction 
    curve from \citet{2000ApJ...533..682C}.
    
    \item For galaxies at $z \geq 6.75$, 
    where H$\alpha$ is not observable
    due to the spectral coverage of
    NIRSpec,
   we instead use the
   H$\gamma$/H$\beta$ ratio with 
   an assumed intrinsic value 
   of H$\gamma$/H$\beta$ = 0.47, 
   corresponding to a temperature
   of \(10^4\) K for Case B 
   recombination \citep{2006agna.book.....O}.
    
    \item If neither
    H$\gamma$ nor H$\alpha$ is detected, 
    we estimate the nebular dust
    attenuation using SED fitting 
    performed on {\it PRISM} 
    spectra using the $\bagpipes$
    code, which incorporates a two-component
    dust attenuation model for
    both nebular and stellar emission
    (see Section 
    \ref{sec:prism_spectra_fitting}).
\end{itemize}

We analyze gas-phase metallicity using
line ratio diagnostics, as described in 
Section \ref{sec:metallicity}. 
We calculate line flux ratios 
using dust-corrected emission lines: 
[$\oii$]$\lambda\lambda$3727,29, 
[$\oiii$]$\lambda\lambda$5007,4959, 
[$\nii$]$\lambda$6584, and 
[$\sii$]$\lambda\lambda$6716,31. 
For inclusion in our analysis, 
each line ratio requires all 
constituent lines to be detected 
with a significance of at 
least $3\sigma$. 
Specifically, we consider the 
following line ratios: 
R3, O32, N2, and S2.



\begin{equation}
\text{R3} = \frac{[\text{O III}]\lambda5007}{H\beta}
\end{equation}


\begin{equation}
\text{O32} = \frac{[\text{O III}]\lambda5007}{[\text{O II}]\lambda\lambda3727,29}
\end{equation}

\begin{equation}
\text{N2} = \frac{[\text{N II}]\lambda6584}{H\alpha}
\end{equation}

\begin{equation}
\text{S2} = \frac{[\text{S II}]\lambda\lambda6717, 31}{H\alpha}
\end{equation}


Given the close proximity of the 
involved lines, the N2 ratios 
are largely unaffected by 
reddening correction.
In contrast, O32 and
S2 ratios are sensitive to reddening. 
Additionally, 
R3 ratio is 
slightly sensitive to reddening due 
to the narrow separation of the
lines involved.

\subsection{AGN contamination and star-forming galaxies}\label{sec:agn_contam}
In this study, we utilize 
line-flux ratio diagnostics 
specifically developed for
measuring gas-phase metallicity 
in star-forming regions and galaxies.
However, ionization driven by AGNs
can compromise these standard
metallicity calibrations by 
introducing inaccuracies 
if AGN emissions contaminate 
the emission line-fluxes. 
To ensure the accuracy of our
metallicity assessments, we 
meticulously scrutinize each galaxies
for potential AGN contamination. 
We have adopted two methods to 
systematically exclude AGN contamination
from our sample, 
thereby enhancing the reliability 
of our findings.

\begin{figure}
    \centering
    \includegraphics[width=.5\textwidth]{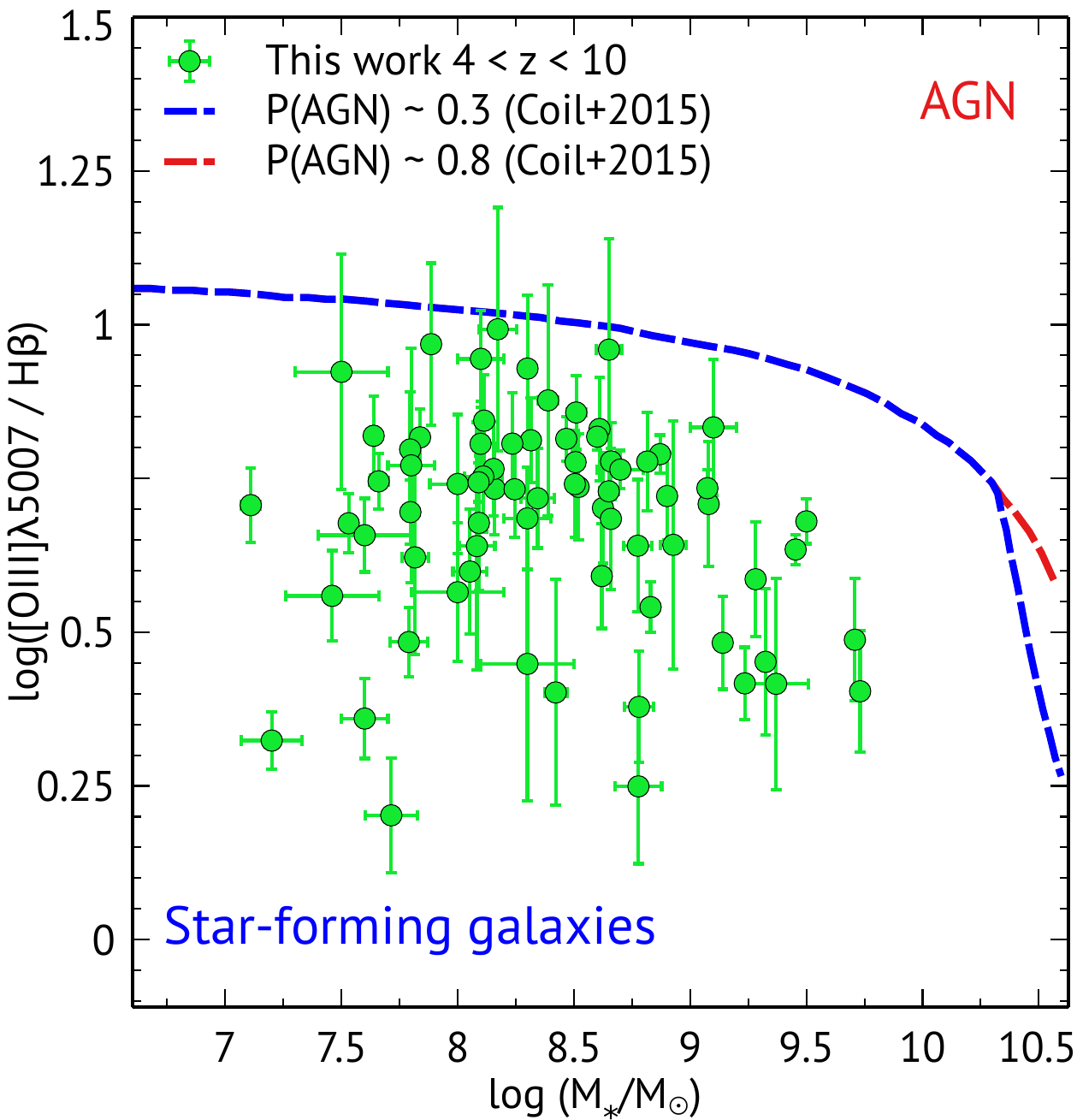}
    \caption{
    The Mass-Excitation (MEx) diagnostic for our full sample, plotting the log($[\oiii]\lambda5007$/H$\beta$) against the log of stellar mass. This diagnostic is adapted from \citet{2024ApJ...960L..13H}, with the MEx curves originally established by \citet{2015ApJ...801...35C}.
    The locations of probable AGN galaxies, with likelihoods of $\sim$ 0.3 and $\sim$ 0.8 are indicated by the blue and red curves, respectively.
   We find no significant AGN contamination in our sample.  }
    \label{fig:agn}
\end{figure}

First, we use the Mass-Excitation 
(MEx) diagnostic diagrams, 
as introduced by 
\citet{2014ApJ...788...88J} and 
refined by \citet{2015ApJ...801...35C}, 
which utilize 
[$\oiii$]~$\lambda5007$/H$\beta$ (R3)
emission-line ratio 
with stellar mass for distinguishing 
between AGNs and star-forming galaxies. 
This diagram serves as an alternative 
to the widely used BPT diagram 
\citep[e.g.,][]{1981PASP...93....5B,2013ApJ...774L..10K}, 
which compares the $[\oiii]\lambda5007$/H$\beta$ to the 
$[\nii]\lambda6584$/H$\alpha$ 
emission line ratios, especially when the 
[$\nii$] or H$\alpha$ lines 
fall out of the visibility window, 
or are blended.
Since we solely measure
emission-line fluxes from the
medium-resolution grating spectra,
we, therefore, use dust-corrected 
$[\oiii]\lambda5007$ and H$\beta$
fluxes for the MEx diagram.
Figure \ref{fig:agn} presents 
our JWST samples in the 
log($[\oiii]\lambda5007$/H$\beta$) --
log($M_{\ast}/M_{\odot}$) plane.

In Figure \ref{fig:agn}, 
the blue and red curves indicate steep 
gradients at P(AGN) $\sim$ 0.3 and
P(AGN) $\sim$ 0.8, respectively. 
These curves represent the probability 
that a galaxy hosts an AGN,
established by 
\citet{2015ApJ...801...35C} for 
$z$ = 2.3 galaxies and AGN from 
the MOSDEF survey
\citep{2024ApJ...960L..13H}.
The positions of our sources 
in the MEx diagram suggest that 
our sample predominantly comprises
star-forming galaxies, 
positioned below or close to 
the boundary line.
Consequently, no possible AGN 
is eliminated and we 
retain all galaxies in our sample.
Similar approach is also
adopted by 
\citet{2024ApJ...960L..13H}
to distinguish star-forming
galaxies from
AGN
in GLASS-JWST sample of galaxies.
Additionally, we visually
inspected spectra for each galaxies
to search for evidence of broad 
emission line regions 
\citep{2021ApJ...907...12S}.

\subsection{Star formation rate for high redshift galaxies}\label{sec:MSFR}
The relationship between 
stellar mass and SFR exhibits a  tight 
correlation across the 
redshift range 
$z \sim 0 - 2$, 
characterized by a slope slightly 
below unity and a scatter generally 
under 0.3 dex 
\citep{2007ApJ...660L..43N, 2007A&A...468...33E, 2007ApJ...670..156D}.
This correlation between 
stellar mass and SFR
in star-forming galaxies is 
commonly referred to as the 
galaxy main sequence (MS).
However, results for higher 
redshift galaxies ($z > 2$)
have shown considerable 
divergence in the literature 
\citep{2014ApJS..214...15S}.

While many studies suggest a 
tight correlation at higher 
redshifts 
\citep{2009ApJ...698L.116P, 2010MNRAS.401.1521M, 2011ApJ...733...99L, 2014ApJ...791L..25S}, 
implying smooth gas accretion and 
aligning well with hydrodynamic 
simulations 
\citep{2006ApJ...639..672F, 2008MNRAS.385..147D}, 
other studies find no 
correlation or high scatter 
in the SFR–stellar mass relationship 
\citep{2006ApJ...644..792R, 2012ApJ...752...66L},
hinting towards bursty star
formation.
This ambiguity is compounded
by the significant challenges in 
establishing the correlation at high 
redshifts, due to various systematic 
uncertainties and selection effects 
inherent in compiling representative 
galaxy samples \citep[e.g.,][]{2015A&A...575A..96G,2020ARA&A..58..661F,2021MNRAS.501.1568F}.

\begin{figure*}
\centering
\includegraphics[width=1.0\textwidth]{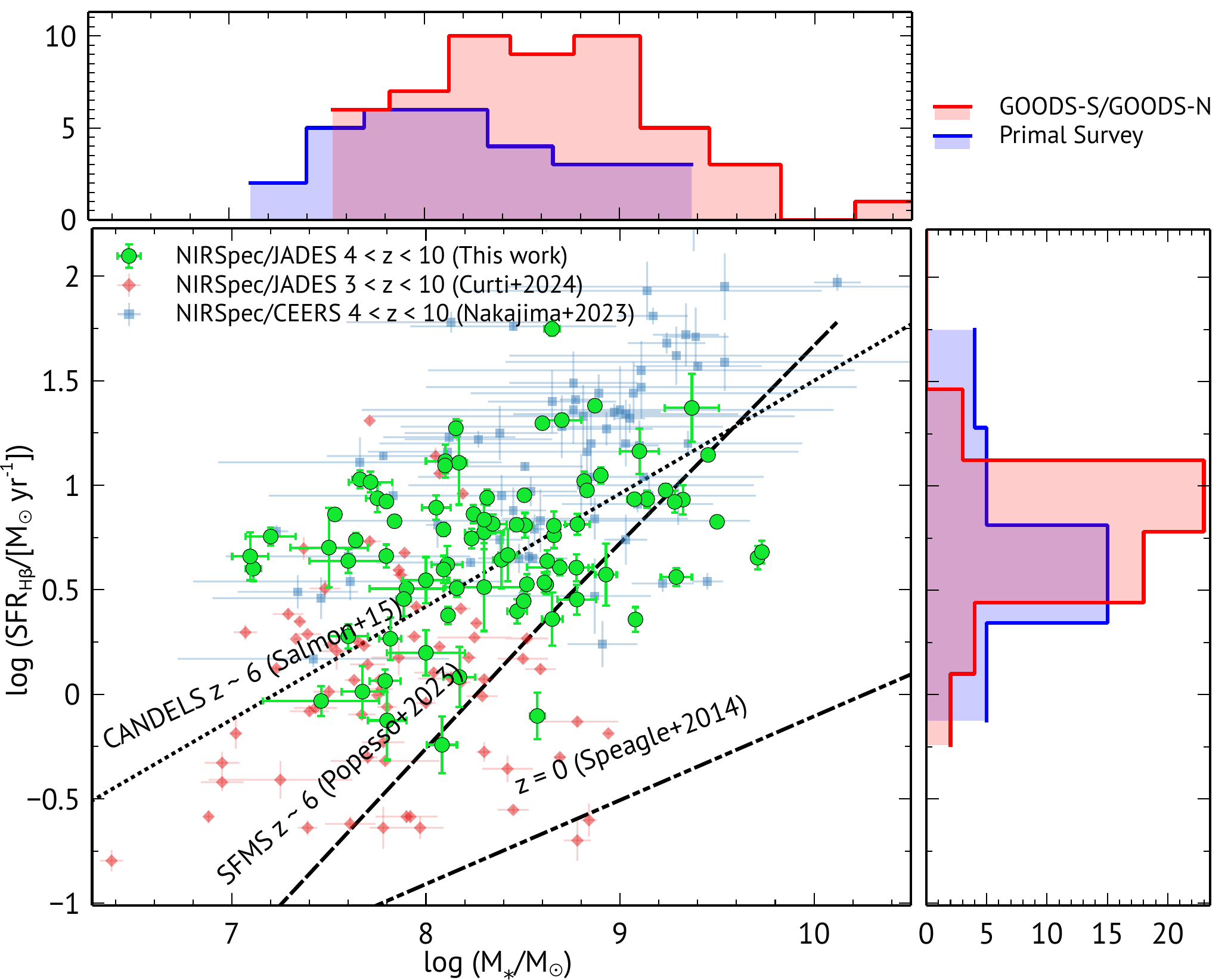}
    \caption{
Stellar mass versus star formation 
rate for our full sample of galaxies 
within the redshift range $4 < z < 10$ 
utilized in this study (green data 
points).
The top and side panels 
respectively display histograms 
of stellar mass and SFR 
distributions. 
The red histogram represents data
from GOODS-N and GOODS-S 
Data Release 3, 
while the blue histogram 
combines data from the PRIMAL survey
\citep{2024arXiv240402211H} with 
additional high-redshift galaxies from 
the literature. We also compared this 
distribution with prior high-redshift 
studies: 
the NIRSpec/CEERS survey for $4 < z < 
10$ galaxies 
\citep{2023ApJS..269...33N}, 
the NIRSpec/JADES survey for $3 < z < 
10$ galaxies 
\citep{2024A&A...684A..75C}. 
The main sequence star formation
rates from the
CANDELS survey at $z \sim 6$ galaxies 
\citep{2015ApJ...799..183S}, 
star-forming galaxies (SFGs) at $z 
\sim 6$ \citep{2023MNRAS.519.1526P}, 
and studies of the local universe at 
$z \sim 0$ 
\citep{2014ApJS..214...15S}. 
The plotted data points from 
\citet{2023ApJS..269...33N} and 
\citet{2024A&A...684A..75C} are based 
on H$\alpha$(H$\beta$) derived SFRs, 
originally calibrated for a 
\citet{2003PASP..115..763C} IMF.
For consistency with our analysis, 
we have scaled these SFRs to match 
the \citet{2002Sci...295...82K} 
IMF used in this study, using a 
scaling factor of 1.06
from \citet{2014ARA&A..52..415M}.
 }
\label{fig:m_star}
\end{figure*}

For this work, 
we use H$\beta$ as an SFR 
tracer in lieu of the commonly
used H$\alpha$ luminosity, 
as 
it is the best indicator for 
ongoing ($\sim$10 Myr) star 
formation activity
\citep[e.g.,][]{2023NatAs...7.1517H,2023ApJS..269...33N}.
This choice is also necessitated
by the spectral coverage 
limitations of NIRSpec,
which do not extend to 
H$\alpha$ at redshifts 
\(z > 6.75\). 
The use of H$\beta$ allows 
us to maintain consistency
in SFR measurements across our
sample. 
The SFR is derived 
based on H$\beta$ luminosity
assuming a Kroupa IMF
\citep{2002Sci...295...82K}
as \citep{2023NatAs...7.1517H}
\begin{equation}\label{eq:sfr}
    {\rm SFR}_{H\beta} (M_{\odot} {\rm yr^{-1}}) = 5.5 \times 10^{-42}
L_{H\beta} ({\rm erg/s}) \times f_{H\alpha/H\beta},
\end{equation}
where $f_{H\alpha/H\beta}$ = 2.86
is
theoretical flux ratio
between $H_{\alpha}$ and 
$H_{\beta}$ assuming a Case B
recombination model at $T_e$ = 10$^4$
K \citep{2006agna.book.....O}.
Figure \ref{fig:m_star} illustrates
the resulting
correlation between stellar
mass and SFR for our entire
sample of 81 galaxies, spanning
redshifts from \(z = 4\) to \(10\). 
The stellar mass is derived from
the $\bagpipes$ fit, as 
discussed in Section 
\ref{sec:prism_spectra_fitting}.
The SFR and stellar masses of
our sample is listed in
Table \ref{tab:goods}, 
\ref{tab:goodss}, \ref{tab:primal},
and \ref{tab:literat}
in Appendix.

Early results from the 
CEERS/NIRCam-selected 
galaxies indicate that the
increasing trend of the 
SFR–\(M_{\star}\) relation persists 
at least out to \(z = 8-9\) 
\citep{2023ApJ...949L..25F}. 
This trend aligns with predictions 
from simulations, which attribute 
it to the increased gas accretion 
rate onto dark matter halos at
higher redshifts 
\citep{2013ApJ...770...57B}.
We contextualize our 
findings by comparing them 
with datasets from several 
earlier surveys. 
This includes the JADES/NIRSpec 
surveys for galaxies within the 
redshift range \(z = 3-10\) 
\citep{2024A&A...684A..75C}, 
the CEERS/NIRSpec survey 
focusing on galaxies at \(z = 4-10\) 
\citep{2023ApJS..269...33N}. 
Additionally, we compare
our results with
the main sequence of star-forming 
galaxies at \(z \sim 6\), 
as documented in 
\citet{2015ApJ...799..183S} 
and \citet{2023MNRAS.519.1526P},
as well as at $z \sim 0$ 
\citep{2014ApJS..214...15S}.

Despite large scatter, 
our results
demonstrate a upward 
trend in SFR as stellar mass 
increases { (Spearman
coefficient of 0.4)}, consistent with 
the previous studies.
The galaxies in our sample exhibit
a distribution along the sequence
of specific star formation rates 
(sSFR) ranging from 10$^{-9}$ to 10$^{-7}$ yr\(^{-1}\). 
This distribution aligns well
with the star formation main 
sequence observed in galaxies
at \(z = 4 - 7\), where typical
sSFR values are between 
\(10^{-8.5}\) and \(10^{-6.5}\) yr\(^{-1}\) 
\citep{2013ApJ...763..129S, 2017ApJ...847...76S, 2023ApJS..269...33N}.

\subsection{Gas-phase Metallicity}\label{sec:metallicity}
We next estimate the gas-phase metallicities
for individual galaxies in our sample.
We use reddening-corrected emission 
line-fluxes
obtained from medium resolution gratings
to
determine the gas-phase metallicity
\citep{2019ARA&A..57..511K}.
There are two widely-used methods to 
determining the gas-phase metallicity:
(1) direct $T_e$-based method, which 
involves
accurate flux measurements of few key
emission
lines, such as [$\oii]\lambda$4363 
\citep{2024ApJ...962...24S},
(2) empirical method, which 
involves 
comparing line-ratios to 
empirical calibrations,
constructed using samples for 
which metallicity
has been derived using 
``direct method''
\citep[e.g.,][]{2020MNRAS.491..944C,2023ApJS..269...33N}.
Several previous studies
presented
empirical 
calibrations,
such as, 
\citet{2004MNRAS.348L..59P,2008A&A...488..463M,2013A&A...559A.114M,2017MNRAS.465.1384C,2018ApJ...859..175B,2020MNRAS.491..944C,2022ApJS..262....3N,2024ApJ...962...24S}

For this paper,
we adopt the revisited calibrations 
derived by 
\citet{2020MNRAS.491..944C}
and later compared with the
calibrations by 
\citet{2022ApJS..262....3N}. 
Both of these calibrators are
widely used
to measure the metallcity at
high-redshift galaxies
\citep[e.g.,][]{2024A&A...684A..75C,2024arXiv240303977V}.
Their methodology involved 
utilizing the largest sample of 
extremely metal-poor galaxies (EMPGs)
and stacking a sample of 
SDSS galaxies to accurately
measure metallicity based entirely
on the reliable measurements of the 
direct $T_e$ method.

We primarily use R3 index
for measuring metallicity for 
individual
galaxies.
A limitation of using 
R3 index as a metallicity 
callibrator is that it can
yield two different metallicity
estimates for the same R3 value 
\citep{2020MNRAS.491..944C,1979A&A....80..155L,2024ApJ...962...24S}.
If the metallicities obtained from 
the R3 index are close, 
typically occurring around the 
peak value near 
12 + log(O/H) = 8.0, we 
adopt the  $1\sigma$ lower limit 
from the low-metallicity solution 
as the lower bound and the $1\sigma$
upper limit from the
high-metallicity solution as
the upper bound, 
following the approach in 
\citet{2023ApJS..269...33N}. 
When encountering two 
well-separated metallicity solutions, 
we utilize the O32 index to 
differentiate between the degenerate
values of metallicities.
If the [$\oii$]$\lambda\lambda$3727,29 
line is not detected with $\geq$ 
3$\sigma$ level, 
it is not possible to calculate
the O32 index,
we instead utilize the N2 and S2 indices 
to
break the degeneracy whenever available.
Similar methods for breaking the
degeneracy between two metallicity
solutions has been adopted by
\citet{2023ApJS..269...33N} and
\citet{2024A&A...684A..75C}.

An alternative technique involves 
using the equivalent width of 
H$\beta$ (EW(H$\beta$)) to estimate 
O32 when [$\oii$]$\lambda\lambda$3727,29
is not detected at a 3$\sigma$ level, 
based on the average relationship identified 
by \citet{2022ApJS..262....3N}. 
This method can help resolve the 
metallicity degeneracy stemming 
from the R3 calibration. 
However, we chose not to employ this 
technique in this work
to avoid introducing potential 
systematic uncertainties in our 
metallicity measurements. 
For galaxies where we could not 
observationally determine O32, N2, 
or S2 due to a non-detection 
of one or more of the lines involved
at 3$\sigma$ level,
we left the metallicities for those
galaxies unconstrained and
excluded them from our 
final sample.
Our final data set
for analyzing the MZR includes
a total of 81 
star-forming galaxies,
comprising 54 galaxies from 
the GOODS-S and GOODS-N fields,
22 galaxies reported in the 
JWST Primal Survey,
and an additional 5 galaxies 
selected from the literature.
Metallicities of our sample
of galaxies are listed in
Table \ref{tab:goods}, 
\ref{tab:goodss}, \ref{tab:primal},
and \ref{tab:literat}
in Appendix.


\section{The mass-metallicity relation}\label{sec:MZR}
\begin{figure*}
    \centering
\includegraphics[width=1.0\textwidth]{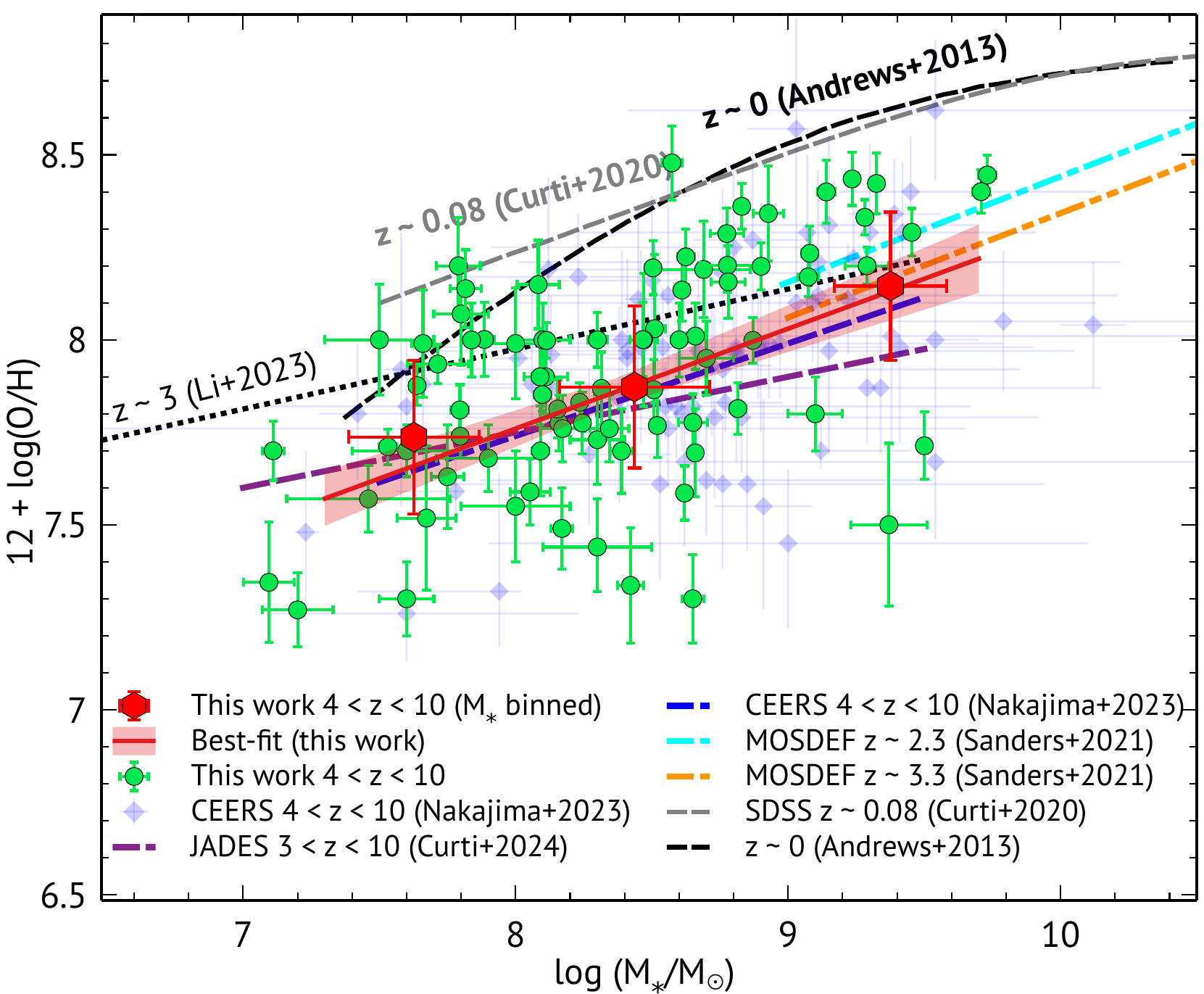}
    \caption{
    Stellar mass–gas-phase metallicity relation for our full sample with $4<z<10$. The green circles represent individual galaxies analyzed in this study.  The green data points represent individual galaxies analyzed in this study. The sample is divided into three stellar mass ranges: $M_{\ast} = 10^7-10^8$, $10^8-10^9$, and $10^9-10^{10}$ $M_{\odot}$, with the average points shown as large red circles. The best-fit regression line for our complete sample, accompanied by its $1\sigma$ uncertainty, is illustrated by the red line and the shaded region around it.
    We overplot our results with prior high-redshift and lower-redshift MZR measurements. Specifically, at high redshift, 
   we reference the NIRSpec/CEERS survey for $4 < z < 10$ galaxies shown with purple inverted squares \citep{2023ApJS..269...33N}, 
the NIRSpec/JADES survey for $3 < z < 10$ galaxies \citep{2024A&A...684A..75C} shown with purple dashed line.  At lower redshifts, we compare with the MZR estimated by \citet{2013ApJ...765..140A} at $z \sim 0$ (black dashed line) and \citet{2020MNRAS.491..944C} at $z \sim 0.08$ (gray dashed line), both using star-forming galaxies from the SDSS. Additionally, we reference \citet{2021ApJ...914...19S}, who estimated the MZR from the MOSDEF survey at $z \sim 2.3$ (blue dashed line) and $z \sim 3.3$ (orange dashed line), as well as \citet{2023ApJ...955L..18L}, who estimated the MZR for dwarf galaxies at $z \sim 3$ (black dotted line).}
\label{fig:massmet}
\end{figure*}

One of the primary objectives of this
paper is to study the evolution of
metallicity in star-forming galaxies.
In this section, we present the 
stellar mass-metallicity relation (or MZR)
at the high-redshift using our final
smaple of galaxies.
Specifically, 
we use the metallicity 
measurements steps discussed in 
Section \ref{sec:metallicity} and the 
stellar mass measurements 
derived from the $\bagpipes$ fit 
(Section \ref{sec:prism_spectra_fitting})
to probe the MZR.
Figure \ref{fig:massmet} 
illustrates our sample of galaxies in
the stellar mass-metallicity plane.
\citet{2021ApJ...914...19S} showed
that the average MZR can be approximated
as 
\begin{equation}\label{eq:metal}
12 + \log(\text{O/H}) = \gamma \times \log\left(\frac{M_\star}{10^{10} M_\odot}\right) + Z_{10}
\end{equation}
where the slope $\gamma$ and offset $Z_{10}$
(the gas-phase metallicity 
at a stellar mass of $10^{10} M_\odot$)
can be estimated by fitting to observed
MZR.

\begin{table}
   \caption{Comparing MZR with 
   different studies following 
   equation \ref{eq:metal}. 
   $Z_{8}$ is converted to 
   $Z_{10}$ using $Z_{10} = 2\gamma + Z_{8}$.}
    \centering
    \setlength{\tabcolsep}{5pt}
    \begin{tabular}{llcc}
        
        Study & $z$ range & $\gamma$ & $Z_{10}$ \\
        \hline
        This work & 4--10 & 0.27 $\pm$ 0.02 & 8.28 $\pm$ 0.08 \\
         & 4--6 & 0.28 $\pm$ 0.03 & 8.37 $\pm$ 0.13 \\
        & 6--8 & 0.23 $\pm$ 0.04 & 8.29 $\pm$ 0.18\\
        & 8--10 & 0.21 $\pm$ 0.04 & 8.09 $\pm$ 0.11\\
        \hline
        \citet{2023ApJS..269...33N} & 4--10 & 0.25 $\pm$ 0.03 & 8.24 $\pm$ 0.05 \\
       \hline
       \citet{curti2023jadesinsightslowmassend} & 3--10 & 0.17 $\pm$ 0.03 & 8.06 $\pm$ 0.18 \\
       & 3--6 & 0.18 $\pm$ 0.03 & 8.11 $\pm$ 0.17 \\
       & 6--10 & 0.18 $\pm$ 0.03 & 7.87 $\pm$ 0.45 \\
        \hline
        \citet{2021ApJ...914...19S} & 0 & 0.28 $\pm$ 0.01 & 8.77 $\pm$ 0.01 \\
       & 2.3 & 0.30 $\pm$ 0.02 & 8.51 $\pm$ 0.02 \\
       & 3.3 & 0.29 $\pm$ 0.02 & 8.41 $\pm$ 0.03 \\
        \hline
        \citet{2023ApJ...955L..18L} &  2 & 0.16 $\pm$ 0.02 & 8.50 $\pm$ 0.13 \\
       & 3 & 0.16 $\pm$ 0.01 & 8.40 $\pm$ 0.06 \\
        \hline
        \citet{2023NatAs...7.1517H} & 7--10 & 0.33 & 7.95 \\
        \hline
        \citet{2024ApJ...960L..13H} & 1.9 & 0.23 $\pm$ 0.03 & 8.54 $\pm$ 0.12 \\
       & 2.88 & 0.26 $\pm$ 0.04 & 8.57 $\pm$ 0.15 \\
        \hline
    \end{tabular}
    \label{tab:mzr_comp}
\end{table}

We divide our full sample 
($4 < z < 10$) into three stellar 
mass ranges: $M_{\ast} = 10^7-10^8$, 
$10^8-10^9$, and $10^9-10^{10}$ solar 
masses.
In Figure \ref{fig:massmet}, 
these average points are 
represented by large red circles. 
For our full sample, we determine 
the best-fit slope of 
$\gamma = 0.27 \pm 0.02$ and 
metallicity intercept 
$Z_{10} = 8.28 \pm 0.08$.
The best-fit regression line, 
along with its 1$\sigma$ uncertainty,
is shown in Figure \ref{fig:massmet}.
We next compare our results with 
that of previous studies of the 
MZR at lower 
redshifts. 
Specifically, 
\citet{2020MNRAS.491..944C}, 
who analyzed SDSS galaxies in the local 
universe using a set of strong-line 
diagnostics calibrated on a fully 
$T_e$-based abundance scale, 
covering the full range of stellar
mass and star formation rates spanned
by SDSS galaxies. 
Additionally, we compare our MZR
findings with those of 
\citet{2021ApJ...914...19S}, who 
investigated the MZR for galaxies over the 
range $z = 0 - 3.3$,
utilizing samples of approximately 
300 galaxies at $z \sim 2.3$ and 
150 galaxies at $z \sim 3.3$ from 
the MOSDEF survey.
Furthermore, we compare our results
with the MZR obtained by 
\citet{2023ApJ...955L..18L} for a 
sample of 51 dwarf galaxies at
$z = 2-3$, using Near-Infrared Imager 
and Slitless Spectrograph (NIRISS) 
grism spectroscopy from JWST observations
in the A2744 and 
SMACS J0723-3732 fields. 
Table \ref{tab:mzr_comp} shows
the comparison of $\gamma$ and $Z_{10}$
with that of several previous works.
The best-fit slope and 
normalizations of our sample of
galaxies are consistent
with that of high-redshift
CEERS galaxies \citep{2023ApJS..269...33N}
and also with low-redshift 
galaxies at $z$ = 0--3.3 by
\citet{2021ApJ...914...19S}
and \citet{2024ApJ...960L..13H}, 
however slightly higher than
previous JADES studies
\citep{2024A&A...684A..75C}.

Figure \ref{fig:massmet}
clearly demonstrates that
the metallicity at a given stellar
mass
in our sample is noticably lower 
compared 
to the reference curve for
\(z \sim 0\) from 
\citet{2013ApJ...765..140A}
and for $z \sim 0.08$ from
\citet{2020MNRAS.491..944C}. 
This difference is dependent on
stellar mass, showing a variation 
of $\sim$ 0.4 dex at \(10^8 \, M_\odot\) 
and  $\sim$ 0.5 dex at \(10^9 \, M_\odot\). 
Compared to the MZR reported
by 
\citet{2023ApJ...955L..18L} for 
galaxies at \(z \sim 2-3\), 
our higher redshift sample exhibits a 
less prominent reduction in metallicity
than the case of   $z \sim 0$. 
The extent of this reduction is also
mass-dependent, varying from $\sim$ 
0.2 dex at \(10^8 \, M_\odot\) 
to $\sim$ 0.08 dex at 
\(10^9 \, M_\odot\).
A comparison with the MZR from 
\citet{2021ApJ...914...19S} at
\(z \sim 2.3\) and \(z \sim 3.3\) 
reveals that the metallicity of
our sample is marginally lower, 
by $\sim$ 0.12 dex at \(z \sim 2.3\) 
and $\sim$ 0.03 dex at \(z \sim 3.3\), 
for a stellar masses of 
\(10^9 \, M_\odot\).

We also compare our results 
with the
previous studies that have
examined the mass-metallicity 
relation at redshifts up
to $z \sim 10$. 
\citet{2023ApJS..269...33N} 
performed an in-depth analysis 
of 135 galaxies using 
JWST/NIRSpec data from the 
ERO, GLASS, and CEERS programs, 
thereby illustrating the evolution 
of the mass-metallicity  relations 
over the redshift range $z = 4-10$. 
\citet{2024A&A...684A..75C} 
utilized JWST/NIRSpec observations
from the JADES deep GOODS-S tier to 
conduct a comprehensive analysis of
the gas-phase metallicity properties
in a sample of 66 low-stellar-mass 
galaxies (log $M_\star/M_\odot \lesssim 9$) 
within the redshift range $3 < z < 10$.
Despite minor variations in 
redshift intervals, 
our best-fit MZR for galaxies 
at \(z = 4-10\) closely aligns with
the MZR observed by 
\citet{2023ApJS..269...33N},
however steeper than that of
\citet{2024A&A...684A..75C}, 
as listed in
Table \ref{tab:mzr_comp}.
\citet{2023ApJS..269...33N} reported 
values of 
\(\gamma = 0.25 \pm 0.03\) and 
\(Z_{10} = 8.24 \pm 0.05\) for
galaxies within \(4 < z < 10\),
while \citet{2024A&A...684A..75C} 
reported 
\(\gamma = 0.17 \pm 0.02\) and 
\(Z_{10} = 8.06 \pm 0.18\) for
galaxies within \(3 < z < 10\), 
derived from their \(\beta\) 
factor slope and gas-phase metallicity 
at a stellar mass of 
\(10^8 \, M_\odot\). 

\begin{figure*}
\centering
\begin{tabular}{cc}
\includegraphics[width=0.5\textwidth]{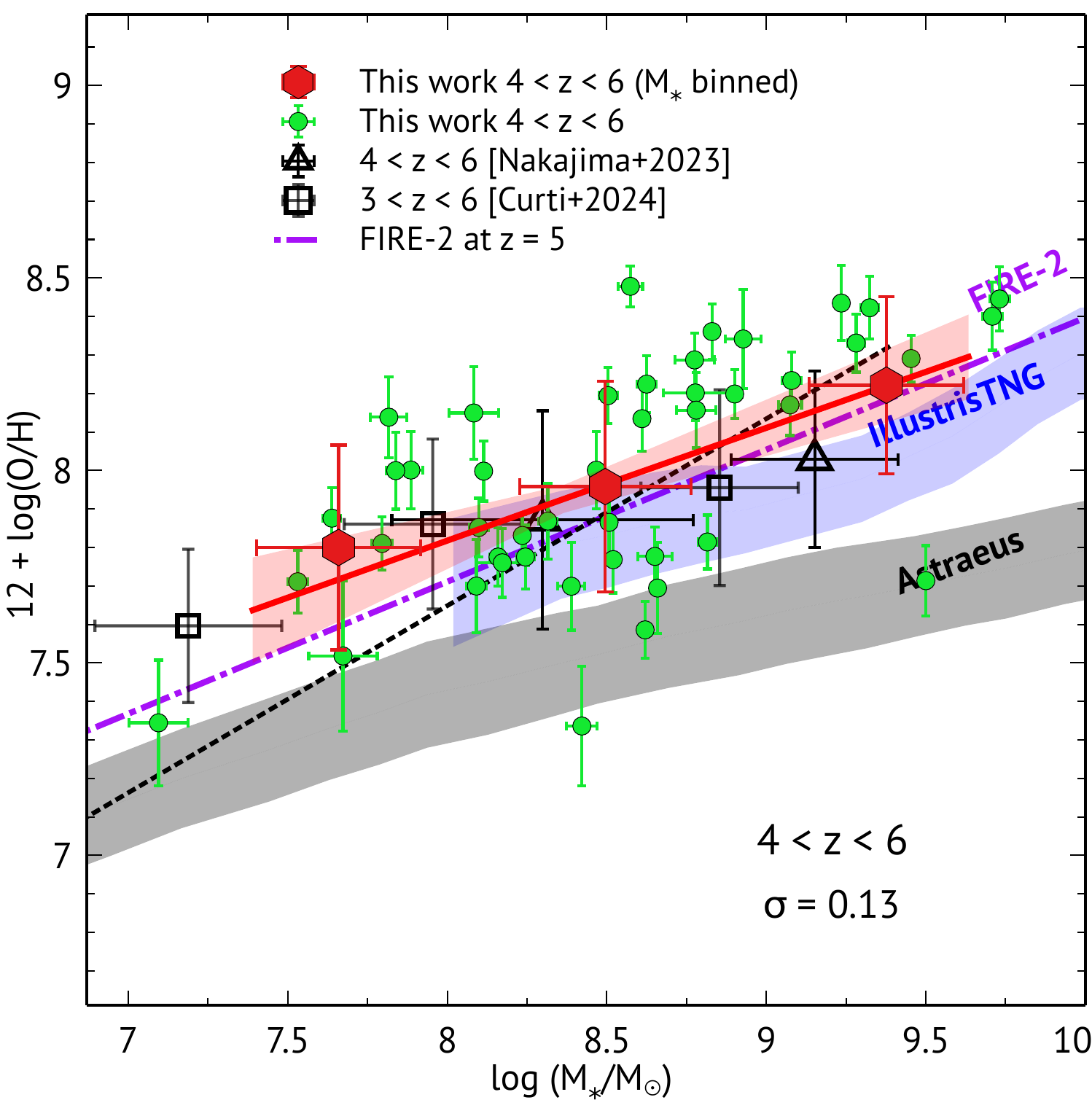} &   \includegraphics[width=0.5\textwidth]{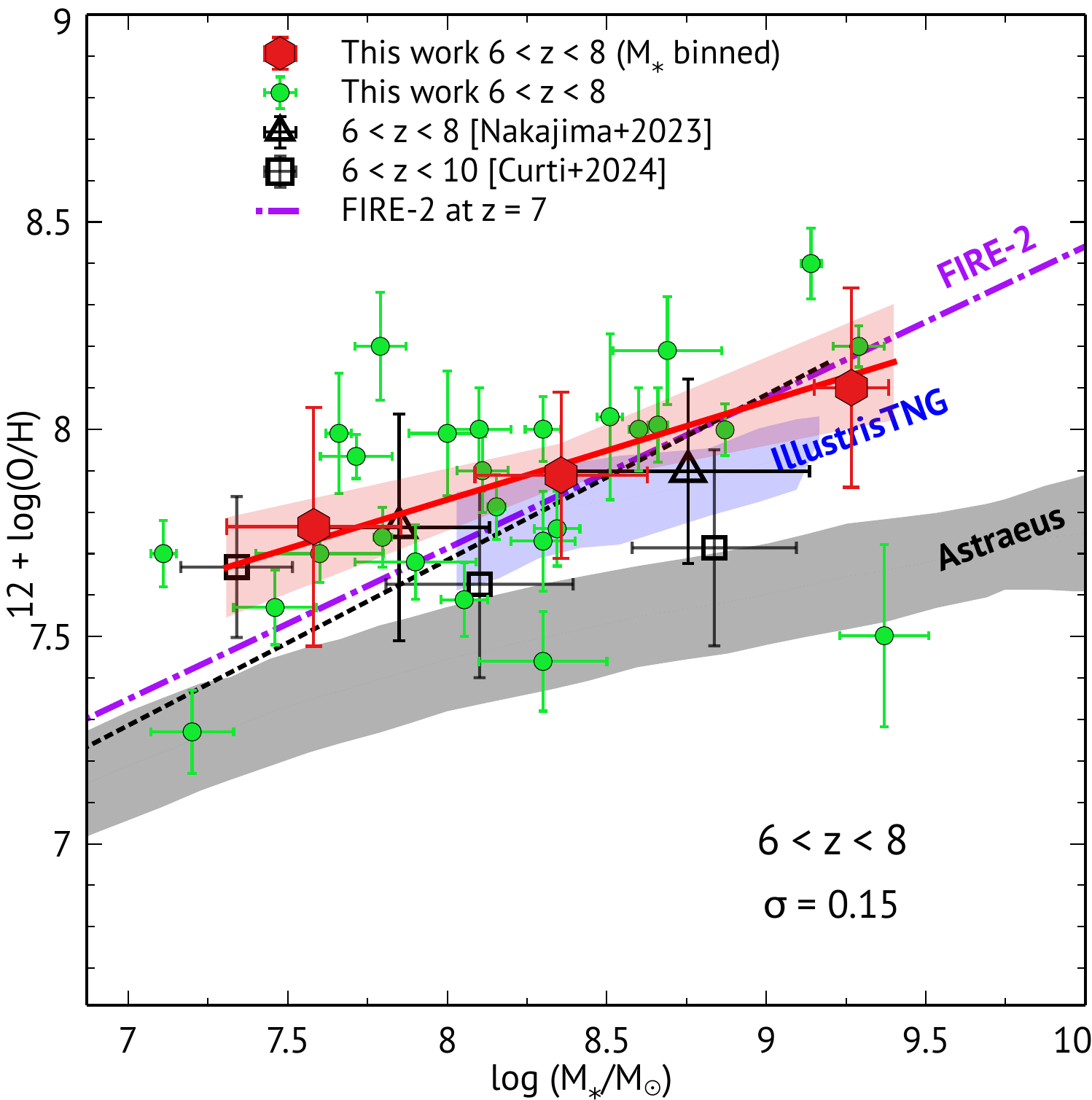}\\
\end{tabular}  
\begin{tabular}{c}
\includegraphics[width=0.5\textwidth]{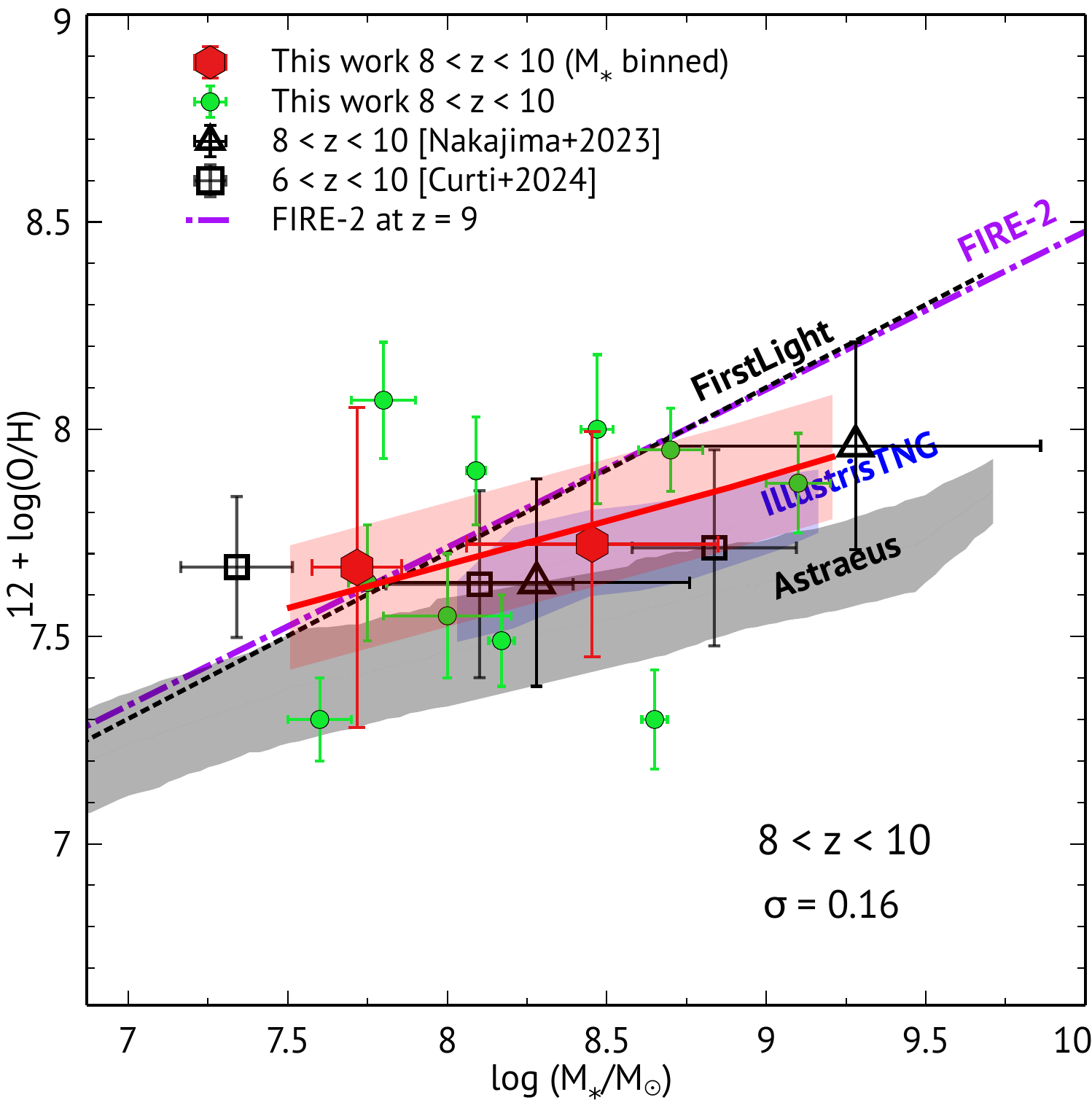}\\
\end{tabular}
    \caption{The mass-metallicity relation (MZR) is shown for three redshift 
    intervals: 4--6, 6--8, and 
    8--10. For the $z$=4--6 and $z$=6--8 
    intervals, the sample is divided into 
    three stellar mass ranges: $M_{\ast} = 
    10^7-10^8$, $10^8-10^9$, and 
    $10^9-10^{10}$ $M_{\odot}$, with the 
    averages depicted as large red 
    circles. In the $z= 8-10$ interval, 
    due to the presence of only a single 
    data point for $M_{\ast} > 10^9$ 
    $M_{\odot}$, we compute averages for 
    the $M_{\ast} = 10^7-10^8$ and 
    $10^8-10^9$ ranges. The red circles 
    indicate the mass-averaged values for 
    each redshift interval, whereas the 
    solid red lines represent the best-fit 
    regression curves for each respective 
    interval. 
    { The standard deviations
    of the residuals between the observations and 
    the best-fit curves
    are represented by $\sigma$.}
    See table \ref{tab:mzr_comp} 
    for the best-fit slopes and 
    metallicity intercepts at each 
    redshift intervals. We incorporate a 
    comparative analysis of the MZR 
    redshift evolution with findings from 
    \citet{2023ApJS..269...33N}, spanning 
    the intervals $4-6$, $6-8$, and $8-10$ 
    (shown with open black triangles), as 
    well as from 
    \citet{2024A&A...684A..75C}, covering 
    the intervals $4-6$ and $6-10$ (shown 
    with open black squares).
    The cosmological simulation outcomes 
     for $z = 5$, 7, and 9 are
     also overplotted in the
     upper-left, upper-right, 
     and bottom panels respectively. 
     These include results from the
     FIRE-2 (dashed purple line, 
     \citealt{2024ApJ...967L..41M}), 
     FirstLight (black dashed line, 
     \citealt{2020MNRAS.494.1988L}), 
     IllustrisTNG (blue shaded region, 
     \citealt{2019MNRAS.484.5587T}), and 
     Astraeus (grey shaded region, 
     \citealt{2023MNRAS.518.3557U}) 
     simulations. 
     Extrapolated 
     IllustrisTNG simulation
     in $8 < z < 10$ are taken
     from \citet{2023ApJS..269...33N}.
     }
\label{fig:massmet_binned}
\end{figure*}

To investigate the redshift evolution
of  MZR within the range of
$4 < z < 10$, 
we divided our sample into three 
subsamples: $z$ = 4--6, 6--8,
and 8--10. 
Our objective was to determine
whether there is a noticeable 
change in the slope \(\gamma\) 
and normalization
of the MZR across these redshift 
intervals.
Table \ref{tab:mzr_comp} compares 
the $\gamma$ and $Z_{10}$ 
values for these
redshift bins, 
as well as for the complete 
sample spanning \(z = 4\) to \(10\).
Figure \ref{fig:massmet_binned} 
presents 
the MZR for the three redshift 
intervals. 
Within these intervals, galaxies are
binned by mass, with the average mass 
in each bin shown with red data points
with error bars.
We observe that the \(\gamma\) value
for the \(z = 4\) to \(6\) interval 
is slightly higher 
(\(\gamma = 0.28 \pm 0.03\)) 
compared to the \(z = 6\) to \(8\) 
(\(\gamma = 0.23 \pm 0.04\))
and $z$ = 8--10 ($\gamma$ = 0.21
$\pm$ 0.04),
indicating a trend towards 
diminished metallicity and 
a flattening of the MZR slope as 
redshift increases.
Such trend of decreasing
metallicity in the highest-redshift
bin is consistent with previous 
JWST
studies by \citet{2023ApJS..269...33N} 
and \citet{2024A&A...684A..75C}, 
who observed similar trends
at higher redshifts.
The evolution of MZR in our
sample is also consistent with
simulations, as further discussed
in Section \ref{sec:simulation}.

Given the limited size of our 
sample, conducting a statistically
robust evaluation of the scatter 
in the scaling relationship remains 
challenging. 
However, we can get an estimate 
of the intrinsic scatter using the 
equation:
\begin{equation}\label{eq:scatter}
    \sigma_{\rm scatter} = \sqrt{\sigma_{\rm obs}^2 - \sigma_{\rm measured}^2}
\end{equation}

where \(\sigma_{\text{obs}}\) 
represents the observed scatter
in the full sample
around the best-fit MZR line, 
and \(\sigma_{\text{measured}}\) 
indicates the average
uncertainty in the 
metallicity measurement.

We estimate \(\sigma_{\text{scatter}}\)
to be approximately 0.16 dex for our 
entire sample of galaxies. 
This aligns with the 
intrinsic scatter 
of 0.16–0.18 dex found in dwarf galaxies
at redshifts \(z = 2\) to \(3\) with
stellar masses between \(10^8\) and
\(10^9 \, M_\odot\) 
\citep{2023ApJ...955L..18L};
however larger than the intrinsic 
scatter of 0.08 dex observed in 
low-stellar-mass 
(log \(M_\star / M_\odot \lesssim 9\)) 
galaxies at redshifts \(3 < z < 10\) 
\citep{2024A&A...684A..75C}.

\section{The Fundamental Metallicity Relation}\label{FMR}

The scaling relations observed in galaxies are central to 
understanding galaxy formation and evolution.
Among them, the SFR-MZ relation is particularly significant
as it relates metallicity with stellar mass
and SFR \citep[e.g.,][]{2008ApJ...672L.107E,2010MNRAS.408.2115M,2010A&A...521L..53L,2013ApJ...765..140A,  2020ApJ...892..125H, 2021ApJ...914...19S, 2023ApJ...955L..18L, 2024A&A...684A..75C}. Using a sample of  43,690  SDSS galaxies, 
\citet{2008ApJ...672L.107E} identified 
that galaxies with higher SFRs systematically
show lower metallicities compared to their
less star-forming counterparts of the same 
stellar mass. Expanding on this, 
\citet{2010MNRAS.408.2115M} and 
\citet{2010A&A...521L..53L} systematically 
demonstrated that incorporating SFR into the MZR 
significantly reduces its scatter.
{ However, 
in a recent study 
using the MaNGA survey 
and TNG50 simulations, 
\citet{2024ApJ...971L..14M} 
show that SFR
alone does not 
significantly reduce 
scatter in the MZR. 
They further demonstrate
that including galaxy size,
along with SFR, 
significantly reduces 
this scatter.
}

Investigations of large galaxy samples
from $z$ = 3 to 0 have shown minimal
redshift evolution in the SFR-MZ relation for
galaxies with masses above
10$^{8} M_{\odot}$
\citep{2013ApJ...776L..27H, 2015ApJ...799..138S, 2020MNRAS.491..944C, 2021ApJ...919..143H}.
{
Au contraire,
\citet{2024A&A...683A.203P}
found
a modest yet statistically 
significant evolution
in the MZR and FMR up to
z $\sim$ 0.63,
highlighting the significance of 
stellar
mass and SFR on gas-phase metallicity 
regardless of cosmic redshift.}
Multiple expressions describe the 
interdependencies of these three
properties \citep{2010MNRAS.408.2115M, 2010A&A...521L..53L}.

We utilize the method of  
\citet{2010MNRAS.408.2115M}  to
parameterize the SFR-MZ relation by determining the 
value of \(\alpha\) that minimizes the
scatter in the oxygen-to-hydrogen ratio (O/H)
at a fixed \(\mu_\alpha\), defined as: 
\begin{equation}
\mu_\alpha = \log\left(\frac{M_{\star}}{M_\odot}\right) - \alpha \log\left(\frac{\text{SFR}}{M_\odot \, \text{yr}^{-1}}\right).
\end{equation}

\begin{figure}
\centering     \includegraphics[width=0.5\textwidth]{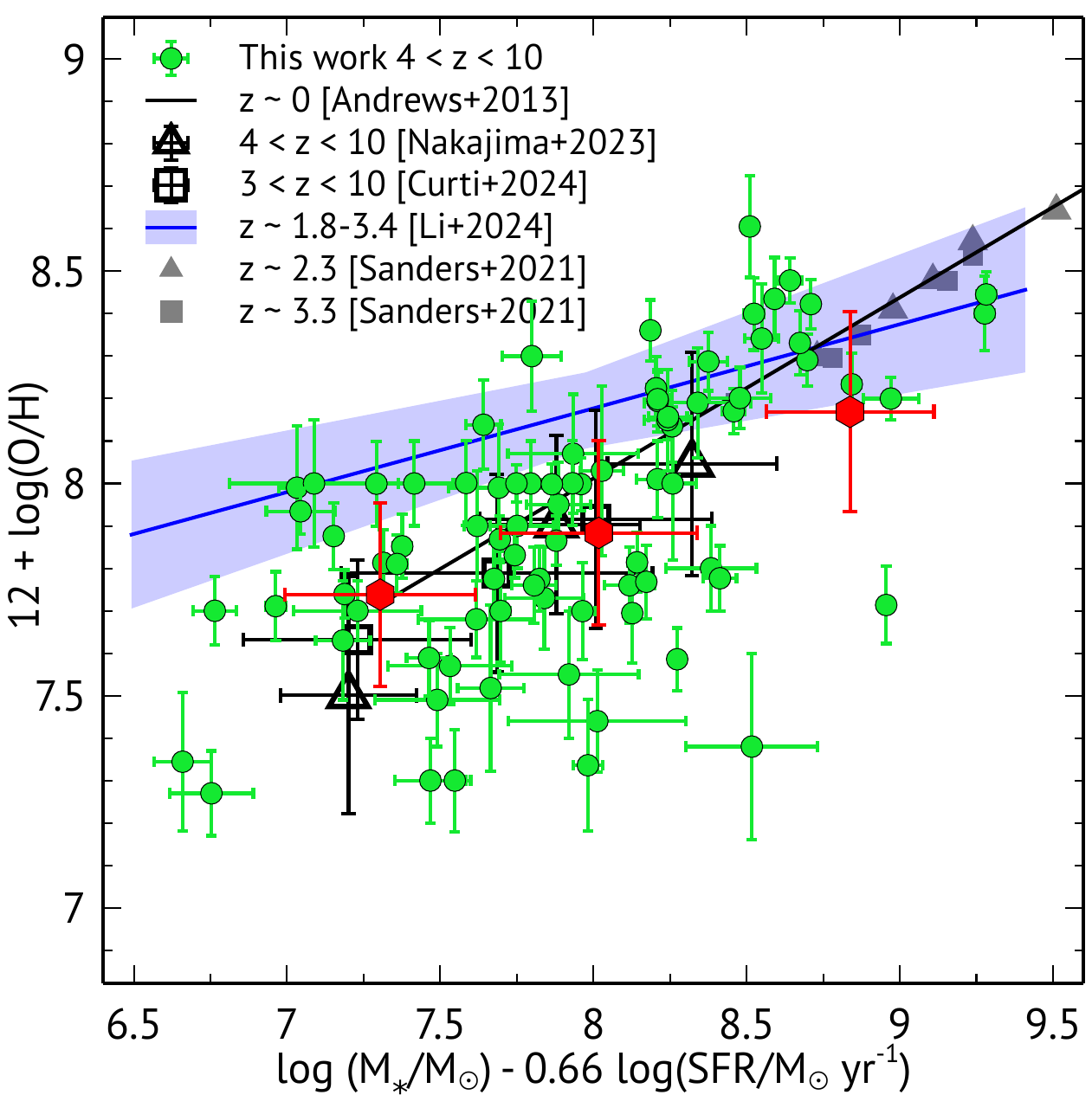} \\
    \caption{
  Projection of the SFR-MZ relation in O/H versus $\mu_{0.66} = \log(M_{\ast}/M_{\odot}) - 0.66 \times \log(\text{SFR}/M_{\odot} \ \text{yr}^{-1})$ plane, where $\alpha$ = 0.66 minimizes the scatter of the local low-metallicity galaxies. The green data points represent our entire sample, while the red hexagons indicate the weighted averages within the stellar mass ranges of $M_{\ast} = 10^7 - 10^8$, $10^8 - 10^9$, and $10^9 - 10^{10} \ M_{\odot}$. The solid black line represents the SFR-MZ relation at $z \sim 0$ from \citet{2013ApJ...765..140A}, while the grey filled triangles and squares indicate the $z \sim 2.3$ and $z \sim 3.3$ SFR-MZ relations from \citet{2021ApJ...914...19S}. The solid blue line depicts the SFR-MZ relation for dwarf galaxies at $z \sim 2-3$ from \citet{2023ApJ...955L..18L}, with the blue shaded region showing its 1$\sigma$ uncertainty.}
\label{fig:fmr}
\end{figure}

 \citet{2013ApJ...765..140A} showed
that
$\alpha$ = 0.66 minimizes the scatter of 
the local low-metallicity galaxies 
with a direct $T_e$-based metallicity in the
$\mu_{\alpha}$-metallicity plane, resulting
in
\begin{equation}\label{eq:am13}
    12 + {\rm log(O/H)} = 0.43 \times \mu_{0.66} + 4.58.
\end{equation}
\citet{2013ApJ...765..140A} measurements extends
down to log~($M_{\ast}/M_{\odot}$) $\sim$ 7.4.
This is advantageous for comparing results 
with our sample since our sample has lower
mass cutoff of around $\sim$ 7.1 $\pm$ 0.1. 
Figure \ref{fig:fmr}  shows the
SFR-MZ relation of galaxies our full sample on the $\mu_{\alpha}$-metallicity 
plane. 
The green data points represent
our full sample, while the red 
hexagon data points indicate the 
weighted average in the stellar mass 
ranges: $M_{\ast} = 10^7 - 10^8$, 
$10^8 - 10^9$, and $10^9 - 10^{10} \ 
M_{\odot}$.
We also show that 
the $z \sim 0$
relation 
by \citet{2013ApJ...765..140A}, as 
defined in Equation \ref{eq:am13}. Our 
results are compared with the MOSDEF 
study at $z \sim 2.3$ and $z \sim 
3.3$, as detailed in  
\citet{2021ApJ...914...19S}. 
Additionally, we compare our findings 
with two recent JWST studies at 
similar redshift ranges, such as CEERS 
($4 < z < 10$; 
\citealt{2023ApJS..269...33N}) and 
JADES+CEERS ($3 < z < 10$; 
\citealt{2024A&A...684A..75C}). 
We find that the average SFR-MZ 
relation of our sample of galaxies 
overlaps with those of previous JWST 
studies, the SFR-MZ relation at $z 
\sim 0$ by 
\citet{2013ApJ...765..140A}, and the 
relation derived from MOSDEF galaxies 
at $z \sim 2.3$ and $z \sim 3.3$ 
\citep{2021ApJ...914...19S}.

Two other forms of the SFR-MZ relation 
have been derived by 
\citet{2020MNRAS.491..944C} and 
\citet{2021ApJ...914...19S} at $z \sim 
0$. 
Specifically, 
\citet{2020MNRAS.491..944C} and 
\citet{2021ApJ...914...19S} identified 
$\alpha = 0.55$ and $\alpha = 0.6$, 
respectively, as yielding the most 
precise 2D projection of the 
fundamental metallicity relation on 
the O/H versus $\mu_{\alpha}$ plane, 
with minimal scatter, derived from 
composite SDSS spectra. We utilize the 
SFR-MZ relation from 
\citet{2013ApJ...765..140A} with 
$\alpha = 0.66$ due to its alignment 
with the parameter space examined in 
this study. 
Specifically, the metallicity 
measurements in 
\citet{2013ApJ...765..140A} span
three decades in stellar mass, from 
$\log(M_{\ast}/M_{\odot})  \simeq  
7.4$ to $10.5$, with 78 out of 81 
galaxies in our sample falling within 
this range.  
In contrast, 
\citet{2020MNRAS.491..944C} and 
\citet{2021ApJ...914...19S} span the 
higher stellar mass range, from 
$\log(M_{\ast}/M_{\odot}) \simeq 8.5$ 
to $11.4$, which is outside the range 
of the majority of our galaxy stellar 
masses.

\begin{figure}
\centering
  \includegraphics[width=0.5\textwidth]{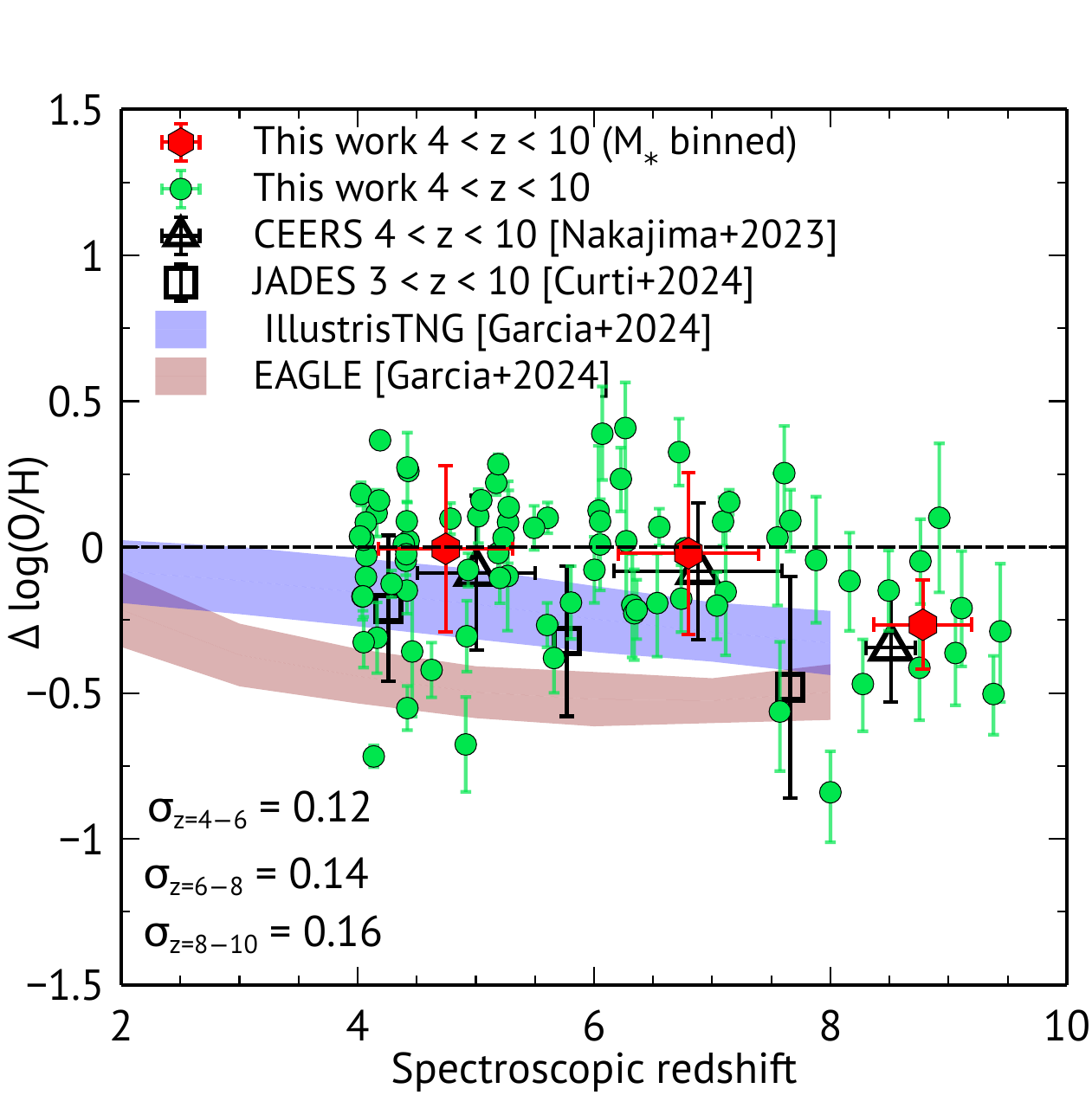}\\

    \caption{ The discrepancies in metallicity between our sample galaxies and the predictions of the local Fundamental Metallicity Relation (FMR) by \citet{2013ApJ...765..140A} as a function of redshift. The green data points represent our full sample, while the red data points indicate the weighted average of the discrepancies in three redshift bins: $z = 4-6$, $z = 6-8$, and $z = 8-10$. 
    { The standard deviations
    of the 
    residuals in each redshift
    bin are denoted by $\sigma$.}
    Redshift-binned deviations from the local FMR, taken from \citet{2023ApJS..269...33N} for three bins of redshift $z = 4-6$, $z = 6-8$, and $z = 8-10$, and from \citet{2024A&A...684A..75C} for three bins of redshift $z = 3-5$, $z = 5-7$, and $z = 7-10$, are shown with open black triangles and open black squares, respectively. This plot demonstrates that metallicity remains relatively constant up to $z \sim 8$, but there is a marked decline in metallicity beyond $z > 8$. We also compared our
    results with IllustrisTNG (blue shade) and EAGLE (dark red shade)
    cosmological simulations.
     }
\label{fig:fmr_resid}
\end{figure}

We next investigate that if there
are any redshift evolution of the
SFR-MZ relations that we adopted
above.  
Figure \ref{fig:fmr_resid} 
shows the metallicity residuals at 
fixed $\mu_{0.66}$ of our $4<z<10$ 
sample compared to the SFR-MZ relation 
of \citet{2013ApJ...765..140A} at z 
$\sim$ 0. The metallicity residual is 
calculated as follows:
\begin{equation}
\Delta {\rm log(O/H)} = (12 + {\rm log(O/H)){_{obs}}} - (12 + {\rm log(O/H)){_{\alpha =0.66}}}.
\end{equation}

The green circles represent the full 
sample of $4 < z < 10$. We also 
divided the $z = 4-10$ sample into 
three redshift bins: $z = 4-6$, $z = 6-
8$, and $z = 8-10$, and calculated 
their weighted-mean offset from the 
local \citet{2013ApJ...765..140A} SFR-
MZ relation, which are shown with red 
hexagons. We find the weighted-mean 
offset for the $z = 4-6$ bin to be 
$\Delta \log(\text{O/H}) \sim -0.01$ 
dex, while for the $z = 6-8$ bin, it 
is $\Delta \log(\text{O/H}) \sim 
-0.02$ dex. This consistency suggests 
that a unified relation between SFR 
and MZ accurately characterizes the 
average properties of galaxy samples 
from $z = 0$ to $z \sim 8$, indicating 
weak/no
significant redshift evolution of the 
SFR-MZ relations up to
$z \sim$ 8. However, a significant 
decrease in metallicity is observed 
beyond $z > 8$ surpassing the error 
margins, with $\Delta \log(\text{O/H}) 
\sim -0.27$ dex. 

A similar conclusion has been reached 
by \citet{2023ApJS..269...33N} for 
their CEERS galaxy sample within the 
redshift range $4 < z < 10$. They 
found no deviation from the  SFR-MZ 
relation of 
\citet{2013ApJ...765..140A} for both 
the $z = 4-6$ and $z = 6-8$ intervals, 
but for the $z = 8-10$ interval, they 
reported $\Delta \log(\text{O/H}) 
\sim -0.3$ dex, showing a 
significant decrease in metallicity 
beyond $z > 8$.  
These findings also align with 
\citet{2023NatAs...7.1517H}, 
who compared their $z > 7$ galaxy 
sample from the Abell 2744, RXJ-2129, 
and CEERS fields, and with the 
observations by 
\citet{2023ApJ...950...67M} of strong 
[O~\textsc{iii}]-emitting galaxies at 
$z = 5-7$, showing no significant 
evolution in metallicity from $z 
\approx 6$ to $z \approx 8$.
In contrast, the JADES+CEERS sample 
study by 
\citet{curti2023jadesinsightslowmassend} 
investigated low-mass 
(log(M$_{\star}$/M$_{\odot}$) 
$\lesssim$ 8.5) galaxies at 
$3 < z < 10$ and found a 
tentative decrease in metallicity and 
a deviation from the local SFR-MZ 
relation of 
\citet{2020MNRAS.491..944C} beyond $z 
> 6$. 
However, the redshift where the 
SFR-MZ relation starts to deviate from 
the local SFR-MZ relation, such as 
at $z \approx 6$ or $z \approx 8$, 
can arise from the use of different 
local SFR-MZ relations with varying 
values of $\alpha$. 
In fact, \citet{2023ApJS..269...33N} 
found that using the formalism of 
\citet{2020MNRAS.491..944C} for
the local SFR-MZ relation, 
instead of the 
\citet{2013ApJ...765..140A}
SFR-MZ relation for the local 
universe, leads to a metallicity 
deviation in the SFR-MZ relation
at $z \sim 6$. 
However, we adhere to the 
\citet{2013ApJ...765..140A} 
SFR-MZ relation at $z \sim 0$
because the mass range of our 
sample is most consistent with
the mass range of their formalism.

{
We have also compared
our results with the
IllustrisTNG
\citep{2018MNRAS.475..676S,2019MNRAS.490.3234N,2019ComAC...6....2N} 
and EAGLE
\citep{2016A&C....15...72M}
simulations,
as depicted in Figure 
\ref{fig:fmr_resid}.
In both simulations, 
there is an observed
negative shift from zero
that increases with 
redshift. 
Specifically, in TNG, 
the deviations from 
the $z = 0$ calibrated 
FMR linearly become more 
negative as redshift increases.
In EAGLE, these deviations 
continue to become more 
negative up to $z \sim 5$ 
and then stabilize from 
$z = 6$ to $z = 8$ 
(for more details, see 
\citealt{2024arXiv240706254G}).
The negative offsets observed in
our galaxy sample are consistent 
with those in the IllustrisTNG 
simulation for $z < 8$,
similar to findings by 
\citet{2023ApJS..269...33N}.
However, it is important to note
that these conclusions strongly 
depend on the chosen local FMR 
relation and the sample 
size of high-redshift galaxies, 
both in simulations and 
observations.
}

\section{Is metallicity redshift invariant?}\label{sec:simulation}

Gas-phase metallicity is a 
crucial indicator of the current 
evolutionary state of galaxies. 
Prior to JWST, the 
MZR and its evolution were
widely studied within
redshift $z \lesssim 3$
\citep[e.g.,][]{2006ApJ...644..813E,2012ApJ...750..120Z,2013ApJ...771L..19Z,2013ApJ...769..148H,2013ApJ...776L..27H,2014ApJ...792....3M,2014ApJ...795..165S,2015ApJ...799..138S}
Specifically, \citet{2013ApJ...771L..19Z} 
demonstrated that for a given 
stellar mass, gas-phase
metallicity is inversely
proportional to redshift in
the range $0 < z < 2.3$.
With smaller galaxy samples
\citet{2008ASPC..396..409M} and
\citet{2009MNRAS.398.1915M}
also confirmed the existence 
of the MZR up to 
$z$ = 3.

Several simulations have also 
explored the MZR and its evolution
at high redshifts, such as
FIRE-1 
\citep{2016MNRAS.456.2140M},
IllustrisTNG 
\citep{2014MNRAS.444.1518V,2019MNRAS.484.5587T,2024MNRAS.531.1398G,2024arXiv240706254G}, 
FirstLight 
\citep{2020MNRAS.494.1988L}, 
SERRA 
\citep{2022MNRAS.513.5621P}, 
ASTRAEUS 
\citep{2023MNRAS.518.3557U}, 
FLARES 
\citep{2023MNRAS.519.3118W},
and 
FIRE-2
\citep{2024ApJ...967L..41M}.
FIRE-1 simulations 
covered redshifts from 
$z$ = 0 to 6.
Recently, FIRE-2 simulations  
extended this study to the
redshift range 
$z$ = 5 to 12
showing that the normalization 
of the MZR evolves weakly for 
$z \geq$  3. 
\citet{2019MNRAS.484.5587T} 
investigated 
the distribution and evolution of 
metals within the IllustrisTNG 
simulation suite, focusing on the gas-
phase MZR and its redshift evolution 
over a stellar-mass range of 
$10^{8} < M_{\ast}/M_{\odot} < 
10^{10.5}$
and redshift range of 
$0 < z < 10$.
IllustrisTNG galaxies broadly
reproduce the MZR slope and 
normalization out to 
$z$ = 2. 
Interestingly, they found a 
declining trend of gas-phase 
metallicity with redshift, with 
metallicity at 
$z$ = 8 being 0.5 dex lower 
than at $z$=0.

\begin{figure*}
    \centering
   \includegraphics[width=0.8\textwidth]{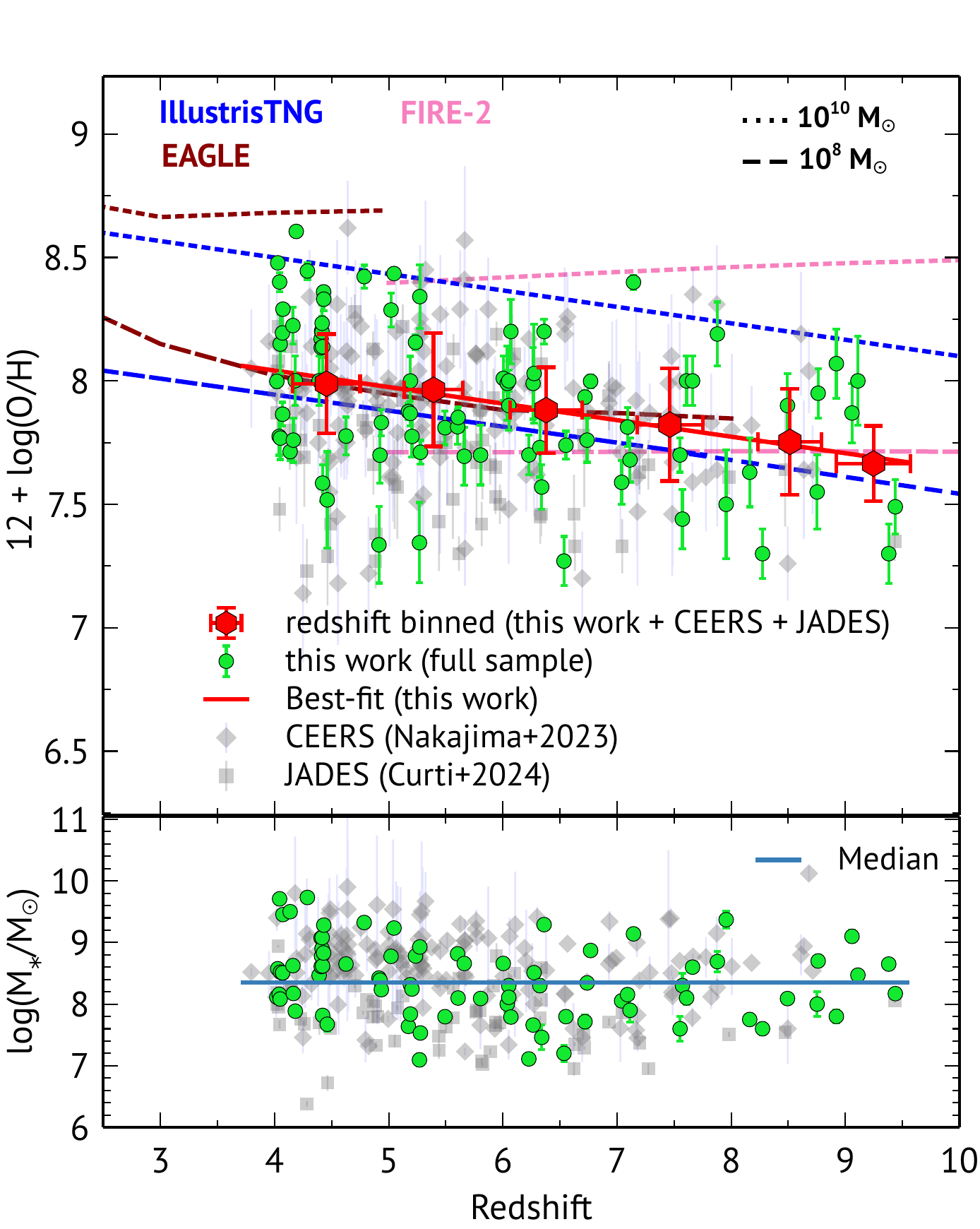} 
    \caption{Top: Comparing redshift
    evolution
    of gas-phase metallicity 
    of our full sample +
    galaxies from earlier
    CEERS and JADES studies
    with
    IllustriTNG 
    \citep{2019MNRAS.484.5587T},
    FIRE-2 
    \citep{2024ApJ...967L..41M},
    and EAGLE 
    \citep{2024arXiv240706254G}
    simulations. 
    Red data points
    present the redshift-averaged
    metallicity in 
    $z$ = 4--10 with 
    $\Delta z$=1 size
    bin for full 263 galaxies
    (this work + CEERS + JADES).
    Red solid line shows the
    best-fit line for redshift
    averaged metallicity.
    The sample from earlier CEERS 
    \citep{2023ApJS..269...33N}
    and JADES 
    \citep{2024A&A...684A..75C}
    studies are shown in
    grey diamond and squares, 
    respectively.
    The iso-stellar mass curves
    at $M_{\ast}$ = 10$^8$ $M_{\odot}$ (dashed) and 
    10$^{10}$ $M_{\odot}$ (dotted) 
    predicted by  simulations
    are shown in blue 
    (IllustrisTNG), 
    pink (FIRE-2),
    and darkred 
    (EAGLE)
    lines.
    { Bottom: Stellar mass vs 
    redshift for the 
    full sample. 
    The light-blue line shows the
    median stellar mass. 
    The data points are well distributed around this
    median
    line, suggesting no 
    significant evolution in 
    stellar masses.}}
    \label{fig:z_vs_metal}
\end{figure*}

We compare the MZR for our full
sample of 81 galaxies with
numerous simulations and
further compared redshift evolution
of gas-phase metallicity with FIRE-2 and illustrisTNG 
simulations, as shown in
Figure \ref{fig:massmet_binned}
and \ref{fig:z_vs_metal}.

Figure \ref{fig:massmet_binned} 
shows the comparison of the MZR 
with simulations across three 
redshift bins: 
$z = 4$--6, 6--8, and 8--10.
As discussed in Section 
\ref{sec:MZR}, 
our mass-averaged metallicities
align with two previous JWST 
observations by 
\citet{2024A&A...684A..75C} 
and \citet{2023ApJS..269...33N}. 
We found that within $z < 6$, 
our best-fit MZR normalization 
and slope are consistent with 
FIRE-2 and IllustrisTNG, but 
significantly higher than the 
Astraeus simulation. 
For the redshift bins $z = 6$--8 and 
8--10, our best-fit MZR slope 
is flatter compared to FIRE-2 
and FirstLight but matches well 
with IllustrisTNG. 
Our best-fit MZR normalizations 
are $\sim$ 0.14 dex and
0.36 dex lower than those 
of FIRE-2 and FirstLight in
the same redshift bins, 
but they are consistent with the 
extrapolated IllustrisTNG 
simulation.

As discussed in Section 
\ref{FMR}, the investigation
of redshift evolution through 
the SFR-MZR relation is
highly dependent on the models used. 
This approach yields 
significantly varied
results across different models.
{
\citet{2024MNRAS.531.1398G} 
showed that the slope
of the anti-correlation of 
offset from the MZR with SFR
changes with redshift
}
We, therefore,
conducted a more detailed 
analysis of the evolution of
metallicities across redshifts 
ranging from 4 to 10. 
In this analysis, we 
integrated our dataset of 81
galaxies with data from two recent
JWST studies by \citet{2023ApJS..269...33N} 
and \citet{2024A&A...684A..75C}, 
which both encompass similar
ranges in redshift and 
stellar mass ($10^{7} \lesssim 
M_{\ast}/M_{\odot} \lesssim 10^{10}$). 
Although the study by 
\citet{2024A&A...684A..75C} includes 
galaxies from a slightly broader
redshift range of 3 to 10, 
for consistency in our analysis, 
we only considered galaxies with
redshifts $z \geq 4$. 
This approach resulted in a 
extensive sample of 
263 star-forming galaxies.
We depicted the metallicities of 
these galaxies on
a 12+log(O/H)--$z$ plane, as shown
in Figure \ref{fig:z_vs_metal}. 
We further calculated the average 
metallicities for the entire
sample (263 galaxies) across each 
redshift interval of $\Delta z = 1$ 
(between $z$ = 4 to 10), 
as shown in 
red
hexagons in Figure \ref{fig:z_vs_metal}.
Additionally, we included 
iso-stellar mass curves from the 
IllustrisTNG, EAGLE,
and FIRE-2 simulations 
for stellar masses of 
$\log(M_{\ast}/M_{\odot}) = 8$ and 10.

Our findings indicate a discernible 
decrease in metallicity as the
redshift increases, a trend that 
aligns with results from 
\citet{2024arXiv240307103R}, 
who observed similar declines in 
their study of stacked metallicities 
for galaxies between redshifts of 5.5
and 9.5.
To quantitatively compare our results
with these simulations, 
we fit a linear regression to our 
full sample. 
The best-fit line, depicted in Figure 
\ref{fig:z_vs_metal}, showed a 
slope of $0.067 \pm 0.013$ and an
intercept of $8.31 \pm 0.10$. 
This slope is more consistent 
with predictions from 
the IllustrisTNG than 
those from FIRE-2
and EAGLE, bolstering
the observation of a steady decrease
in metallicity with increasing
redshift.
These results corroborate our
earlier findings, which suggested
lower mass-averaged MZR
at $z > 8$.
{
While a detailed comparison
between IllustrisTNG and 
FIRE-2 is beyond the 
scope of this paper,
the key difference lies
in their ISM treatment: 
TNG uses sub-grid prescriptions, 
resulting in smooth star formation 
and feedback, while FIRE models 
these processes locally and 
explicitly, leading to rapid,
bursty star formation and feedback
\citep[e.g.,][]{2017MNRAS.466...88S}. 
This bursty behavior is thought 
to regulate early galaxy growth and 
may convert dark matter cusps into 
cores, making it critical for 
understanding both galaxy formation 
and dark matter
\citep{2017MNRAS.465.1682H,2018MNRAS.473.3717F}. 
JWST observations of 
high-redshift galaxies
show a continued metallicity 
decrease, which aligns more with 
TNG's smooth feedback model than 
FIRE's bursty predictions.
}

{
We also examine whether
the redshift evolution
of metallicity
observed in our 
analysis is driven by the 
intrinsic evolution of stellar 
mass within our sample. 
The bottom panel of Figure 
\ref{fig:z_vs_metal} presents 
stellar mass as a function of 
redshift for the 263 galaxies. 
The stellar masses are well 
distributed around a median
value of 8.35 within the redshift 
range $4 < z < 10$, 
indicating no significant 
evolution in stellar mass. 
This confirms that the 
redshift evolution
of metallicity observed
in this work has a physical 
origin, warranting further 
investigation.
}

\begin{figure*}
    \centering
   \begin{tabular}{cc}
       \includegraphics[width=0.5\textwidth]{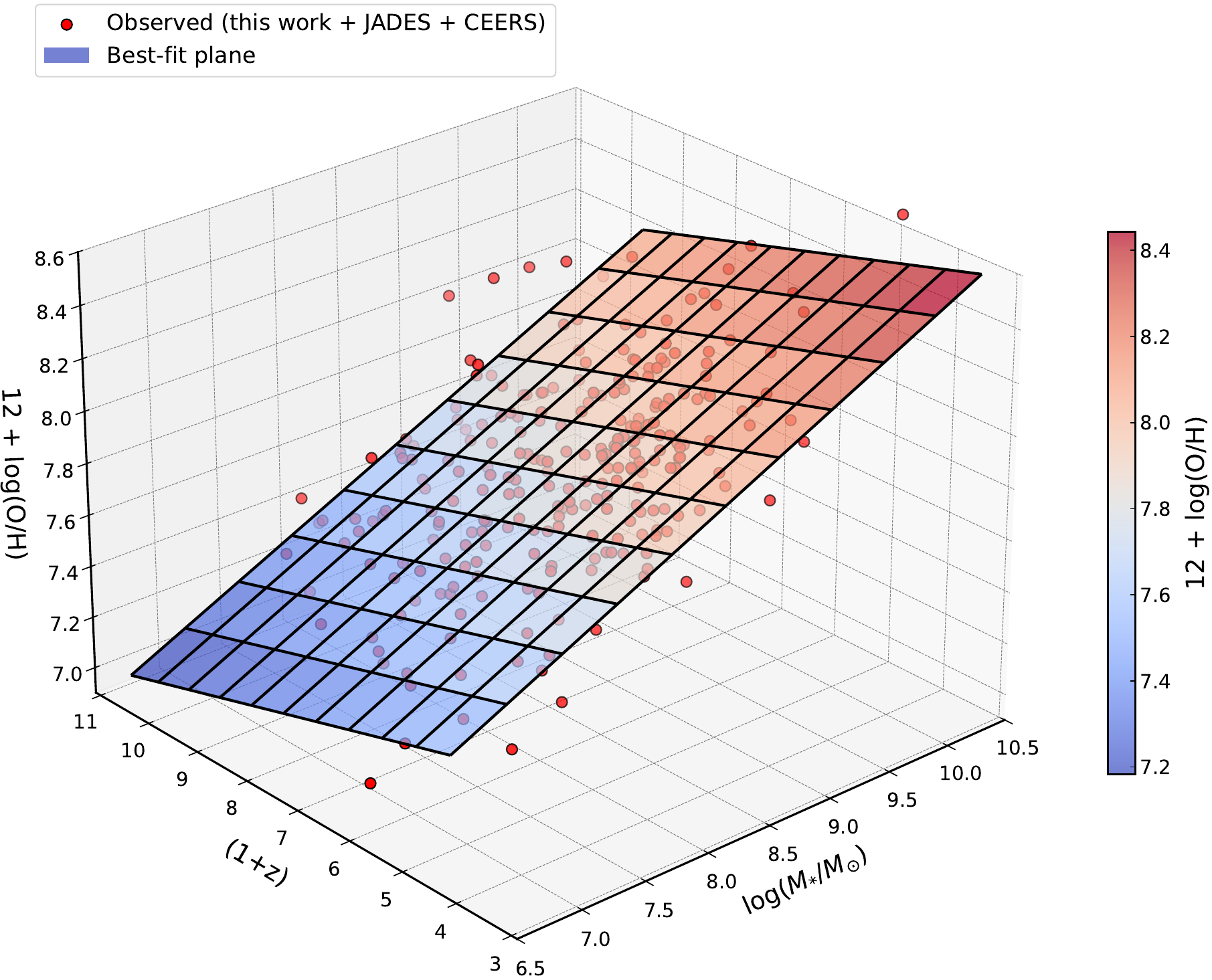} & \includegraphics[width=0.45\textwidth]{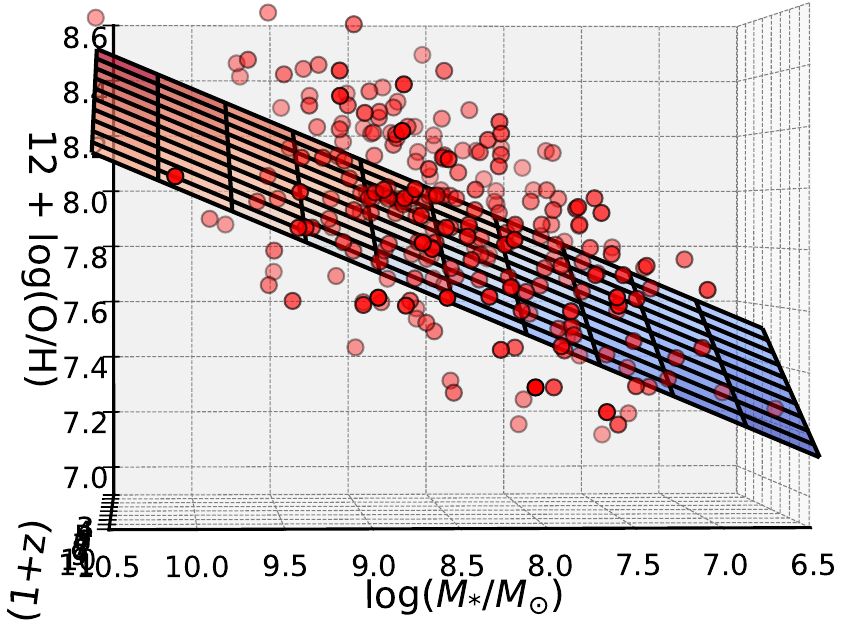}
   \end{tabular} 
      
      \begin{tabular}{cc}
       \includegraphics[width=0.45\textwidth]{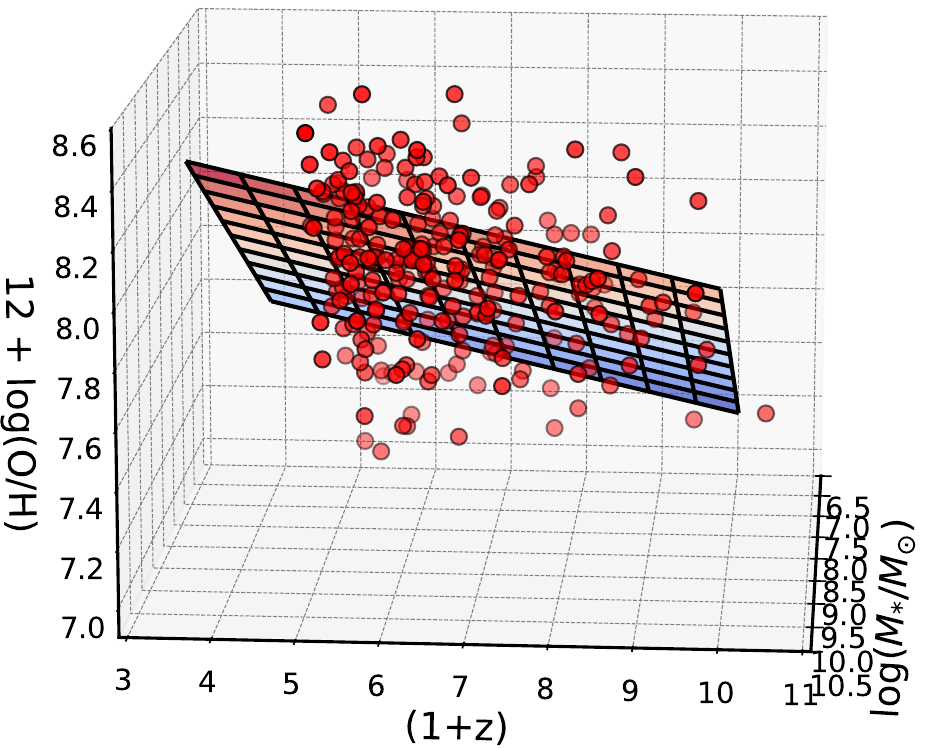} & \includegraphics[width=0.5\textwidth]{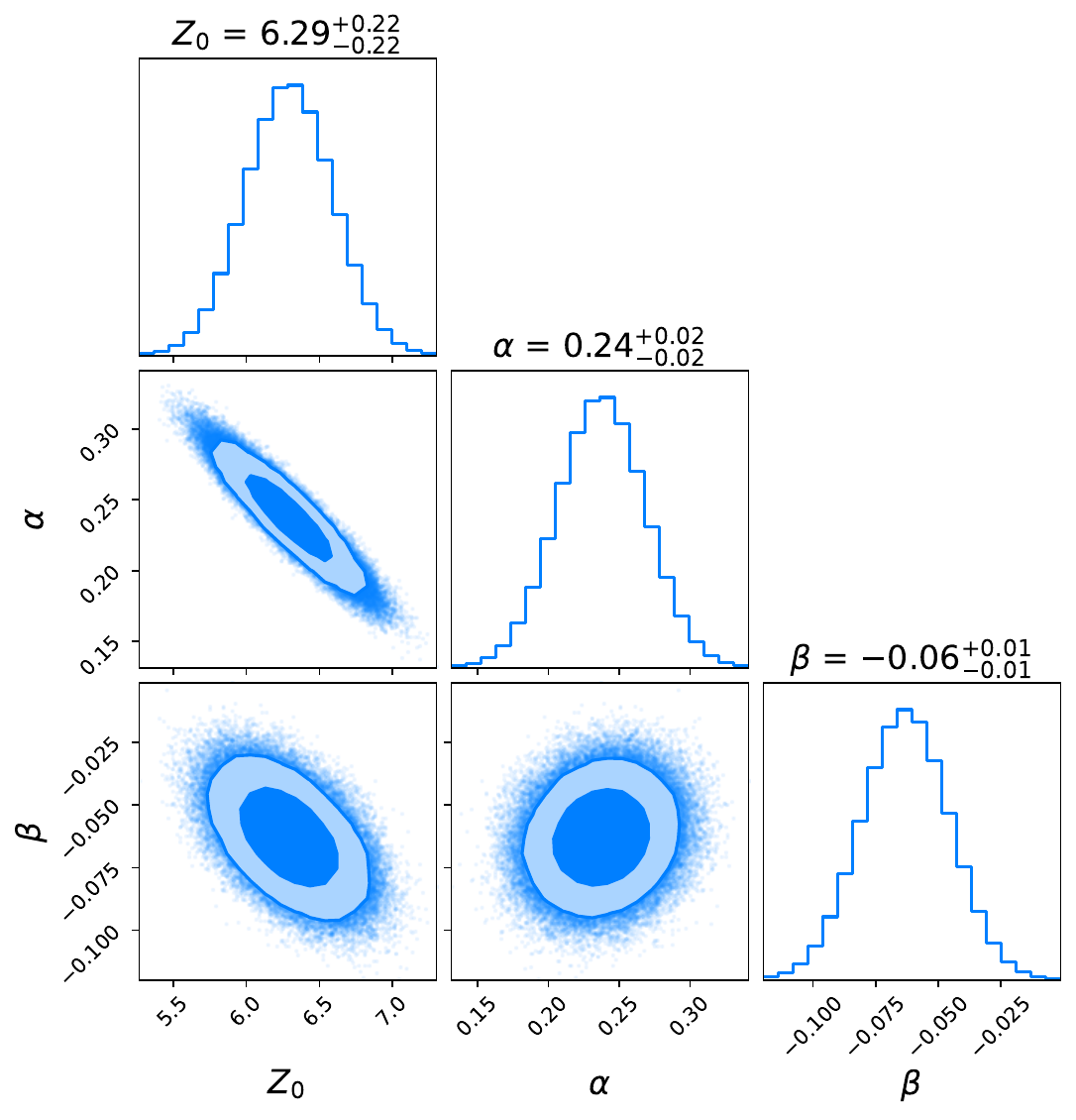}
   \end{tabular} 
    \caption{Top-right to bottom-left anticlockwise: 3D visualization 
     the mass–metallicity–redshift
     (MZ--$z$) relation for our
     full sample of 263 galaxies,
     including 81 galaxies 
     from this work,
     135 galaxies from CEERS
     \citep{2023ApJS..269...33N},
     47 galaxies from JADES
\citep{2024A&A...684A..75C}. 
The surface representing the
best-fitting MZ-$z$ relation 
is displayed, and the
histogram of metallicity
dispersion for individual
galaxies relative to this 
surface is illustrated in
the colorbar. Bottom-right: posterior distribution of the
    best-fit parameters,
    $Z_0$, $\alpha$, and
    $\beta$ of the
mass–metallicity–redshift
     (MZ--$z$) relation. }
    \label{fig:z_vs_mass_vs_metal}
\end{figure*}

We explore an empirical 
relationship among gas-phase 
metallicity, stellar mass, 
and redshift. 
Figure 
\ref{fig:z_vs_mass_vs_metal} 
presents a 
3D representation of the 
metallicities of 263 galaxies, 
shown as functions of both 
stellar mass and redshift.
We introduce an empirical
relation with the following
functional form: 
\begin{equation}\label{eq:empi}
    12+{\rm log(O/H)} = Z_{0} + \alpha{\rm log}\left(\frac{M_{\ast}}{M_{\odot}}\right) 
    - \beta(1+z),
\end{equation}
We find the best-fit values
for our full sample to be,
$Z_{0} = 6.29 \pm 0.22$,
$\alpha = 0.237 \pm 0.023$,
and 
$\beta = 0.06 \pm 0.01$.
{ The corresponding 
best-fit surface 
and the posterior 
distribution of the
fitting parameters are
shown in Figure 
\ref{fig:z_vs_mass_vs_metal}.}
It is important to note that
while our sample spans a
stellar mass range from
$10^{7} < M_{\ast}/M_{\odot} 
< 10^{10}$ and and a redshift
range from
$4 < z < 10$, this simple 
empirical model does not
account for any potential 
flattening in metallicity,
as previously shown in 
\citet{2020MNRAS.491..944C}
and \citet{2021ApJ...914...19S}.
Our best-fit model 
effectively captures the 
observed trends, 
demonstrating an increase
in metallicity with stellar
mass and a decrease with
rising redshift,
providing a well represented
MZ--redshift relation.
We also note that current 
JWST observations 
predominantly target massive, 
high-redshift galaxies
($z > 8$), which generally
exhibit higher metallicity
compared to lower-mass,
metal-poor galaxies. 
This selection bias may
influence the observed slope
for redshift evolution, 
potentially over-estimating it. 
We anticipate that
the inclusion of high-$z$,
low-mass galaxies in 
future JWST observations could
reveal a steeper slope for
this evolutionary trend.

The primary factors driving the
evolution of gas-phase
metallicity
is currently a ``Grand Challenge''
problem in cosmology. 
Galactic metallicity 
evolutionary trends
with redshift or mass have
been explained by competing 
scenarios, such as changes 
in metal retention efficiency or 
changes in the gas fractions of
galaxies
\citep[e.g.,][]{2016MNRAS.456.2140M,2019MNRAS.484.5587T,2020MNRAS.494.1988L,2024MNRAS.532L..14B}.
A relatively high efficiency 
of metal ejection via AGN/stellar
feedback processes has been found
in high-mass galaxies, which in turn 
reduces the retained metal
budget of these galaxies
\citep{2015MNRAS.448..895S}. 
\citet{2019MNRAS.484.5587T} found
that metal retention 
efficiency increases with redshift 
in IllustrisTNG galaxies,
which directly contradicts the 
observed decline in metallicity
with increasing redshift.
However, they also found that 
the gas fraction, defined as the
ratio between ISM
gas mass and stellar mass, 
increases with redshift in 
high-redshift galaxies, 
serving as a significant 
contributing factor to the
observed MZR evolution. 
Increased gas fractions due 
to gas inflow dilute the 
metal-enriched gas in the ISM, 
effectively 
decreasing the metallicity. 
The MZR for our sample of 
galaxies shows a similar
declining trend 
as seen 
in IllustrisTNG, 
suggesting that increased gas 
fractions (e.g., galactic inflows)
may drive the low MZR normalization
in high-redshift galaxies.
We need more theoretical studies
and baseline models to understand
the physics behind the decreasing
metallicity trend towards
early universe.

\section{Summary}\label{summary}

In this study, we investigated the 
evolution of the mass-metallicity 
relation (MZR) and the fundamental 
metallicity relation (FMR) using a 
sample of 81 galaxies observed by 
JWST/NIRSpec spanning a stellar mass 
range of $10^7 < M_{\ast}/M_{\odot} < 
10^{10}$ and a redshift range of $4 < 
z < 10$. The sample comprises galaxies 
from the JADES GOODS-N and GOODS-S 
fields, the JWST-PRIMAL Legacy Survey, 
and additional galaxies from the 
literature in the Abell 2744, 
SMACS-0723, RXJ2129, BDF, COSMOS, and 
MACS1149 fields, which were previously 
not utilized for these scaling 
relations.

Our main findings are 
summarized below:

 \begin{itemize}
\item
For metallicity determination, our analysis focused on the emission lines observed with the medium-resolution grating: [O III]$\lambda5007$, [O II]$\lambda\lambda 3727,29$,
H\(\beta\),
H$\alpha$,
[$\nii$]$\lambda$6584, and 
[$\sii$]$\lambda\lambda$6716,31. 
Prior to conducting the metallicity analysis, we 
corrected for dust reddening
by estimating
the Balmer 
decrement using the H$\alpha$/H$\beta$ 
and H$\gamma$/H$\beta$ ratios, employing 
the reddening curve from 
\citet{2000ApJ...533..682C}. 
We then utilized the 
reddening-corrected line ratios
R3 and O32/N2/S2, along with
the calibrations developed by 
\citet{2020MNRAS.491..944C}, 
to determine metallicities through
line ratio diagnostics. 
For the primary metallicity
calibration, we employed 
R3 indices. 
To address potential degeneracy 
inherent in these calibrations, we 
further utilized the O32/N2/S2 
line ratios to refine and accurately 
determine the metallicities.
To ensure precise metallicity 
measurements, we excluded sources with 
potential AGN emission, as AGN-driven 
ionization disrupts the standard 
calibrations for star-forming 
galaxies, by employing the Mass-
Excitation diagnostic diagrams 
introduced by 
\citet{2014ApJ...788...88J} and 
refined by \citet{2015ApJ...801...35C}.

\item We estimated SFRs 
from the reddening-corrected
H$\beta$ luminosity using
calibrations provided by 
\citet{2022A&A...667A..29F} for 
star-forming galaxies identified 
through the Mass-Excitation (MEx)
diagnostic diagram.
We examine the relationship 
between stellar mass and star 
formation rate for our
full sample, 
which demonstrates a positive
correlation throughout the 
redshift range of \(z \sim 4-10\), 
consistent
with previous observations
and main sequence 
galaxies.

\item
We examined the stellar
mass-metallicity
relation (MZR) for high-redshift
galaxies using
81 galaxies spanning 
a redshift range of
$z$ = 4 -- 10 
and the stellar mass 
range of $M_{\ast}$ = 10$^{7}$ -- 
10$^{10}$ $M_\odot$.
We found 
the best-fit  
$Z_{10} = 8.28 \pm 0.08$ and 
$\gamma = 0.27 \pm 0.02$ for 
our full sample. 
When comparing our MZR with those 
of lower redshift galaxies, we
found weak evolution in the MZ 
relation. 
Compared to the MZR documented 
by \citet{2023ApJ...955L..18L} 
for galaxies at $z \sim 2-3$, 
our $4<z<10$ sample shows a small 
reduction in metallicity,
ranging within \(\Delta 
\log(\text{O/H})\) $\sim$ 0.2 
at \(10^8 \, M_\odot\) and 
\(\Delta \log(\text{O/H})\) 
$\sim$ 0.07 at \(10^9 \, M_\odot\). 
In comparison to the MZR at 
\(z \sim 3.3\) from 
\citet{2021ApJ...914...19S} 
extrapolated to the lower-mass regime, 
our sample  shows a slight average 
reduction in \(\log(\text{O/H})\), 
with a \(\Delta \log(\text{O/H})\) 
of 0.03 dex. 
We also compared our MZR with 
recent JWST studies of 
high-redshift galaxies. 
Despite variations in redshift 
and mass intervals, our 
best-fit MZR for galaxies at 
\(z = 4 - 10\) closely aligns
with earlier studies by 
\citet{2023ApJS..269...33N}, 
\citet{2023NatAs...7.1517H}, 
\citet{2023ApJ...950L...1S}, and 
\citet{2024A&A...684A..75C},
with their MZR slopes/predictions 
consistent with our findings. 

\item To investigate the 
redshift evolution of the MZR from
\(z = 4\) to \(z = 10\),
we segmented our sample into 
three distinct groups: 
\(z = 4\) to 6, \(z = 6\) to 8, 
and \(z = 8\) to 10. 
We find a steeper MZR slope 
(\(\gamma = 0.28 \pm 0.03\)) for
the \(z = 4\) to 6 compared
to the \(z = 6\) to 8 
(\(\gamma = 0.23 \pm 0.04\)) 
and the \(z = 8\) to 10
(\(\gamma = 0.21 \pm 0.04\)).
This observation suggests a
gentle decline in metallicity.
The observed trend of decreasing 
metallicity at higher redshifts is 
consistent with the findings of 
\citet{2023ApJS..269...33N} and 
\citet{2024A&A...684A..75C}, who 
reported similar trends in 
their studies.

\item 
We investigated the SFR-MZ relation, 
also known as the Fundamental 
Metallicity Relation (FMR), 
for our entire sample using the 
parameterization from
\citet{2013ApJ...765..140A}, 
with 
$\alpha$ = 0.66 in $\mu_{\alpha}$.
Within redshift range $z \sim$ 4--8,  
we observed no evolution in SFR-MZ
relation, suggesting that
galaxies maintain
metallicity equilibrium 
through star formation, 
the inflow of pristine gas,
and the outflow of
metal-enriched gas.
However, we found a significant 
decrease in metallicity 
($\Delta$log(O/H) 
= 0.27 dex) at $z > 8$,
consistent with previous 
JWST observations by 
\citet{2023ApJS..269...33N}.
This suggests that early-stage
galaxies may not yet have 
reached a state of metallicity 
equilibrium, 
likely due to their nascent 
stage of formation.

\item We further explored
the redshift evolution of
metallicity within the range
$4 < z < 10$ by integrating samples
from two major JWST studies
into our analysis. 
This includes 135 galaxies from 
CEERS \citep{2023ApJS..269...33N} 
and 47 galaxies from JADES 
\citep{2024A&A...684A..75C}.
Our analysis revealed
a
distinct decreasing trend in
redshift-averaged metallicities,
with a slope of
$0.067 \pm 0.013$.
This trend aligns with 
predictions from the 
IllustrisTNG simulations 
\citep{2019MNRAS.484.5587T}
and corroborates previous
JWST observations in the 
redshift range $5.5 < z < 9.5$
\citep{2024arXiv240307103R}.

\item We also introduce an empirical 
mass-metallicity-redshift 
(MZ--$z$) relation that captures
the observed trends, 
including increasing
metallicity with rising
stellar mass and
decreasing metallicity
with increasing redshift.
We find the best-fit MZ--$z$ 
for our full sample of galaxies:
$12+{\rm log(O/H)} = (6.29\pm 0.10) + (0.237\pm0.023)\times{\rm log}\left(\frac{M_{\ast}}{M_{\odot}}\right) 
    - (0.06\pm0.01)\times(1+z)$.
The observed decline in
metallicity at higher redshifts
points to complex, 
possibly unexplored, 
physical processes that
modulate star formation, 
gas inflow, and outflow, 
thereby impacting the 
ISM
metal content in galaxies.

 \end{itemize}

\begin{acknowledgments}
We sincerely thank the anonymous
referee for their insightful 
comments and suggestions.
A.S acknowledges support from 
NASA via sub-award SV2-82023 
from the Chandra X-ray Center 
and via sub-award S001534-NASA from 
the Pennsylvania State University.
The authors also
acknowledge the 
MIT SuperCloud and 
Lincoln Laboratory Supercomputing 
Center for providing (HPC, database, consultation) 
resources that have contributed
to the research results reported 
within this paper.
PT and AMG acknowledge support from NSF-AST 2346977.
This work is based on observations made with the NASA/ESA/CSA James Webb Space Telescope. The data were obtained from the Mikulski Archive for Space Telescopes at the Space Telescope Science Institute, which is operated by the Association of Universities for Research in Astronomy, Inc., under NASA contract NAS 5-03127 for JWST.
JADES data products presented
herein were retrieved 
from the data release 3
archive \url{https://jades-survey.github.io/scientists/data.html}.
Primal Survey data products presented herein were retrieved from the DAWN JWST Archive (DJA). DJA is an initiative of the Cosmic Dawn Center, which is funded by the Danish National Research Foundation under grant DNRF140.

\end{acknowledgments}
\bibliography{sample631}{}
\bibliographystyle{aasjournal}

\appendix


\begin{table}[h!!]
   \caption{GOODS-S and GOODS-N JWST/NIRSpec Sample of Galaxies
   Analysed in this Paper
   \citep{2024arXiv240406531D}.}
    \centering
    \setlength{\tabcolsep}{5pt}
    \begin{tabular}{cccccccc}
    NIRSpec ID & RA & Dec & $z$ & log($M_{\ast}/M_{\odot}$) &  log(SFR$_{\rm H\beta}$) &  12+log(O/H)\\
\hline
\hline
42453 & 53.1173859 & -27.8033791 & 4.46 & 7.67 $\pm$ 0.11 & 0.01 $\pm$ 0.12 & 7.52 $\pm$ 0.20 \\ 
103483 & 53.188310688 & -27.81283361 & 5.27 & 7.09 $\pm$ 0.09 & 0.66 $\pm$ 0.11 & 7.34 $\pm$ 0.16 \\ 
10004736 & 53.1640716 & -27.8191086 & 4.03 & 8.57 $\pm$ 0.04 & -0.10 $\pm$ 0.11 & 8.48 $\pm$ 0.01 \\ 
110 & 189.146378721 & 62.21550823 & 4.07 & 9.45 $\pm$ 0.02 & 1.14 $\pm$ 0.01 & 8.29 $\pm$ 0.01 \\ 
604 & 189.117202094 & 62.2214337 & 4.16 & 8.62 $\pm$ 0.03 & 0.64 $\pm$ 0.05 & 8.22 $\pm$ 0.07 \\ 
607 & 189.116947265 & 62.22207876 & 5.19 & 8.32 $\pm$ 0.03 & 0.94 $\pm$ 0.04 & 7.87 $\pm$ 0.05 \\ 
666 & 189.113413603 & 62.2276564 & 4.92 & 8.39 $\pm$ 0.04 & 0.64 $\pm$ 0.14 & 7.70 $\pm$ 0.11 \\ 
795 & 189.191792532 & 62.24204531 & 4.79 & 9.32 $\pm$ 0.03 & 0.93 $\pm$ 0.07 & 8.42 $\pm$ 0.05 \\ 
834 & 189.09701079 & 62.24698089 & 4.91 & 8.42 $\pm$ 0.05 & 0.67 $\pm$ 0.13 & 7.34 $\pm$ 0.16 \\ 
896 & 189.082651931 & 62.25247723 & 6.77 & 8.87 $\pm$ 0.02 & 1.38 $\pm$ 0.02 & 8.00 $\pm$ 0.03 \\ 
902 & 189.193276303 & 62.25372707 & 4.07 & 8.51 $\pm$ 0.02 & 0.95 $\pm$ 0.02 & 7.87 $\pm$ 0.05 \\ 
910 & 189.113442988 & 62.25480338 & 4.42 & 8.62 $\pm$ 0.02 & 0.53 $\pm$ 0.05 & 7.59 $\pm$ 0.07 \\ 
917 & 189.080143561 & 62.25539861 & 4.41 & 8.61 $\pm$ 0.02 & 0.53 $\pm$ 0.05 & 8.13 $\pm$ 0.04 \\ 
964 & 189.13723537 & 62.26063625 & 5.61 & 8.10 $\pm$ 0.03 & 1.09 $\pm$ 0.04 & 7.85 $\pm$ 0.04 \\ 
971 & 189.130932109 & 62.26199976 & 4.43 & 9.28 $\pm$ 0.03 & 0.92 $\pm$ 0.05 & 8.33 $\pm$ 0.05 \\ 
993 & 189.081903696 & 62.26492219 & 4.18 & 7.89 $\pm$ 0.04 & 0.46 $\pm$ 0.08 & 8.00 $\pm$ 0.10 \\ 
1044 & 189.111041015 & 62.27190179 & 4.04 & 8.16 $\pm$ 0.03 & 0.51 $\pm$ 0.04 & 7.78 $\pm$ 0.08 \\ 
1129 & 189.179752709 & 62.28238705 & 7.09 & 8.15 $\pm$ 0.03 & 1.27 $\pm$ 0.04 & 7.81 $\pm$ 0.08 \\ 
1139 & 189.176550776 & 62.28372336 & 4.16 & 8.17 $\pm$ 0.08 & 0.08 $\pm$ 0.14 & 7.76 $\pm$ 0.09 \\ 
1560 & 189.100309169 & 62.23085249 & 5.20 & 8.24 $\pm$ 0.03 & 0.86 $\pm$ 0.04 & 7.78 $\pm$ 0.08 \\ 
1674 & 189.131144715 & 62.27035136 & 4.05 & 8.52 $\pm$ 0.01 & 0.53 $\pm$ 0.05 & 7.77 $\pm$ 0.09 \\ 
1926 & 189.140718886 & 62.27724124 & 4.04 & 9.71 $\pm$ 0.03 & 0.65 $\pm$ 0.05 & 8.40 $\pm$ 0.04 \\ 
1931 & 189.069641344 & 62.28101917 & 7.04 & 8.05 $\pm$ 0.07 & 0.89 $\pm$ 0.06 & 7.59 $\pm$ 0.09 \\ 
1948 & 189.177312132 & 62.29105785 & 6.74 & 8.34 $\pm$ 0.07 & 0.82 $\pm$ 0.04 & 7.76 $\pm$ 0.09 \\ 
2000 & 189.175947335 & 62.31153443 & 5.66 & 8.66 $\pm$ 0.01 & 0.81 $\pm$ 0.07 & 7.69 $\pm$ 0.12 \\ 
2113 & 189.170329427 & 62.22949872 & 6.72 & 7.71 $\pm$ 0.11 & 1.02 $\pm$ 0.05 & 7.93 $\pm$ 0.03 \\ 
3012 & 189.120110053 & 62.30436157 & 5.27 & 8.93 $\pm$ 0.06 & 0.57 $\pm$ 0.15 & 8.34 $\pm$ 0.13 \\ 
3608 & 189.117937758 & 62.2355185 & 5.28 & 7.53 $\pm$ 0.03 & 0.86 $\pm$ 0.02 & 7.71 $\pm$ 0.05 \\ 
4545 & 189.185252674 & 62.23876092 & 4.05 & 8.08 $\pm$ 0.08 & -0.24 $\pm$ 0.14 & 8.15 $\pm$ 0.12 \\ 
7351 & 189.108182935 & 62.24714628 & 6.05 & 8.30 $\pm$ 0.04 & 0.84 $\pm$ 0.07 & 8.00 $\pm$ 0.04 \\ 
12067 & 189.207450251 & 62.26445323 & 4.07 & 8.50 $\pm$ 0.03 & 0.45 $\pm$ 0.05 & 8.19 $\pm$ 0.07 \\ 
13410 & 189.188837292 & 62.2699137 & 5.02 & 8.78 $\pm$ 0.06 & 0.61 $\pm$ 0.06 & 8.29 $\pm$ 0.07 \\ 
16553 & 189.143602845 & 62.28054547 & 4.39 & 8.47 $\pm$ 0.02 & 0.81 $\pm$ 0.02 & 8.00 $\pm$ 0.10 \\ 
17722 & 189.108883186 & 62.28421706 & 4.94 & 8.24 $\pm$ 0.02 & 0.75 $\pm$ 0.05 & 7.83 $\pm$ 0.05 \\ 
24819 & 189.136473583 & 62.22340297 & 7.14 & 9.14 $\pm$ 0.03 & 0.93 $\pm$ 0.04 & 8.40 $\pm$ 0.03 \\ 
25356 & 189.147811293 & 62.23045386 & 4.42 & 9.08 $\pm$ 0.03 & 0.36 $\pm$ 0.06 & 8.23 $\pm$ 0.07 \\ 
27003 & 189.014595403 & 62.26820484 & 5.60 & 8.82 $\pm$ 0.03 & 1.02 $\pm$ 0.04 & 7.81 $\pm$ 0.07 \\ 
28174 & 189.227700921 & 62.25176394 & 4.41 & 9.07 $\pm$ 0.04 & 0.93 $\pm$ 0.01 & 8.17 $\pm$ 0.03 \\ 
28229 & 189.225841987 & 62.25193981 & 4.41 & 8.78 $\pm$ 0.10 & 0.45 $\pm$ 0.07 & 8.20 $\pm$ 0.04 \\ 
28746 & 189.17607762 & 62.2563246 & 4.43 & 8.83 $\pm$ 0.01 & 0.98 $\pm$ 0.02 & 8.36 $\pm$ 0.02 \\ 
32944 & 189.089677913 & 62.3048876 & 4.29 & 9.73 $\pm$ 0.03 & 0.68 $\pm$ 0.05 & 8.45 $\pm$ 0.03 \\ 
38849 & 189.155065954 & 62.25900074 & 4.42 & 7.82 $\pm$ 0.06 & 0.27 $\pm$ 0.10 & 8.14 $\pm$ 0.11 \\  
\hline
    \end{tabular}
    \label{tab:goods}
\end{table}

   \begin{table} 
   \caption{Continued.}
    \centering
    \setlength{\tabcolsep}{5pt}
    \begin{tabular}{cccccccc}
    NIRSpec ID & RA & Dec & $z$ & log($M_{\ast}/M_{\odot}$) &  log(SFR$_{\rm H\beta}$) &  12+log(O/H)\\
    \hline
    \hline
52923 & 189.286013502 & 62.19204378 & 4.02 & 8.11 $\pm$ 0.01 & 0.38 $\pm$ 0.04 & 8.00 $\pm$ 0.02 \\ 
58561 & 189.221830211 & 62.20735743 & 6.55 & 7.80 $\pm$ 0.03 & 0.92 $\pm$ 0.03 & 7.74 $\pm$ 0.06 \\ 
59156 & 189.306169277 & 62.20920468 & 5.23 & 8.78 $\pm$ 0.06 & 0.81 $\pm$ 0.05 & 8.16 $\pm$ 0.03 \\ 
62309 & 189.248977223 & 62.21835017 & 5.17 & 7.64 $\pm$ 0.03 & 0.74 $\pm$ 0.04 & 7.88 $\pm$ 0.03 \\ 
71093 & 189.275721336 & 62.16168901 & 5.05 & 9.24 $\pm$ 0.02 & 0.98 $\pm$ 0.03 & 8.43 $\pm$ 0.02 \\ 
73488 & 189.197395939 & 62.17723313 & 4.14 & 9.50 $\pm$ 0.01 & 0.83 $\pm$ 0.02 & 7.71 $\pm$ 0.03 \\ 
79349 & 189.20968225 & 62.20725204 & 5.19 & 7.84 $\pm$ 0.03 & 0.83 $\pm$ 0.02 & 8.00 $\pm$ 0.10 \\ 
80185 & 189.14847252 & 62.21165963 & 5.49 & 7.80 $\pm$ 0.03 & 0.66 $\pm$ 0.05 & 7.81 $\pm$ 0.07 \\ 
80391 & 189.194218105 & 62.21250554 & 4.63 & 8.65 $\pm$ 0.05 & 0.36 $\pm$ 0.13 & 7.78 $\pm$ 0.08 \\ 
82830 & 189.237876124 & 62.2340822 & 5.80 & 8.09 $\pm$ 0.03 & 0.60 $\pm$ 0.07 & 7.70 $\pm$ 0.12 \\ 
10000885 & 189.270594 & 62.25156684 & 4.42 & 8.90 $\pm$ 0.02 & 1.05 $\pm$ 0.04 & 8.20 $\pm$ 0.06 \\ 
\hline
    \end{tabular}
    \label{tab:goodss}
\end{table}

   \begin{table}
   \caption{JWST/NIRSpec Sample of Galaxies from Primal Survey
   \citep{2024arXiv240402211H}}
    \centering
    \setlength{\tabcolsep}{5pt}
    \begin{tabular}{cccccccc}
    NIRSpec ID & RA & Dec & $z$ & log($M_{\ast}/M_{\odot}$) &  log(SFR$_{\rm H\beta}$) &  12+log(O/H)\\
    \hline
    \hline
9457 & 53.17324 & -27.795674 & 6.00 & 8.66 $\pm$ 0.03 & 0.76 $\pm$ 0.06 & 8.01 $\pm$ 0.09 \\ 
50010 & 189.2881531 & 62.1837804 & 6.04 & 8.00 $\pm$ 0.12 & 0.55 $\pm$ 0.11 & 7.99 $\pm$ 0.15 \\ 
13887 & 53.1958751 & -27.7684324 & 6.05 & 8.11 $\pm$ 0.08 & 0.62 $\pm$ 0.09 & 7.90 $\pm$ 0.10 \\ 
99302 & 53.125818 & -27.818228 & 6.07 & 7.79 $\pm$ 0.08 & 0.06 $\pm$ 0.05 & 8.20 $\pm$ 0.13 \\ 
617 & 214.860024 & 52.898125 & 6.23 & 7.11 $\pm$ 0.04 & 0.60 $\pm$ 0.06 & 7.70 $\pm$ 0.08 \\ 
15265 & 53.0831133 & -27.786351 & 6.27 & 7.66 $\pm$ 0.04 & 1.03 $\pm$ 0.04 & 7.99 $\pm$ 0.15 \\ 
42988 & 53.0906811 & -27.7442159 & 6.27 & 8.51 $\pm$ 0.04 & 0.81 $\pm$ 0.06 & 8.03 $\pm$ 0.20 \\ 
1817 & 189.3963575 & 62.2301725 & 6.32 & 8.30 $\pm$ 0.10 & 0.78 $\pm$ 0.08 & 7.73 $\pm$ 0.12 \\ 
106197 & 53.131047 & -27.8090823 & 6.34 & 7.46 $\pm$ 1.30 & -0.03 $\pm$ 0.07 & 7.57 $\pm$ 0.09 \\ 
662 & 214.877883 & 52.897675 & 6.54 & 7.20 $\pm$ 0.13 & 0.75 $\pm$ 0.04 & 7.27 $\pm$ 0.10 \\ 
449 & 214.940489 & 52.932556 & 7.55 & 7.60 $\pm$ 0.20 & 0.64 $\pm$ 0.06 & 7.70 $\pm$ 0.07 \\ 
10038687 & 189.2030733 & 62.1439285 & 7.57 & 8.30 $\pm$ 0.20 & 0.51 $\pm$ 0.21 & 7.44 $\pm$ 0.12 \\ 
9074 & 53.0863067 & -27.8239613 & 7.61 & 8.10 $\pm$ 0.10 & 1.12 $\pm$ 0.08 & 8.00 $\pm$ 0.10 \\ 
12637 & 53.133469 & -27.760373 & 7.66 & 8.60 $\pm$ 0.03 & 1.30 $\pm$ 0.02 & 8.00 $\pm$ 0.10 \\ 
3626 & 53.0873836 & -27.8603113 & 7.96 & 9.37 $\pm$ 0.14 & 1.37 $\pm$ 0.16 & 7.50 $\pm$ 0.22 \\ 
20198852 & 53.1077607 & -27.812944 & 8.27 & 7.60 $\pm$ 0.10 & 0.28 $\pm$ 0.06 & 7.30 $\pm$ 0.10 \\ 
20213084 & 53.1589064 & -27.765076 & 8.49 & 8.09 $\pm$ 0.03 & 0.79 $\pm$ 0.03 & 7.90 $\pm$ 0.13 \\ 
20100293 & 53.1687381 & -27.8169753 & 8.75 & 8.00 $\pm$ 0.20 & 0.20 $\pm$ 0.11 & 7.55 $\pm$ 0.15 \\ 
28 & 214.938642 & 52.911749 & 8.76 & 8.70 $\pm$ 0.10 & 1.31 $\pm$ 0.03 & 7.95 $\pm$ 0.10 \\ 
20110306 & 53.1691329 & -27.8029208 & 8.92 & 7.80 $\pm$ 0.10 & -0.12 $\pm$ 0.19 & 8.07 $\pm$ 0.14 \\ 
10278 & 53.13916658 & -27.8484843 & 9.06 & 9.10 $\pm$ 0.10 & 1.16 $\pm$ 0.11 & 7.87 $\pm$ 0.12 \\ 
3990 & 189.0169954 & 62.2415817 & 9.38 & 8.65 $\pm$ 0.04 & 1.75 $\pm$ 0.03 & 7.30 $\pm$ 0.12 \\ 
\hline
    \end{tabular}
    \label{tab:primal}
\end{table}

   \begin{table}
   \caption{JWST Sample of Galaxies from literature}
    \centering
    \setlength{\tabcolsep}{3pt}
    \begin{tabular}{cccccccc}
    Target & RA & Dec & $z$ & log($M_{\ast}/M_{\odot}$) &  log(SFR$_{\rm H\beta}$) &  12+log(O/H) & Source\\
    \hline
    \hline
A2744-YD4 & 3.60375 & -30.38225 & 7.88 & 8.69 $\pm$ 0.17 & 0.61 $\pm$ 0.04 & 8.19 $\pm$ 0.13 & \citet{2024arXiv240303977V}\\ 
BDF-3299-a & 337.05 & -35.166 & 7.11 & 7.90 $\pm$ 0.19 & 0.51 $\pm$ 0.01 & 7.68 $\pm$ 0.09 & \citet{2024arXiv240303977V}\\ 
COSMOS24108-a & 150.1972 & 2.47865 & 6.36 & 9.29 $\pm$ 0.08 & 0.56 $\pm$ 0.04 & 8.20 $\pm$ 0.05 & \citet{2024arXiv240303977V}\\ 
RX2129-z8HeII & 322.416266 & 0.099675 & 8.16 & 7.75 $\pm$ 0.06 & 0.94 $\pm$ 0.07 & 7.63 $\pm$ 0.14 & \citet{2024ApJ...967L..42W}\\  
MACS1149-JD1 & 177.389945 & 22.412722 & 9.11 & 8.47 $\pm$ 0.05 & 0.40 $\pm$ 0.06 & 8.00 $\pm$ 0.18 & 
\citet{2024arXiv240708616M}\\
\hline
    \end{tabular}
    \label{tab:literat}
\end{table}









\end{document}

%% file: macros.tex
\newcommand{\Cloudy}{\textsc{Cloudy}}
\newcommand{\bagpipes}{\textsc{Bagpipes}}

\font\manual=manfnt at 7pt \def\dbend{\hbox{\raise0.9ex\hbox{\manual\char127\hspace{0.6em}}}}

\providecommand{\e}[1]{\ensuremath{\times 10^{#1}}}

\newcounter{INTERNALionstage}

\def\gtsim{\mathrel{\hbox{\rlap{\hbox{\lower4pt\hbox{$\sim$}}}\hbox{$>$}}}}
\def\lesssim{\mathrel{\hbox{\rlap{\hbox{\lower4pt\hbox{$\sim$}}}\hbox{$<$}}}}

\def\A{{\rm\thinspace \AA}}

\def\g{{\rm\thinspace g}}

\def\K{{\rm\thinspace K}}

\def\m{{\rm\thinspace m}}

\def\s{{\rm\thinspace s}}

\def\W{{\rm\thinspace W}}

\def\hii{\mbox{{\rm H~{\sc ii}}}}

\def\nii{\mbox{{\rm N~{\sc ii}}}}

\def\oii{\mbox{{\rm O~{\sc ii}}}}
\def\oiii{\mbox{{\rm O~{\sc iii}}}}

\def\sii{\mbox{{\rm S~{\sc ii}}}}

\def\h0{\mbox{{\rm H}$^0$}}
\DeclareMathAlphabet{\vib}{OML}{cmm}{m}{it}



%% file: sample631.bbl
\begin{thebibliography}{}
\expandafter\ifx\csname natexlab\endcsname\relax\def\natexlab#1{#1}\fi
\providecommand{\url}[1]{\href{#1}{#1}}
\providecommand{\dodoi}[1]{doi:~\href{http://doi.org/#1}{\nolinkurl{#1}}}
\providecommand{\doeprint}[1]{\href{http://ascl.net/#1}{\nolinkurl{http://ascl.net/#1}}}
\providecommand{\doarXiv}[1]{\href{https://arxiv.org/abs/#1}{\nolinkurl{https://arxiv.org/abs/#1}}}

\bibitem[{{Abazajian} {et~al.}(2003){Abazajian}, {Adelman-McCarthy}, {Ag{\"u}eros}, {Allam}, {Anderson}, {Annis}, {Bahcall}, {Baldry}, {Bastian}, {Berlind}, {Bernardi}, {Blanton}, {Blythe}, {Bochanski}, {Boroski}, {Brewington}, {Briggs}, {Brinkmann}, {Brunner}, {Budav{\'a}ri}, {Carey}, {Carr}, {Castander}, {Chiu}, {Collinge}, {Connolly}, {Covey}, {Csabai}, {Dalcanton}, {Dodelson}, {Doi}, {Dong}, {Eisenstein}, {Evans}, {Fan}, {Feldman}, {Finkbeiner}, {Friedman}, {Frieman}, {Fukugita}, {Gal}, {Gillespie}, {Glazebrook}, {Gonzalez}, {Gray}, {Grebel}, {Grodnicki}, {Gunn}, {Gurbani}, {Hall}, {Hao}, {Harbeck}, {Harris}, {Harris}, {Harvanek}, {Hawley}, {Heckman}, {Helmboldt}, {Hendry}, {Hennessy}, {Hindsley}, {Hogg}, {Holmgren}, {Holtzman}, {Homer}, {Hui}, {Ichikawa}, {Ichikawa}, {Inkmann}, {Ivezi{\'c}}, {Jester}, {Johnston}, {Jordan}, {Jordan}, {Jorgensen}, {Juri{\'c}}, {Kauffmann}, {Kent}, {Kleinman}, {Knapp}, {Kniazev}, {Kron}, {Krzesi{\'n}ski}, {Kunszt}, {Kuropatkin}, {Lamb}, {Lampeitl}, {Laubscher}, {Lee},
  {Leger}, {Li}, {Lidz}, {Lin}, {Loh}, {Long}, {Loveday}, {Lupton}, {Malik}, {Margon}, {McGehee}, {McKay}, {Meiksin}, {Miknaitis}, {Moorthy}, {Munn}, {Murphy}, {Nakajima}, {Narayanan}, {Nash}, {Neilsen}, {Newberg}, {Newman}, {Nichol}, {Nicinski}, {Nieto-Santisteban}, {Nitta}, {Odenkirchen}, {Okamura}, {Ostriker}, {Owen}, {Padmanabhan}, {Peoples}, {Pier}, {Pindor}, {Pope}, {Quinn}, {Rafikov}, {Raymond}, {Richards}, {Richmond}, {Rix}, {Rockosi}, {Schaye}, {Schlegel}, {Schneider}, {Schroeder}, {Scranton}, {Sekiguchi}, {Seljak}, {Sergey}, {Sesar}, {Sheldon}, {Shimasaku}, {Siegmund}, {Silvestri}, {Sinisgalli}, {Sirko}, {Smith}, {Smol{\v{c}}i{\'c}}, {Snedden}, {Stebbins}, {Steinhardt}, {Stinson}, {Stoughton}, {Strateva}, {Strauss}, {SubbaRao}, {Szalay}, {Szapudi}, {Szkody}, {Tasca}, {Tegmark}, {Thakar}, {Tremonti}, {Tucker}, {Uomoto}, {Vanden Berk}, {Vandenberg}, {Vogeley}, {Voges}, {Vogt}, {Walkowicz}, {Weinberg}, {West}, {White}, {Wilhite}, {Willman}, {Xu}, {Yanny}, {Yarger}, {Yasuda}, {Yip}, {Yocum}, {York},
  {Zakamska}, {Zehavi}, {Zheng}, {Zibetti}, \& {Zucker}}]{2003AJ....126.2081A}
{Abazajian}, K., {Adelman-McCarthy}, J.~K., {Ag{\"u}eros}, M.~A., {et~al.} 2003, \aj, 126, 2081, \dodoi{10.1086/378165}

\bibitem[{{Anders} \& {Grevesse}(1989)}]{1989GeCoA..53..197A}
{Anders}, E., \& {Grevesse}, N. 1989, \gca, 53, 197, \dodoi{10.1016/0016-7037(89)90286-X}

\bibitem[{{Andrews} \& {Martini}(2013)}]{2013ApJ...765..140A}
{Andrews}, B.~H., \& {Martini}, P. 2013, \apj, 765, 140, \dodoi{10.1088/0004-637X/765/2/140}

\bibitem[{{Baldwin} {et~al.}(1981){Baldwin}, {Phillips}, \& {Terlevich}}]{1981PASP...93....5B}
{Baldwin}, J.~A., {Phillips}, M.~M., \& {Terlevich}, R. 1981, \pasp, 93, 5, \dodoi{10.1086/130766}

\bibitem[{{Bassini} {et~al.}(2024){Bassini}, {Feldmann}, {Gensior}, {Faucher-Gigu{\`e}re}, {Cenci}, {Moreno}, {Bernardini}, \& {Liang}}]{2024MNRAS.532L..14B}
{Bassini}, L., {Feldmann}, R., {Gensior}, J., {et~al.} 2024, \mnras, 532, L14, \dodoi{10.1093/mnrasl/slae036}

\bibitem[{{Behroozi} {et~al.}(2013){Behroozi}, {Wechsler}, \& {Conroy}}]{2013ApJ...770...57B}
{Behroozi}, P.~S., {Wechsler}, R.~H., \& {Conroy}, C. 2013, \apj, 770, 57, \dodoi{10.1088/0004-637X/770/1/57}

\bibitem[{{Bezanson} {et~al.}(2022){Bezanson}, {Labbe}, {Whitaker}, {Leja}, {Price}, {Franx}, {Brammer}, {Marchesini}, {Zitrin}, {Wang}, {Weaver}, {Furtak}, {Atek}, {Coe}, {Cutler}, {Dayal}, {van Dokkum}, {Feldmann}, {Forster Schreiber}, {Fujimoto}, {Geha}, {Glazebrook}, {de Graaff}, {Greene}, {Juneau}, {Kassin}, {Kriek}, {Khullar}, {Maseda}, {Mowla}, {Muzzin}, {Nanayakkara}, {Nelson}, {Oesch}, {Pacifici}, {Pan}, {Papovich}, {Setton}, {Shapley}, {Smit}, {Stefanon}, {Taylor}, \& {Williams}}]{2022arXiv221204026B}
{Bezanson}, R., {Labbe}, I., {Whitaker}, K.~E., {et~al.} 2022, arXiv e-prints, arXiv:2212.04026, \dodoi{10.48550/arXiv.2212.04026}

\bibitem[{{Bian} {et~al.}(2018){Bian}, {Kewley}, \& {Dopita}}]{2018ApJ...859..175B}
{Bian}, F., {Kewley}, L.~J., \& {Dopita}, M.~A. 2018, \apj, 859, 175, \dodoi{10.3847/1538-4357/aabd74}

\bibitem[{{Bruzual} \& {Charlot}(2003)}]{2003MNRAS.344.1000B}
{Bruzual}, G., \& {Charlot}, S. 2003, \mnras, 344, 1000, \dodoi{10.1046/j.1365-8711.2003.06897.x}

\bibitem[{{Bunker} {et~al.}(2023){Bunker}, {Cameron}, {Curtis-Lake}, {Jakobsen}, {Carniani}, {Curti}, {Witstok}, {Maiolino}, {D'Eugenio}, {Looser}, {Willott}, {Bonaventura}, {Hainline}, {Uebler}, {Willmer}, {Saxena}, {Smit}, {Alberts}, {Arribas}, {Baker}, {Baum}, {Bhatawdekar}, {Bowler}, {Boyett}, {Charlot}, {Chen}, {Chevallard}, {Circosta}, {DeCoursey}, {de Graaff}, {Egami}, {Eisenstein}, {Endsley}, {Ferruit}, {Giardino}, {Hausen}, {Helton}, {Hviding}, {Ji}, {Johnson}, {Jones}, {Kumari}, {Laseter}, {Luetzgendorf}, {Maseda}, {Nelson}, {Parlanti}, {Perna}, {Rawle}, {Rix}, {Rieke}, {Robertson}, {Rodriguez Del Pino}, {Sandles}, {Scholtz}, {Sharpe}, {Skarbinski}, {Stark}, {Sun}, {Tacchella}, {Topping}, {Villanueva}, {Wallace}, {Williams}, \& {Woodrum}}]{2023arXiv230602467B}
{Bunker}, A.~J., {Cameron}, A.~J., {Curtis-Lake}, E., {et~al.} 2023, arXiv e-prints, arXiv:2306.02467, \dodoi{10.48550/arXiv.2306.02467}

\bibitem[{{Calzetti} {et~al.}(2000){Calzetti}, {Armus}, {Bohlin}, {Kinney}, {Koornneef}, \& {Storchi-Bergmann}}]{2000ApJ...533..682C}
{Calzetti}, D., {Armus}, L., {Bohlin}, R.~C., {et~al.} 2000, \apj, 533, 682, \dodoi{10.1086/308692}

\bibitem[{{Calzetti} {et~al.}(1994){Calzetti}, {Kinney}, \& {Storchi-Bergmann}}]{1994ApJ...429..582C}
{Calzetti}, D., {Kinney}, A.~L., \& {Storchi-Bergmann}, T. 1994, \apj, 429, 582, \dodoi{10.1086/174346}

\bibitem[{{Carnall} {et~al.}(2018){Carnall}, {McLure}, {Dunlop}, \& {Dav{\'e}}}]{2018MNRAS.480.4379C}
{Carnall}, A.~C., {McLure}, R.~J., {Dunlop}, J.~S., \& {Dav{\'e}}, R. 2018, \mnras, 480, 4379, \dodoi{10.1093/mnras/sty2169}

\bibitem[{{Chabrier}(2003)}]{2003PASP..115..763C}
{Chabrier}, G. 2003, \pasp, 115, 763, \dodoi{10.1086/376392}

\bibitem[{{Chakraborty} {et~al.}(2024){Chakraborty}, {Sarkar}, {Wolk}, {Schneider}, {Brickhouse}, {Lanzetta}, {Foster}, \& {Smith}}]{2024arXiv240605306C}
{Chakraborty}, P., {Sarkar}, A., {Wolk}, S., {et~al.} 2024, arXiv e-prints, arXiv:2406.05306, \dodoi{10.48550/arXiv.2406.05306}

\bibitem[{{Chatzikos} {et~al.}(2023){Chatzikos}, {Bianchi}, {Camilloni}, {Chakraborty}, {Gunasekera}, {Guzm{\'a}n}, {Milby}, {Sarkar}, {Shaw}, {van Hoof}, \& {Ferland}}]{2023RMxAA..59..327C}
{Chatzikos}, M., {Bianchi}, S., {Camilloni}, F., {et~al.} 2023, \rmxaa, 59, 327, \dodoi{10.22201/ia.01851101p.2023.59.02.12}

\bibitem[{{Coil} {et~al.}(2015){Coil}, {Aird}, {Reddy}, {Shapley}, {Kriek}, {Siana}, {Mobasher}, {Freeman}, {Price}, \& {Shivaei}}]{2015ApJ...801...35C}
{Coil}, A.~L., {Aird}, J., {Reddy}, N., {et~al.} 2015, \apj, 801, 35, \dodoi{10.1088/0004-637X/801/1/35}

\bibitem[{{Cresci} \& {Maiolino}(2018)}]{2018NatAs...2..179C}
{Cresci}, G., \& {Maiolino}, R. 2018, Nature Astronomy, 2, 179, \dodoi{10.1038/s41550-018-0404-5}

\bibitem[{{Cresci} {et~al.}(2019){Cresci}, {Mannucci}, \& {Curti}}]{2019A&A...627A..42C}
{Cresci}, G., {Mannucci}, F., \& {Curti}, M. 2019, \aap, 627, A42, \dodoi{10.1051/0004-6361/201834637}

\bibitem[{{Curti} {et~al.}(2017){Curti}, {Cresci}, {Mannucci}, {Marconi}, {Maiolino}, \& {Esposito}}]{2017MNRAS.465.1384C}
{Curti}, M., {Cresci}, G., {Mannucci}, F., {et~al.} 2017, \mnras, 465, 1384, \dodoi{10.1093/mnras/stw2766}

\bibitem[{{Curti} {et~al.}(2020){Curti}, {Mannucci}, {Cresci}, \& {Maiolino}}]{2020MNRAS.491..944C}
{Curti}, M., {Mannucci}, F., {Cresci}, G., \& {Maiolino}, R. 2020, \mnras, 491, 944, \dodoi{10.1093/mnras/stz2910}

\bibitem[{Curti {et~al.}(2023)Curti, Maiolino, Curtis-Lake, Chevallard, Carniani, D'Eugenio, Looser, Scholtz, Charlot, Cameron, Übler, Witstok, Boyett, Laseter, Sandles, Arribas, Bunker, Giardino, Maseda, Rawle, Pino, Smit, Willott, Eisenstein, Hausen, Johnson, Rieke, Robertson, Tacchella, Williams, Willmer, Baker, Bhatawdekar, Egami, Helton, Ji, Kumari, Perna, Shivaei, \& Sun}]{curti2023jadesinsightslowmassend}
Curti, M., Maiolino, R., Curtis-Lake, E., {et~al.} 2023, JADES: Insights on the low-mass end of the mass--metallicity--star-formation rate relation at $3 < z < 10$ from deep JWST/NIRSpec spectroscopy.
\newblock \doarXiv{2304.08516}

\bibitem[{{Curti} {et~al.}(2024){Curti}, {Maiolino}, {Curtis-Lake}, {Chevallard}, {Carniani}, {D'Eugenio}, {Looser}, {Scholtz}, {Charlot}, {Cameron}, {{\"U}bler}, {Witstok}, {Boyett}, {Laseter}, {Sandles}, {Arribas}, {Bunker}, {Giardino}, {Maseda}, {Rawle}, {Rodr{\'\i}guez Del Pino}, {Smit}, {Willott}, {Eisenstein}, {Hausen}, {Johnson}, {Rieke}, {Robertson}, {Tacchella}, {Williams}, {Willmer}, {Baker}, {Bhatawdekar}, {Egami}, {Helton}, {Ji}, {Kumari}, {Perna}, {Shivaei}, \& {Sun}}]{2024A&A...684A..75C}
{Curti}, M., {Maiolino}, R., {Curtis-Lake}, E., {et~al.} 2024, \aap, 684, A75, \dodoi{10.1051/0004-6361/202346698}

\bibitem[{{Curtis-Lake} {et~al.}(2023){Curtis-Lake}, {Carniani}, {Cameron}, {Charlot}, {Jakobsen}, {Maiolino}, {Bunker}, {Witstok}, {Smit}, {Chevallard}, {Willott}, {Ferruit}, {Arribas}, {Bonaventura}, {Curti}, {D'Eugenio}, {Franx}, {Giardino}, {Looser}, {L{\"u}tzgendorf}, {Maseda}, {Rawle}, {Rix}, {Rodr{\'\i}guez del Pino}, {{\"U}bler}, {Sirianni}, {Dressler}, {Egami}, {Eisenstein}, {Endsley}, {Hainline}, {Hausen}, {Johnson}, {Rieke}, {Robertson}, {Shivaei}, {Stark}, {Tacchella}, {Williams}, {Willmer}, {Bhatawdekar}, {Bowler}, {Boyett}, {Chen}, {de Graaff}, {Helton}, {Hviding}, {Jones}, {Kumari}, {Lyu}, {Nelson}, {Perna}, {Sandles}, {Saxena}, {Suess}, {Sun}, {Topping}, {Wallace}, \& {Whitler}}]{2023NatAs...7..622C}
{Curtis-Lake}, E., {Carniani}, S., {Cameron}, A., {et~al.} 2023, Nature Astronomy, 7, 622, \dodoi{10.1038/s41550-023-01918-w}

\bibitem[{{Daddi} {et~al.}(2007){Daddi}, {Dickinson}, {Morrison}, {Chary}, {Cimatti}, {Elbaz}, {Frayer}, {Renzini}, {Pope}, {Alexander}, {Bauer}, {Giavalisco}, {Huynh}, {Kurk}, \& {Mignoli}}]{2007ApJ...670..156D}
{Daddi}, E., {Dickinson}, M., {Morrison}, G., {et~al.} 2007, \apj, 670, 156, \dodoi{10.1086/521818}

\bibitem[{{Dav{\'e}}(2008)}]{2008MNRAS.385..147D}
{Dav{\'e}}, R. 2008, \mnras, 385, 147, \dodoi{10.1111/j.1365-2966.2008.12866.x}

\bibitem[{{Dav{\'e}} {et~al.}(2011{\natexlab{a}}){Dav{\'e}}, {Finlator}, \& {Oppenheimer}}]{2011MNRAS.416.1354D}
{Dav{\'e}}, R., {Finlator}, K., \& {Oppenheimer}, B.~D. 2011{\natexlab{a}}, \mnras, 416, 1354, \dodoi{10.1111/j.1365-2966.2011.19132.x}

\bibitem[{{Dav{\'e}} {et~al.}(2011{\natexlab{b}}){Dav{\'e}}, {Oppenheimer}, \& {Finlator}}]{2011MNRAS.415...11D}
{Dav{\'e}}, R., {Oppenheimer}, B.~D., \& {Finlator}, K. 2011{\natexlab{b}}, \mnras, 415, 11, \dodoi{10.1111/j.1365-2966.2011.18680.x}

\bibitem[{{Dayal} {et~al.}(2013){Dayal}, {Ferrara}, \& {Dunlop}}]{2013MNRAS.430.2891D}
{Dayal}, P., {Ferrara}, A., \& {Dunlop}, J.~S. 2013, \mnras, 430, 2891, \dodoi{10.1093/mnras/stt083}

\bibitem[{{De Rossi} {et~al.}(2015){De Rossi}, {Theuns}, {Font}, \& {McCarthy}}]{2015MNRAS.452..486D}
{De Rossi}, M.~E., {Theuns}, T., {Font}, A.~S., \& {McCarthy}, I.~G. 2015, \mnras, 452, 486, \dodoi{10.1093/mnras/stv1287}

\bibitem[{{D'Eugenio} {et~al.}(2024){D'Eugenio}, {Cameron}, {Scholtz}, {Carniani}, {Willott}, {Curtis-Lake}, {Bunker}, {Parlanti}, {Maiolino}, {Willmer}, {Jakobsen}, {Robertson}, {Johnson}, {Tacchella}, {Cargile}, {Rawle}, {Arribas}, {Chevallard}, {Curti}, {Egami}, {Eisenstein}, {Kumari}, {Looser}, {Rieke}, {Rodr{\'\i}guez Del Pino}, {Saxena}, {{\"U}bler}, {Venturi}, {Witstok}, {Baker}, {Bhatawdekar}, {Bonaventura}, {Boyett}, {Charlot}, {Danhaive}, {Hainline}, {Hausen}, {Helton}, {Ji}, {Ji}, {Jones}, {Joud{\v{z}}balis}, {Maseda}, {P{\'e}rez-Gonz{\'a}lez}, {Perna}, {Pusk{\'a}s}, {Shivaei}, {Silcock}, {Simmonds}, {Smit}, {Sun}, {Villanueva}, {Williams}, \& {Zhu}}]{2024arXiv240406531D}
{D'Eugenio}, F., {Cameron}, A.~J., {Scholtz}, J., {et~al.} 2024, arXiv e-prints, arXiv:2404.06531, \dodoi{10.48550/arXiv.2404.06531}

\bibitem[{{Diemer} {et~al.}(2018){Diemer}, {Stevens}, {Forbes}, {Marinacci}, {Hernquist}, {Lagos}, {Sternberg}, {Pillepich}, {Nelson}, {Popping}, {Villaescusa-Navarro}, {Torrey}, \& {Vogelsberger}}]{2018ApJS..238...33D}
{Diemer}, B., {Stevens}, A. R.~H., {Forbes}, J.~C., {et~al.} 2018, \apjs, 238, 33, \dodoi{10.3847/1538-4365/aae387}

\bibitem[{{Diemer} {et~al.}(2019){Diemer}, {Stevens}, {Lagos}, {Calette}, {Tacchella}, {Hernquist}, {Marinacci}, {Nelson}, {Pillepich}, {Rodriguez-Gomez}, {Villaescusa-Navarro}, \& {Vogelsberger}}]{2019MNRAS.487.1529D}
{Diemer}, B., {Stevens}, A. R.~H., {Lagos}, C. d.~P., {et~al.} 2019, \mnras, 487, 1529, \dodoi{10.1093/mnras/stz1323}

\bibitem[{{Dopita} {et~al.}(2000){Dopita}, {Kewley}, {Heisler}, \& {Sutherland}}]{2000ApJ...542..224D}
{Dopita}, M.~A., {Kewley}, L.~J., {Heisler}, C.~A., \& {Sutherland}, R.~S. 2000, \apj, 542, 224, \dodoi{10.1086/309538}

\bibitem[{{Eisenstein} {et~al.}(2023){Eisenstein}, {Willott}, {Alberts}, {Arribas}, {Bonaventura}, {Bunker}, {Cameron}, {Carniani}, {Charlot}, {Curtis-Lake}, {D'Eugenio}, {Endsley}, {Ferruit}, {Giardino}, {Hainline}, {Hausen}, {Jakobsen}, {Johnson}, {Maiolino}, {Rieke}, {Rieke}, {Rix}, {Robertson}, {Stark}, {Tacchella}, {Williams}, {Willmer}, {Baker}, {Baum}, {Bhatawdekar}, {Boyett}, {Chen}, {Chevallard}, {Circosta}, {Curti}, {Danhaive}, {DeCoursey}, {de Graaff}, {Dressler}, {Egami}, {Helton}, {Hviding}, {Ji}, {Jones}, {Kumari}, {L{\"u}tzgendorf}, {Laseter}, {Looser}, {Lyu}, {Maseda}, {Nelson}, {Parlanti}, {Perna}, {Pusk{\'a}s}, {Rawle}, {Rodr{\'\i}guez Del Pino}, {Sandles}, {Saxena}, {Scholtz}, {Sharpe}, {Shivaei}, {Silcock}, {Simmonds}, {Skarbinski}, {Smit}, {Stone}, {Suess}, {Sun}, {Tang}, {Topping}, {{\"U}bler}, {Villanueva}, {Wallace}, {Whitler}, {Witstok}, \& {Woodrum}}]{2023arXiv230602465E}
{Eisenstein}, D.~J., {Willott}, C., {Alberts}, S., {et~al.} 2023, arXiv e-prints, arXiv:2306.02465, \dodoi{10.48550/arXiv.2306.02465}

\bibitem[{{Elbaz} {et~al.}(2007){Elbaz}, {Daddi}, {Le Borgne}, {Dickinson}, {Alexander}, {Chary}, {Starck}, {Brandt}, {Kitzbichler}, {MacDonald}, {Nonino}, {Popesso}, {Stern}, \& {Vanzella}}]{2007A&A...468...33E}
{Elbaz}, D., {Daddi}, E., {Le Borgne}, D., {et~al.} 2007, \aap, 468, 33, \dodoi{10.1051/0004-6361:20077525}

\bibitem[{{Ellison} {et~al.}(2008){Ellison}, {Patton}, {Simard}, \& {McConnachie}}]{2008ApJ...672L.107E}
{Ellison}, S.~L., {Patton}, D.~R., {Simard}, L., \& {McConnachie}, A.~W. 2008, \apjl, 672, L107, \dodoi{10.1086/527296}

\bibitem[{{Erb} {et~al.}(2006{\natexlab{a}}){Erb}, {Shapley}, {Pettini}, {Steidel}, {Reddy}, \& {Adelberger}}]{2006ApJ...644..813E}
{Erb}, D.~K., {Shapley}, A.~E., {Pettini}, M., {et~al.} 2006{\natexlab{a}}, \apj, 644, 813, \dodoi{10.1086/503623}

\bibitem[{{Erb} {et~al.}(2006{\natexlab{b}}){Erb}, {Steidel}, {Shapley}, {Pettini}, {Reddy}, \& {Adelberger}}]{2006ApJ...646..107E}
{Erb}, D.~K., {Steidel}, C.~C., {Shapley}, A.~E., {et~al.} 2006{\natexlab{b}}, \apj, 646, 107, \dodoi{10.1086/504891}

\bibitem[{{Faucher-Gigu{\`e}re}(2018)}]{2018MNRAS.473.3717F}
{Faucher-Gigu{\`e}re}, C.-A. 2018, \mnras, 473, 3717, \dodoi{10.1093/mnras/stx2595}

\bibitem[{{Feroz} \& {Hobson}(2008)}]{2008MNRAS.384..449F}
{Feroz}, F., \& {Hobson}, M.~P. 2008, \mnras, 384, 449, \dodoi{10.1111/j.1365-2966.2007.12353.x}

\bibitem[{{Ferruit} {et~al.}(2022){Ferruit}, {Jakobsen}, {Giardino}, {Rawle}, {Alves de Oliveira}, {Arribas}, {Beck}, {Birkmann}, {B{\"o}ker}, {Bunker}, {Charlot}, {de Marchi}, {Franx}, {Henry}, {Karakla}, {Kassin}, {Kumari}, {L{\'o}pez-Caniego}, {L{\"u}tzgendorf}, {Maiolino}, {Manjavacas}, {Marston}, {Moseley}, {Muzerolle}, {Pirzkal}, {Rauscher}, {Rix}, {Sabbi}, {Sirianni}, {te Plate}, {Valenti}, {Willott}, \& {Zeidler}}]{2022A&A...661A..81F}
{Ferruit}, P., {Jakobsen}, P., {Giardino}, G., {et~al.} 2022, \aap, 661, A81, \dodoi{10.1051/0004-6361/202142673}

\bibitem[{{Figueira} {et~al.}(2022){Figueira}, {Pollo}, {Ma{\l}ek}, {Buat}, {Boquien}, {Pistis}, {Cassar{\`a}}, {Vergani}, {Hamed}, \& {Salim}}]{2022A&A...667A..29F}
{Figueira}, M., {Pollo}, A., {Ma{\l}ek}, K., {et~al.} 2022, \aap, 667, A29, \dodoi{10.1051/0004-6361/202141701}

\bibitem[{{Finkelstein} {et~al.}(2022){Finkelstein}, {Bagley}, {Arrabal Haro}, {Dickinson}, {Ferguson}, {Kartaltepe}, {Papovich}, {Burgarella}, {Kocevski}, {Huertas-Company}, {Iyer}, {Koekemoer}, {Larson}, {P{\'e}rez-Gonz{\'a}lez}, {Rose}, {Tacchella}, {Wilkins}, {Chworowsky}, {Medrano}, {Morales}, {Somerville}, {Yung}, {Fontana}, {Giavalisco}, {Grazian}, {Grogin}, {Kewley}, {Kirkpatrick}, {Kurczynski}, {Lotz}, {Pentericci}, {Pirzkal}, {Ravindranath}, {Ryan}, {Trump}, {Yang}, {Almaini}, {Amor{\'\i}n}, {Annunziatella}, {Backhaus}, {Barro}, {Behroozi}, {Bell}, {Bhatawdekar}, {Bisigello}, {Bromm}, {Buat}, {Buitrago}, {Calabr{\`o}}, {Casey}, {Castellano}, {Ch{\'a}vez Ortiz}, {Ciesla}, {Cleri}, {Cohen}, {Cole}, {Cooke}, {Cooper}, {Cooray}, {Costantin}, {Cox}, {Croton}, {Daddi}, {Dav{\'e}}, {de La Vega}, {Dekel}, {Elbaz}, {Estrada-Carpenter}, {Faber}, {Fern{\'a}ndez}, {Finkelstein}, {Freundlich}, {Fujimoto}, {Garc{\'\i}a-Argum{\'a}nez}, {Gardner}, {Gawiser}, {G{\'o}mez-Guijarro}, {Guo}, {Hamblin}, {Hamilton},
  {Hathi}, {Holwerda}, {Hirschmann}, {Hutchison}, {Jaskot}, {Jha}, {Jogee}, {Juneau}, {Jung}, {Kassin}, {Le Bail}, {Leung}, {Lucas}, {Magnelli}, {Mantha}, {Matharu}, {McGrath}, {McIntosh}, {Merlin}, {Mobasher}, {Newman}, {Nicholls}, {Pandya}, {Rafelski}, {Ronayne}, {Santini}, {Seill{\'e}}, {Shah}, {Shen}, {Simons}, {Snyder}, {Stanway}, {Straughn}, {Teplitz}, {Vanderhoof}, {Vega-Ferrero}, {Wang}, {Weiner}, {Willmer}, {Wuyts}, {Zavala}, \& {Ceers Team}}]{2022ApJ...940L..55F}
{Finkelstein}, S.~L., {Bagley}, M.~B., {Arrabal Haro}, P., {et~al.} 2022, \apjl, 940, L55, \dodoi{10.3847/2041-8213/ac966e}

\bibitem[{{Finkelstein} {et~al.}(2023){Finkelstein}, {Bagley}, {Ferguson}, {Wilkins}, {Kartaltepe}, {Papovich}, {Yung}, {Arrabal Haro}, {Behroozi}, {Dickinson}, {Kocevski}, {Koekemoer}, {Larson}, {Le Bail}, {Morales}, {P{\'e}rez-Gonz{\'a}lez}, {Burgarella}, {Dav{\'e}}, {Hirschmann}, {Somerville}, {Wuyts}, {Bromm}, {Casey}, {Fontana}, {Fujimoto}, {Gardner}, {Giavalisco}, {Grazian}, {Grogin}, {Hathi}, {Hutchison}, {Jha}, {Jogee}, {Kewley}, {Kirkpatrick}, {Long}, {Lotz}, {Pentericci}, {Pierel}, {Pirzkal}, {Ravindranath}, {Ryan}, {Trump}, {Yang}, {Bhatawdekar}, {Bisigello}, {Buat}, {Calabr{\`o}}, {Castellano}, {Cleri}, {Cooper}, {Croton}, {Daddi}, {Dekel}, {Elbaz}, {Franco}, {Gawiser}, {Holwerda}, {Huertas-Company}, {Jaskot}, {Leung}, {Lucas}, {Mobasher}, {Pandya}, {Tacchella}, {Weiner}, \& {Zavala}}]{2023ApJ...946L..13F}
{Finkelstein}, S.~L., {Bagley}, M.~B., {Ferguson}, H.~C., {et~al.} 2023, \apjl, 946, L13, \dodoi{10.3847/2041-8213/acade4}

\bibitem[{{Finlator} {et~al.}(2006){Finlator}, {Dav{\'e}}, {Papovich}, \& {Hernquist}}]{2006ApJ...639..672F}
{Finlator}, K., {Dav{\'e}}, R., {Papovich}, C., \& {Hernquist}, L. 2006, \apj, 639, 672, \dodoi{10.1086/499349}

\bibitem[{{F{\"o}rster Schreiber} \& {Wuyts}(2020)}]{2020ARA&A..58..661F}
{F{\"o}rster Schreiber}, N.~M., \& {Wuyts}, S. 2020, \araa, 58, 661, \dodoi{10.1146/annurev-astro-032620-021910}

\bibitem[{{Fujimoto} {et~al.}(2023){Fujimoto}, {Arrabal Haro}, {Dickinson}, {Finkelstein}, {Kartaltepe}, {Larson}, {Burgarella}, {Bagley}, {Behroozi}, {Chworowsky}, {Hirschmann}, {Trump}, {Wilkins}, {Yung}, {Koekemoer}, {Papovich}, {Pirzkal}, {Ferguson}, {Fontana}, {Grogin}, {Grazian}, {Kewley}, {Kocevski}, {Lotz}, {Pentericci}, {Ravindranath}, {Somerville}, {Wilkins}, {Amor{\'\i}n}, {Backhaus}, {Calabr{\`o}}, {Casey}, {Cooper}, {Fern{\'a}ndez}, {Franco}, {Giavalisco}, {Hathi}, {Harish}, {Hutchison}, {Iyer}, {Jung}, {Lucas}, \& {Zavala}}]{2023ApJ...949L..25F}
{Fujimoto}, S., {Arrabal Haro}, P., {Dickinson}, M., {et~al.} 2023, \apjl, 949, L25, \dodoi{10.3847/2041-8213/acd2d9}

\bibitem[{{Furtak} {et~al.}(2021){Furtak}, {Atek}, {Lehnert}, {Chevallard}, \& {Charlot}}]{2021MNRAS.501.1568F}
{Furtak}, L.~J., {Atek}, H., {Lehnert}, M.~D., {Chevallard}, J., \& {Charlot}, S. 2021, \mnras, 501, 1568, \dodoi{10.1093/mnras/staa3760}

\bibitem[{{Garcia} {et~al.}(2024{\natexlab{a}}){Garcia}, {Torrey}, {Ellison}, {Grasha}, {Chen}, {Hemler}, {Zimmerman}, {Wright}, {Zovaro}, {Nelson}, {Sanders}, {Kewley}, \& {Hernquist}}]{2024arXiv240706254G}
{Garcia}, A.~M., {Torrey}, P., {Ellison}, S.~L., {et~al.} 2024{\natexlab{a}}, arXiv e-prints, arXiv:2407.06254, \dodoi{10.48550/arXiv.2407.06254}

\bibitem[{{Garcia} {et~al.}(2024{\natexlab{b}}){Garcia}, {Torrey}, {Ellison}, {Grasha}, {Hernquist}, {Zovaro}, {Chen}, {Hemler}, {Kewley}, {Nelson}, \& {Wright}}]{2024MNRAS.531.1398G}
{Garcia}, A.~M., {Torrey}, P., {Ellison}, S., {et~al.} 2024{\natexlab{b}}, \mnras, 531, 1398, \dodoi{10.1093/mnras/stae1252}

\bibitem[{{Garnett} \& {Shields}(1987)}]{1987ApJ...317...82G}
{Garnett}, D.~R., \& {Shields}, G.~A. 1987, \apj, 317, 82, \dodoi{10.1086/165257}

\bibitem[{{Grazian} {et~al.}(2015){Grazian}, {Fontana}, {Santini}, {Dunlop}, {Ferguson}, {Castellano}, {Amorin}, {Ashby}, {Barro}, {Behroozi}, {Boutsia}, {Caputi}, {Chary}, {Dekel}, {Dickinson}, {Faber}, {Fazio}, {Finkelstein}, {Galametz}, {Giallongo}, {Giavalisco}, {Grogin}, {Guo}, {Kocevski}, {Koekemoer}, {Koo}, {Lee}, {Lu}, {Merlin}, {Mobasher}, {Nonino}, {Papovich}, {Paris}, {Pentericci}, {Reddy}, {Renzini}, {Salmon}, {Salvato}, {Sommariva}, {Song}, \& {Vanzella}}]{2015A&A...575A..96G}
{Grazian}, A., {Fontana}, A., {Santini}, P., {et~al.} 2015, \aap, 575, A96, \dodoi{10.1051/0004-6361/201424750}

\bibitem[{{Hainline} {et~al.}(2020){Hainline}, {Hviding}, {Rieke}, {Shivaei}, {Endsley}, {Curtis-Lake}, {Smit}, {Williams}, {Alberts}, {K Boyett}, {Bunker}, {Egami}, {Maseda}, {Tacchella}, \& {Willmer}}]{2020ApJ...892..125H}
{Hainline}, K.~N., {Hviding}, R.~E., {Rieke}, M., {et~al.} 2020, \apj, 892, 125, \dodoi{10.3847/1538-4357/ab7dc3}

\bibitem[{{Hayward} \& {Hopkins}(2017)}]{2017MNRAS.465.1682H}
{Hayward}, C.~C., \& {Hopkins}, P.~F. 2017, \mnras, 465, 1682, \dodoi{10.1093/mnras/stw2888}

\bibitem[{{He} {et~al.}(2024){He}, {Wang}, {Jones}, {Treu}, {Glazebrook}, {Malkan}, {Vulcani}, {Metha}, {Brada{\v{c}}}, {Brammer}, {Roberts-Borsani}, {Strait}, {Bonchi}, {Castellano}, {Fontana}, {Mason}, {Merlin}, {Morishita}, {Paris}, {Santini}, {Trenti}, {Boyett}, \& {Grasha}}]{2024ApJ...960L..13H}
{He}, X., {Wang}, X., {Jones}, T., {et~al.} 2024, \apjl, 960, L13, \dodoi{10.3847/2041-8213/ad12cd}

\bibitem[{{Heintz} {et~al.}(2023){Heintz}, {Brammer}, {Gim{\'e}nez-Arteaga}, {Strait}, {del P. Lagos}, {Vijayan}, {Matthee}, {Watson}, {Mason}, {Hutter}, {Toft}, {Fynbo}, \& {Oesch}}]{2023NatAs...7.1517H}
{Heintz}, K.~E., {Brammer}, G.~B., {Gim{\'e}nez-Arteaga}, C., {et~al.} 2023, Nature Astronomy, 7, 1517, \dodoi{10.1038/s41550-023-02078-7}

\bibitem[{{Heintz} {et~al.}(2024){Heintz}, {Brammer}, {Watson}, {Oesch}, {Keating}, {Hayes}, {Abdurro'uf}, {Arellano-C{\'o}rdova}, {Carnall}, {Christiansen}, {Cullen}, {Dav{\'e}}, {Dayal}, {Ferrara}, {Finlator}, {Fynbo}, {Flury}, {Gelli}, {Gillman}, {Gottumukkala}, {Gould}, {Greve}, {Hardin}, {Y. -Y Hsiao}, {Hutter}, {Jakobsson}, {Killi}, {Khosravaninezhad}, {Laursen}, {Lee}, {Magdis}, {Matthee}, {Naidu}, {Narayanan}, {Pollock}, {Prescott}, {Rusakov}, {Shuntov}, {Sneppen}, {Smit}, {Tanvir}, {Terp}, {Toft}, {Valentino}, {Vijayan}, {Weaver}, {Wise}, \& {Witstok}}]{2024arXiv240402211H}
{Heintz}, K.~E., {Brammer}, G.~B., {Watson}, D., {et~al.} 2024, arXiv e-prints, arXiv:2404.02211, \dodoi{10.48550/arXiv.2404.02211}

\bibitem[{{Henry} {et~al.}(2013{\natexlab{a}}){Henry}, {Martin}, {Finlator}, \& {Dressler}}]{2013ApJ...769..148H}
{Henry}, A., {Martin}, C.~L., {Finlator}, K., \& {Dressler}, A. 2013{\natexlab{a}}, \apj, 769, 148, \dodoi{10.1088/0004-637X/769/2/148}

\bibitem[{{Henry} {et~al.}(2013{\natexlab{b}}){Henry}, {Scarlata}, {Dom{\'\i}nguez}, {Malkan}, {Martin}, {Siana}, {Atek}, {Bedregal}, {Colbert}, {Rafelski}, {Ross}, {Teplitz}, {Bunker}, {Dressler}, {Hathi}, {Masters}, {McCarthy}, \& {Straughn}}]{2013ApJ...776L..27H}
{Henry}, A., {Scarlata}, C., {Dom{\'\i}nguez}, A., {et~al.} 2013{\natexlab{b}}, \apjl, 776, L27, \dodoi{10.1088/2041-8205/776/2/L27}

\bibitem[{{Henry} {et~al.}(2021){Henry}, {Rafelski}, {Sunnquist}, {Pirzkal}, {Pacifici}, {Atek}, {Bagley}, {Baronchelli}, {Barro}, {Bunker}, {Colbert}, {Dai}, {Elmegreen}, {Elmegreen}, {Finkelstein}, {Kocevski}, {Koekemoer}, {Malkan}, {Martin}, {Mehta}, {Pahl}, {Papovich}, {Rutkowski}, {S{\'a}nchez Almeida}, {Scarlata}, {Snyder}, \& {Teplitz}}]{2021ApJ...919..143H}
{Henry}, A., {Rafelski}, M., {Sunnquist}, B., {et~al.} 2021, \apj, 919, 143, \dodoi{10.3847/1538-4357/ac1105}

\bibitem[{{Jakobsen} {et~al.}(2022){Jakobsen}, {Ferruit}, {Alves de Oliveira}, {Arribas}, {Bagnasco}, {Barho}, {Beck}, {Birkmann}, {B{\"o}ker}, {Bunker}, {Charlot}, {de Jong}, {de Marchi}, {Ehrenwinkler}, {Falcolini}, {Fels}, {Franx}, {Franz}, {Funke}, {Giardino}, {Gnata}, {Holota}, {Honnen}, {Jensen}, {Jentsch}, {Johnson}, {Jollet}, {Karl}, {Kling}, {K{\"o}hler}, {Kolm}, {Kumari}, {Lander}, {Lemke}, {L{\'o}pez-Caniego}, {L{\"u}tzgendorf}, {Maiolino}, {Manjavacas}, {Marston}, {Maschmann}, {Maurer}, {Messerschmidt}, {Moseley}, {Mosner}, {Mott}, {Muzerolle}, {Pirzkal}, {Pittet}, {Plitzke}, {Posselt}, {Rapp}, {Rauscher}, {Rawle}, {Rix}, {R{\"o}del}, {Rumler}, {Sabbi}, {Salvignol}, {Schmid}, {Sirianni}, {Smith}, {Strada}, {te Plate}, {Valenti}, {Wettemann}, {Wiehe}, {Wiesmayer}, {Willott}, {Wright}, {Zeidler}, \& {Zincke}}]{2022A&A...661A..80J}
{Jakobsen}, P., {Ferruit}, P., {Alves de Oliveira}, C., {et~al.} 2022, \aap, 661, A80, \dodoi{10.1051/0004-6361/202142663}

\bibitem[{{Juneau} {et~al.}(2014){Juneau}, {Bournaud}, {Charlot}, {Daddi}, {Elbaz}, {Trump}, {Brinchmann}, {Dickinson}, {Duc}, {Gobat}, {Jean-Baptiste}, {Le Floc'h}, {Lehnert}, {Pacifici}, {Pannella}, \& {Schreiber}}]{2014ApJ...788...88J}
{Juneau}, S., {Bournaud}, F., {Charlot}, S., {et~al.} 2014, \apj, 788, 88, \dodoi{10.1088/0004-637X/788/1/88}

\bibitem[{{Kewley} {et~al.}(2013){Kewley}, {Maier}, {Yabe}, {Ohta}, {Akiyama}, {Dopita}, \& {Yuan}}]{2013ApJ...774L..10K}
{Kewley}, L.~J., {Maier}, C., {Yabe}, K., {et~al.} 2013, \apjl, 774, L10, \dodoi{10.1088/2041-8205/774/1/L10}

\bibitem[{{Kewley} {et~al.}(2019){Kewley}, {Nicholls}, \& {Sutherland}}]{2019ARA&A..57..511K}
{Kewley}, L.~J., {Nicholls}, D.~C., \& {Sutherland}, R.~S. 2019, \araa, 57, 511, \dodoi{10.1146/annurev-astro-081817-051832}

\bibitem[{{Kroupa}(2002)}]{2002Sci...295...82K}
{Kroupa}, P. 2002, Science, 295, 82, \dodoi{10.1126/science.1067524}

\bibitem[{{Langan} {et~al.}(2020){Langan}, {Ceverino}, \& {Finlator}}]{2020MNRAS.494.1988L}
{Langan}, I., {Ceverino}, D., \& {Finlator}, K. 2020, \mnras, 494, 1988, \dodoi{10.1093/mnras/staa880}

\bibitem[{{Langeroodi} \& {Hjorth}(2023)}]{2023arXiv230706336L}
{Langeroodi}, D., \& {Hjorth}, J. 2023, arXiv e-prints, arXiv:2307.06336, \dodoi{10.48550/arXiv.2307.06336}

\bibitem[{{Langeroodi} {et~al.}(2023){Langeroodi}, {Hjorth}, {Chen}, {Kelly}, {Williams}, {Lin}, {Scarlata}, {Zitrin}, {Broadhurst}, {Diego}, {Huang}, {Filippenko}, {Foley}, {Jha}, {Koekemoer}, {Oguri}, {Perez-Fournon}, {Pierel}, {Poidevin}, \& {Strolger}}]{2023ApJ...957...39L}
{Langeroodi}, D., {Hjorth}, J., {Chen}, W., {et~al.} 2023, \apj, 957, 39, \dodoi{10.3847/1538-4357/acdbc1}

\bibitem[{{Lara-L{\'o}pez} {et~al.}(2010){Lara-L{\'o}pez}, {Cepa}, {Bongiovanni}, {P{\'e}rez Garc{\'\i}a}, {Ederoclite}, {Casta{\~n}eda}, {Fern{\'a}ndez Lorenzo}, {Povi{\'c}}, \& {S{\'a}nchez-Portal}}]{2010A&A...521L..53L}
{Lara-L{\'o}pez}, M.~A., {Cepa}, J., {Bongiovanni}, A., {et~al.} 2010, \aap, 521, L53, \dodoi{10.1051/0004-6361/201014803}

\bibitem[{{Lee} {et~al.}(2011){Lee}, {Dey}, {Reddy}, {Brown}, {Gonzalez}, {Jannuzi}, {Cooper}, {Fan}, {Bian}, {Glikman}, {Stern}, {Brodwin}, \& {Cooray}}]{2011ApJ...733...99L}
{Lee}, K.-S., {Dey}, A., {Reddy}, N., {et~al.} 2011, \apj, 733, 99, \dodoi{10.1088/0004-637X/733/2/99}

\bibitem[{{Lee} {et~al.}(2012){Lee}, {Ferguson}, {Wiklind}, {Dahlen}, {Dickinson}, {Giavalisco}, {Grogin}, {Papovich}, {Messias}, {Guo}, \& {Lin}}]{2012ApJ...752...66L}
{Lee}, K.-S., {Ferguson}, H.~C., {Wiklind}, T., {et~al.} 2012, \apj, 752, 66, \dodoi{10.1088/0004-637X/752/1/66}

\bibitem[{{Lequeux} {et~al.}(1979){Lequeux}, {Peimbert}, {Rayo}, {Serrano}, \& {Torres-Peimbert}}]{1979A&A....80..155L}
{Lequeux}, J., {Peimbert}, M., {Rayo}, J.~F., {Serrano}, A., \& {Torres-Peimbert}, S. 1979, \aap, 80, 155

\bibitem[{{Li} {et~al.}(2023){Li}, {Cai}, {Bian}, {Lin}, {Li}, {Wu}, {Sun}, {Zhang}, {Golden-Marx}, {Sun}, {Zou}, {Fan}, {Egami}, {Charlot}, {Bruzual}, \& {Chevallard}}]{2023ApJ...955L..18L}
{Li}, M., {Cai}, Z., {Bian}, F., {et~al.} 2023, \apjl, 955, L18, \dodoi{10.3847/2041-8213/acf470}

\bibitem[{{Lilly} {et~al.}(2013){Lilly}, {Carollo}, {Pipino}, {Renzini}, \& {Peng}}]{2013ApJ...772..119L}
{Lilly}, S.~J., {Carollo}, C.~M., {Pipino}, A., {Renzini}, A., \& {Peng}, Y. 2013, \apj, 772, 119, \dodoi{10.1088/0004-637X/772/2/119}

\bibitem[{{Lower} {et~al.}(2020){Lower}, {Narayanan}, {Leja}, {Johnson}, {Conroy}, \& {Dav{\'e}}}]{2020ApJ...904...33L}
{Lower}, S., {Narayanan}, D., {Leja}, J., {et~al.} 2020, \apj, 904, 33, \dodoi{10.3847/1538-4357/abbfa7}

\bibitem[{{Ma} {et~al.}(2024){Ma}, {Wang}, {Wang}, {Peng}, {Jiang}, {Yu}, {Jia}, {Chen}, {Li}, \& {Kong}}]{2024ApJ...971L..14M}
{Ma}, C., {Wang}, K., {Wang}, E., {et~al.} 2024, \apjl, 971, L14, \dodoi{10.3847/2041-8213/ad675f}

\bibitem[{{Ma} {et~al.}(2016){Ma}, {Hopkins}, {Faucher-Gigu{\`e}re}, {Zolman}, {Muratov}, {Kere{\v{s}}}, \& {Quataert}}]{2016MNRAS.456.2140M}
{Ma}, X., {Hopkins}, P.~F., {Faucher-Gigu{\`e}re}, C.-A., {et~al.} 2016, \mnras, 456, 2140, \dodoi{10.1093/mnras/stv2659}

\bibitem[{{Madau} \& {Dickinson}(2014)}]{2014ARA&A..52..415M}
{Madau}, P., \& {Dickinson}, M. 2014, \araa, 52, 415, \dodoi{10.1146/annurev-astro-081811-125615}

\bibitem[{{Magdis} {et~al.}(2010){Magdis}, {Rigopoulou}, {Huang}, \& {Fazio}}]{2010MNRAS.401.1521M}
{Magdis}, G.~E., {Rigopoulou}, D., {Huang}, J.~S., \& {Fazio}, G.~G. 2010, \mnras, 401, 1521, \dodoi{10.1111/j.1365-2966.2009.15779.x}

\bibitem[{{Maier} {et~al.}(2014){Maier}, {Lilly}, {Ziegler}, {Contini}, {P{\'e}rez Montero}, {Peng}, \& {Balestra}}]{2014ApJ...792....3M}
{Maier}, C., {Lilly}, S.~J., {Ziegler}, B.~L., {et~al.} 2014, \apj, 792, 3, \dodoi{10.1088/0004-637X/792/1/3}

\bibitem[{{Maiolino} {et~al.}(2008{\natexlab{a}}){Maiolino}, {Nagao}, {Grazian}, {Cocchia}, {Marconi}, {Mannucci}, {Cimatti}, {Pipino}, {Ballero}, {Calura}, {Chiappini}, {Fontana}, {Granato}, {Matteucci}, {Pastorini}, {Pentericci}, {Risaliti}, {Salvati}, \& {Silva}}]{2008A&A...488..463M}
{Maiolino}, R., {Nagao}, T., {Grazian}, A., {et~al.} 2008{\natexlab{a}}, \aap, 488, 463, \dodoi{10.1051/0004-6361:200809678}

\bibitem[{{Maiolino} {et~al.}(2008{\natexlab{b}}){Maiolino}, {Nagao}, {Grazian}, {Cocchia}, {Marconi}, {Mannucci}, {Cimatti}, {Pipino}, {Fontana}, {Granato}, {Matteucci}, {Pentericci}, {Risaliti}, {Salvati}, \& {Silva}}]{2008ASPC..396..409M}
{Maiolino}, R., {Nagao}, T., {Grazian}, A., {et~al.} 2008{\natexlab{b}}, in Astronomical Society of the Pacific Conference Series, Vol. 396, Formation and Evolution of Galaxy Disks, ed. J.~G. {Funes} \& E.~M. {Corsini}, 409

\bibitem[{{Mannucci} {et~al.}(2010){Mannucci}, {Cresci}, {Maiolino}, {Marconi}, \& {Gnerucci}}]{2010MNRAS.408.2115M}
{Mannucci}, F., {Cresci}, G., {Maiolino}, R., {Marconi}, A., \& {Gnerucci}, A. 2010, \mnras, 408, 2115, \dodoi{10.1111/j.1365-2966.2010.17291.x}

\bibitem[{{Mannucci} {et~al.}(2011){Mannucci}, {Salvaterra}, \& {Campisi}}]{2011MNRAS.414.1263M}
{Mannucci}, F., {Salvaterra}, R., \& {Campisi}, M.~A. 2011, \mnras, 414, 1263, \dodoi{10.1111/j.1365-2966.2011.18459.x}

\bibitem[{{Mannucci} {et~al.}(2009){Mannucci}, {Cresci}, {Maiolino}, {Marconi}, {Pastorini}, {Pozzetti}, {Gnerucci}, {Risaliti}, {Schneider}, {Lehnert}, \& {Salvati}}]{2009MNRAS.398.1915M}
{Mannucci}, F., {Cresci}, G., {Maiolino}, R., {et~al.} 2009, \mnras, 398, 1915, \dodoi{10.1111/j.1365-2966.2009.15185.x}

\bibitem[{{Marconcini} {et~al.}(2024){Marconcini}, {D'Eugenio}, {Maiolino}, {Arribas}, {Bunker}, {Carniani}, {Charlot}, {Perna}, {Rodriguez Del Pino}, {Ubler}, {Willott}, {Boker}, {Cresci}, {Curti}, {Jones}, {Lamperti}, {Parlanti}, \& {Venturi}}]{2024arXiv240708616M}
{Marconcini}, C., {D'Eugenio}, F., {Maiolino}, R., {et~al.} 2024, arXiv e-prints, arXiv:2407.08616, \dodoi{10.48550/arXiv.2407.08616}

\bibitem[{{Marino} {et~al.}(2013){Marino}, {Rosales-Ortega}, {S{\'a}nchez}, {Gil de Paz}, {V{\'\i}lchez}, {Miralles-Caballero}, {Kehrig}, {P{\'e}rez-Montero}, {Stanishev}, {Iglesias-P{\'a}ramo}, {D{\'\i}az}, {Castillo-Morales}, {Kennicutt}, {L{\'o}pez-S{\'a}nchez}, {Galbany}, {Garc{\'\i}a-Benito}, {Mast}, {Mendez-Abreu}, {Monreal-Ibero}, {Husemann}, {Walcher}, {Garc{\'\i}a-Lorenzo}, {Masegosa}, {Del Olmo Orozco}, {Mour{\~a}o}, {Ziegler}, {Moll{\'a}}, {Papaderos}, {S{\'a}nchez-Bl{\'a}zquez}, {Gonz{\'a}lez Delgado}, {Falc{\'o}n-Barroso}, {Roth}, {van de Ven}, \& {CALIFA Team}}]{2013A&A...559A.114M}
{Marino}, R.~A., {Rosales-Ortega}, F.~F., {S{\'a}nchez}, S.~F., {et~al.} 2013, \aap, 559, A114, \dodoi{10.1051/0004-6361/201321956}

\bibitem[{{Marszewski} {et~al.}(2024){Marszewski}, {Sun}, {Faucher-Gigu{\`e}re}, {Hayward}, \& {Feldmann}}]{2024ApJ...967L..41M}
{Marszewski}, A., {Sun}, G., {Faucher-Gigu{\`e}re}, C.-A., {Hayward}, C.~C., \& {Feldmann}, R. 2024, \apjl, 967, L41, \dodoi{10.3847/2041-8213/ad4cee}

\bibitem[{{Matthee} {et~al.}(2023){Matthee}, {Mackenzie}, {Simcoe}, {Kashino}, {Lilly}, {Bordoloi}, \& {Eilers}}]{2023ApJ...950...67M}
{Matthee}, J., {Mackenzie}, R., {Simcoe}, R.~A., {et~al.} 2023, \apj, 950, 67, \dodoi{10.3847/1538-4357/acc846}

\bibitem[{{McAlpine} {et~al.}(2016){McAlpine}, {Helly}, {Schaller}, {Trayford}, {Qu}, {Furlong}, {Bower}, {Crain}, {Schaye}, {Theuns}, {Dalla Vecchia}, {Frenk}, {McCarthy}, {Jenkins}, {Rosas-Guevara}, {White}, {Baes}, {Camps}, \& {Lemson}}]{2016A&C....15...72M}
{McAlpine}, S., {Helly}, J.~C., {Schaller}, M., {et~al.} 2016, Astronomy and Computing, 15, 72, \dodoi{10.1016/j.ascom.2016.02.004}

\bibitem[{{Nakajima} {et~al.}(2023){Nakajima}, {Ouchi}, {Isobe}, {Harikane}, {Zhang}, {Ono}, {Umeda}, \& {Oguri}}]{2023ApJS..269...33N}
{Nakajima}, K., {Ouchi}, M., {Isobe}, Y., {et~al.} 2023, \apjs, 269, 33, \dodoi{10.3847/1538-4365/acd556}

\bibitem[{{Nakajima} {et~al.}(2022){Nakajima}, {Ouchi}, {Xu}, {Rauch}, {Harikane}, {Nishigaki}, {Isobe}, {Kusakabe}, {Nagao}, {Ono}, {Onodera}, {Sugahara}, {Kim}, {Komiyama}, {Lee}, \& {Zahedy}}]{2022ApJS..262....3N}
{Nakajima}, K., {Ouchi}, M., {Xu}, Y., {et~al.} 2022, \apjs, 262, 3, \dodoi{10.3847/1538-4365/ac7710}

\bibitem[{{Nelson} {et~al.}(2019{\natexlab{a}}){Nelson}, {Pillepich}, {Springel}, {Pakmor}, {Weinberger}, {Genel}, {Torrey}, {Vogelsberger}, {Marinacci}, \& {Hernquist}}]{2019MNRAS.490.3234N}
{Nelson}, D., {Pillepich}, A., {Springel}, V., {et~al.} 2019{\natexlab{a}}, \mnras, 490, 3234, \dodoi{10.1093/mnras/stz2306}

\bibitem[{{Nelson} {et~al.}(2019{\natexlab{b}}){Nelson}, {Springel}, {Pillepich}, {Rodriguez-Gomez}, {Torrey}, {Genel}, {Vogelsberger}, {Pakmor}, {Marinacci}, {Weinberger}, {Kelley}, {Lovell}, {Diemer}, \& {Hernquist}}]{2019ComAC...6....2N}
{Nelson}, D., {Springel}, V., {Pillepich}, A., {et~al.} 2019{\natexlab{b}}, Computational Astrophysics and Cosmology, 6, 2, \dodoi{10.1186/s40668-019-0028-x}

\bibitem[{{Noeske} {et~al.}(2007){Noeske}, {Weiner}, {Faber}, {Papovich}, {Koo}, {Somerville}, {Bundy}, {Conselice}, {Newman}, {Schiminovich}, {Le Floc'h}, {Coil}, {Rieke}, {Lotz}, {Primack}, {Barmby}, {Cooper}, {Davis}, {Ellis}, {Fazio}, {Guhathakurta}, {Huang}, {Kassin}, {Martin}, {Phillips}, {Rich}, {Small}, {Willmer}, \& {Wilson}}]{2007ApJ...660L..43N}
{Noeske}, K.~G., {Weiner}, B.~J., {Faber}, S.~M., {et~al.} 2007, \apjl, 660, L43, \dodoi{10.1086/517926}

\bibitem[{{Oke} \& {Gunn}(1983)}]{1983ApJ...266..713O}
{Oke}, J.~B., \& {Gunn}, J.~E. 1983, \apj, 266, 713, \dodoi{10.1086/160817}

\bibitem[{{Onodera} {et~al.}(2016){Onodera}, {Carollo}, {Lilly}, {Renzini}, {Arimoto}, {Capak}, {Daddi}, {Scoville}, {Tacchella}, {Tatehora}, \& {Zamorani}}]{2016ApJ...822...42O}
{Onodera}, M., {Carollo}, C.~M., {Lilly}, S., {et~al.} 2016, \apj, 822, 42, \dodoi{10.3847/0004-637X/822/1/42}

\bibitem[{{Osterbrock} \& {Ferland}(2006)}]{2006agna.book.....O}
{Osterbrock}, D.~E., \& {Ferland}, G.~J. 2006, {Astrophysics of gaseous nebulae and active galactic nuclei}

\bibitem[{{Pallottini} {et~al.}(2022){Pallottini}, {Ferrara}, {Gallerani}, {Behrens}, {Kohandel}, {Carniani}, {Vallini}, {Salvadori}, {Gelli}, {Sommovigo}, {D'Odorico}, {Di Mascia}, \& {Pizzati}}]{2022MNRAS.513.5621P}
{Pallottini}, A., {Ferrara}, A., {Gallerani}, S., {et~al.} 2022, \mnras, 513, 5621, \dodoi{10.1093/mnras/stac1281}

\bibitem[{{Pannella} {et~al.}(2009){Pannella}, {Carilli}, {Daddi}, {McCracken}, {Owen}, {Renzini}, {Strazzullo}, {Civano}, {Koekemoer}, {Schinnerer}, {Scoville}, {Smol{\v{c}}i{\'c}}, {Taniguchi}, {Aussel}, {Kneib}, {Ilbert}, {Mellier}, {Salvato}, {Thompson}, \& {Willott}}]{2009ApJ...698L.116P}
{Pannella}, M., {Carilli}, C.~L., {Daddi}, E., {et~al.} 2009, \apjl, 698, L116, \dodoi{10.1088/0004-637X/698/2/L116}

\bibitem[{{Pettini} \& {Pagel}(2004)}]{2004MNRAS.348L..59P}
{Pettini}, M., \& {Pagel}, B. E.~J. 2004, \mnras, 348, L59, \dodoi{10.1111/j.1365-2966.2004.07591.x}

\bibitem[{{Pistis} {et~al.}(2024){Pistis}, {Pollo}, {Figueira}, {Vergani}, {Hamed}, {Ma{\l}ek}, {Durkalec}, {Donevski}, {Salim}, {Iovino}, {Pearson}, {Romano}, \& {Scodeggio}}]{2024A&A...683A.203P}
{Pistis}, F., {Pollo}, A., {Figueira}, M., {et~al.} 2024, \aap, 683, A203, \dodoi{10.1051/0004-6361/202346943}

\bibitem[{{Planck Collaboration} {et~al.}(2020){Planck Collaboration}, {Aghanim}, {Akrami}, {Ashdown}, {Aumont}, {Baccigalupi}, {Ballardini}, {Banday}, {Barreiro}, {Bartolo}, {Basak}, {Battye}, {Benabed}, {Bernard}, {Bersanelli}, {Bielewicz}, {Bock}, {Bond}, {Borrill}, {Bouchet}, {Boulanger}, {Bucher}, {Burigana}, {Butler}, {Calabrese}, {Cardoso}, {Carron}, {Challinor}, {Chiang}, {Chluba}, {Colombo}, {Combet}, {Contreras}, {Crill}, {Cuttaia}, {de Bernardis}, {de Zotti}, {Delabrouille}, {Delouis}, {Di Valentino}, {Diego}, {Dor{\'e}}, {Douspis}, {Ducout}, {Dupac}, {Dusini}, {Efstathiou}, {Elsner}, {En{\ss}lin}, {Eriksen}, {Fantaye}, {Farhang}, {Fergusson}, {Fernandez-Cobos}, {Finelli}, {Forastieri}, {Frailis}, {Fraisse}, {Franceschi}, {Frolov}, {Galeotta}, {Galli}, {Ganga}, {G{\'e}nova-Santos}, {Gerbino}, {Ghosh}, {Gonz{\'a}lez-Nuevo}, {G{\'o}rski}, {Gratton}, {Gruppuso}, {Gudmundsson}, {Hamann}, {Handley}, {Hansen}, {Herranz}, {Hildebrandt}, {Hivon}, {Huang}, {Jaffe}, {Jones}, {Karakci}, {Keih{\"a}nen},
  {Keskitalo}, {Kiiveri}, {Kim}, {Kisner}, {Knox}, {Krachmalnicoff}, {Kunz}, {Kurki-Suonio}, {Lagache}, {Lamarre}, {Lasenby}, {Lattanzi}, {Lawrence}, {Le Jeune}, {Lemos}, {Lesgourgues}, {Levrier}, {Lewis}, {Liguori}, {Lilje}, {Lilley}, {Lindholm}, {L{\'o}pez-Caniego}, {Lubin}, {Ma}, {Mac{\'\i}as-P{\'e}rez}, {Maggio}, {Maino}, {Mandolesi}, {Mangilli}, {Marcos-Caballero}, {Maris}, {Martin}, {Martinelli}, {Mart{\'\i}nez-Gonz{\'a}lez}, {Matarrese}, {Mauri}, {McEwen}, {Meinhold}, {Melchiorri}, {Mennella}, {Migliaccio}, {Millea}, {Mitra}, {Miville-Desch{\^e}nes}, {Molinari}, {Montier}, {Morgante}, {Moss}, {Natoli}, {N{\o}rgaard-Nielsen}, {Pagano}, {Paoletti}, {Partridge}, {Patanchon}, {Peiris}, {Perrotta}, {Pettorino}, {Piacentini}, {Polastri}, {Polenta}, {Puget}, {Rachen}, {Reinecke}, {Remazeilles}, {Renzi}, {Rocha}, {Rosset}, {Roudier}, {Rubi{\~n}o-Mart{\'\i}n}, {Ruiz-Granados}, {Salvati}, {Sandri}, {Savelainen}, {Scott}, {Shellard}, {Sirignano}, {Sirri}, {Spencer}, {Sunyaev}, {Suur-Uski}, {Tauber}, {Tavagnacco},
  {Tenti}, {Toffolatti}, {Tomasi}, {Trombetti}, {Valenziano}, {Valiviita}, {Van Tent}, {Vibert}, {Vielva}, {Villa}, {Vittorio}, {Wandelt}, {Wehus}, {White}, {White}, {Zacchei}, \& {Zonca}}]{2020A&A...641A...6P}
{Planck Collaboration}, {Aghanim}, N., {Akrami}, Y., {et~al.} 2020, \aap, 641, A6, \dodoi{10.1051/0004-6361/201833910}

\bibitem[{{Pontoppidan} {et~al.}(2022){Pontoppidan}, {Barrientes}, {Blome}, {Braun}, {Brown}, {Carruthers}, {Coe}, {DePasquale}, {Espinoza}, {Marin}, {Gordon}, {Henry}, {Hustak}, {James}, {Jenkins}, {Koekemoer}, {LaMassa}, {Law}, {Lockwood}, {Moro-Martin}, {Mullally}, {Pagan}, {Player}, {Proffitt}, {Pulliam}, {Ramsay}, {Ravindranath}, {Reid}, {Robberto}, {Sabbi}, {Ubeda}, {Balogh}, {Flanagan}, {Gardner}, {Hasan}, {Meinke}, \& {Nota}}]{2022ApJ...936L..14P}
{Pontoppidan}, K.~M., {Barrientes}, J., {Blome}, C., {et~al.} 2022, \apjl, 936, L14, \dodoi{10.3847/2041-8213/ac8a4e}

\bibitem[{{Popesso} {et~al.}(2023){Popesso}, {Concas}, {Cresci}, {Belli}, {Rodighiero}, {Inami}, {Dickinson}, {Ilbert}, {Pannella}, \& {Elbaz}}]{2023MNRAS.519.1526P}
{Popesso}, P., {Concas}, A., {Cresci}, G., {et~al.} 2023, \mnras, 519, 1526, \dodoi{10.1093/mnras/stac3214}

\bibitem[{{Reddy} {et~al.}(2006){Reddy}, {Steidel}, {Fadda}, {Yan}, {Pettini}, {Shapley}, {Erb}, \& {Adelberger}}]{2006ApJ...644..792R}
{Reddy}, N.~A., {Steidel}, C.~C., {Fadda}, D., {et~al.} 2006, \apj, 644, 792, \dodoi{10.1086/503739}

\bibitem[{{Roberts-Borsani} {et~al.}(2024){Roberts-Borsani}, {Treu}, {Shapley}, {Fontana}, {Pentericci}, {Castellano}, {Morishita}, {Bergamini}, \& {Rosati}}]{2024arXiv240307103R}
{Roberts-Borsani}, G., {Treu}, T., {Shapley}, A., {et~al.} 2024, arXiv e-prints, arXiv:2403.07103, \dodoi{10.48550/arXiv.2403.07103}

\bibitem[{{Robertson}(2022)}]{2022ARA&A..60..121R}
{Robertson}, B.~E. 2022, \araa, 60, 121, \dodoi{10.1146/annurev-astro-120221-044656}

\bibitem[{{Salim} {et~al.}(2014){Salim}, {Lee}, {Ly}, {Brinchmann}, {Dav{\'e}}, {Dickinson}, {Salzer}, \& {Charlot}}]{2014ApJ...797..126S}
{Salim}, S., {Lee}, J.~C., {Ly}, C., {et~al.} 2014, \apj, 797, 126, \dodoi{10.1088/0004-637X/797/2/126}

\bibitem[{{Salmon} {et~al.}(2015){Salmon}, {Papovich}, {Finkelstein}, {Tilvi}, {Finlator}, {Behroozi}, {Dahlen}, {Dav{\'e}}, {Dekel}, {Dickinson}, {Ferguson}, {Giavalisco}, {Long}, {Lu}, {Mobasher}, {Reddy}, {Somerville}, \& {Wechsler}}]{2015ApJ...799..183S}
{Salmon}, B., {Papovich}, C., {Finkelstein}, S.~L., {et~al.} 2015, \apj, 799, 183, \dodoi{10.1088/0004-637X/799/2/183}

\bibitem[{{S{\'a}nchez} {et~al.}(2017){S{\'a}nchez}, {Barrera-Ballesteros}, {S{\'a}nchez-Menguiano}, {Walcher}, {Marino}, {Galbany}, {Bland-Hawthorn}, {Cano-D{\'\i}az}, {Garc{\'\i}a-Benito}, {L{\'o}pez-Cob{\'a}}, {Zibetti}, {Vilchez}, {Igl{\'e}sias-P{\'a}ramo}, {Kehrig}, {L{\'o}pez S{\'a}nchez}, {Duarte Puertas}, \& {Ziegler}}]{2017MNRAS.469.2121S}
{S{\'a}nchez}, S.~F., {Barrera-Ballesteros}, J.~K., {S{\'a}nchez-Menguiano}, L., {et~al.} 2017, \mnras, 469, 2121, \dodoi{10.1093/mnras/stx808}

\bibitem[{{Sanders} {et~al.}(2024){Sanders}, {Shapley}, {Topping}, {Reddy}, \& {Brammer}}]{2024ApJ...962...24S}
{Sanders}, R.~L., {Shapley}, A.~E., {Topping}, M.~W., {Reddy}, N.~A., \& {Brammer}, G.~B. 2024, \apj, 962, 24, \dodoi{10.3847/1538-4357/ad15fc}

\bibitem[{{Sanders} {et~al.}(2015){Sanders}, {Shapley}, {Kriek}, {Reddy}, {Freeman}, {Coil}, {Siana}, {Mobasher}, {Shivaei}, {Price}, \& {de Groot}}]{2015ApJ...799..138S}
{Sanders}, R.~L., {Shapley}, A.~E., {Kriek}, M., {et~al.} 2015, \apj, 799, 138, \dodoi{10.1088/0004-637X/799/2/138}

\bibitem[{{Sanders} {et~al.}(2021){Sanders}, {Shapley}, {Jones}, {Reddy}, {Kriek}, {Siana}, {Coil}, {Mobasher}, {Shivaei}, {Dav{\'e}}, {Azadi}, {Price}, {Leung}, {Freeman}, {Fetherolf}, {de Groot}, {Zick}, \& {Barro}}]{2021ApJ...914...19S}
{Sanders}, R.~L., {Shapley}, A.~E., {Jones}, T., {et~al.} 2021, \apj, 914, 19, \dodoi{10.3847/1538-4357/abf4c1}

\bibitem[{{Santini} {et~al.}(2017){Santini}, {Fontana}, {Castellano}, {Di Criscienzo}, {Merlin}, {Amorin}, {Cullen}, {Daddi}, {Dickinson}, {Dunlop}, {Grazian}, {Lamastra}, {McLure}, {Micha{\l}owski}, {Pentericci}, \& {Shu}}]{2017ApJ...847...76S}
{Santini}, P., {Fontana}, A., {Castellano}, M., {et~al.} 2017, \apj, 847, 76, \dodoi{10.3847/1538-4357/aa8874}

\bibitem[{{Sarkar} {et~al.}(2021){Sarkar}, {Ferland}, {Chatzikos}, {Guzm{\'a}n}, {van Hoof}, {Smyth}, {Ramsbottom}, {Keenan}, \& {Ballance}}]{2021ApJ...907...12S}
{Sarkar}, A., {Ferland}, G.~J., {Chatzikos}, M., {et~al.} 2021, \apj, 907, 12, \dodoi{10.3847/1538-4357/abcaa6}

\bibitem[{{Shapley} {et~al.}(2023){Shapley}, {Reddy}, {Sanders}, {Topping}, \& {Brammer}}]{2023ApJ...950L...1S}
{Shapley}, A.~E., {Reddy}, N.~A., {Sanders}, R.~L., {Topping}, M.~W., \& {Brammer}, G.~B. 2023, \apjl, 950, L1, \dodoi{10.3847/2041-8213/acd939}

\bibitem[{{Sparre} {et~al.}(2017){Sparre}, {Hayward}, {Feldmann}, {Faucher-Gigu{\`e}re}, {Muratov}, {Kere{\v{s}}}, \& {Hopkins}}]{2017MNRAS.466...88S}
{Sparre}, M., {Hayward}, C.~C., {Feldmann}, R., {et~al.} 2017, \mnras, 466, 88, \dodoi{10.1093/mnras/stw3011}

\bibitem[{{Speagle} {et~al.}(2014){Speagle}, {Steinhardt}, {Capak}, \& {Silverman}}]{2014ApJS..214...15S}
{Speagle}, J.~S., {Steinhardt}, C.~L., {Capak}, P.~L., \& {Silverman}, J.~D. 2014, \apjs, 214, 15, \dodoi{10.1088/0067-0049/214/2/15}

\bibitem[{{Springel} {et~al.}(2018){Springel}, {Pakmor}, {Pillepich}, {Weinberger}, {Nelson}, {Hernquist}, {Vogelsberger}, {Genel}, {Torrey}, {Marinacci}, \& {Naiman}}]{2018MNRAS.475..676S}
{Springel}, V., {Pakmor}, R., {Pillepich}, A., {et~al.} 2018, \mnras, 475, 676, \dodoi{10.1093/mnras/stx3304}

\bibitem[{{Stark} {et~al.}(2013){Stark}, {Schenker}, {Ellis}, {Robertson}, {McLure}, \& {Dunlop}}]{2013ApJ...763..129S}
{Stark}, D.~P., {Schenker}, M.~A., {Ellis}, R., {et~al.} 2013, \apj, 763, 129, \dodoi{10.1088/0004-637X/763/2/129}

\bibitem[{{Steidel} {et~al.}(2014){Steidel}, {Rudie}, {Strom}, {Pettini}, {Reddy}, {Shapley}, {Trainor}, {Erb}, {Turner}, {Konidaris}, {Kulas}, {Mace}, {Matthews}, \& {McLean}}]{2014ApJ...795..165S}
{Steidel}, C.~C., {Rudie}, G.~C., {Strom}, A.~L., {et~al.} 2014, \apj, 795, 165, \dodoi{10.1088/0004-637X/795/2/165}

\bibitem[{{Steinhardt} {et~al.}(2014){Steinhardt}, {Speagle}, {Capak}, {Silverman}, {Carollo}, {Dunlop}, {Hashimoto}, {Hsieh}, {Ilbert}, {Le Fevre}, {Le Floc'h}, {Lee}, {Lin}, {Lin}, {Masters}, {McCracken}, {Nagao}, {Petric}, {Salvato}, {Sanders}, {Scoville}, {Sheth}, {Strauss}, \& {Taniguchi}}]{2014ApJ...791L..25S}
{Steinhardt}, C.~L., {Speagle}, J.~S., {Capak}, P., {et~al.} 2014, \apjl, 791, L25, \dodoi{10.1088/2041-8205/791/2/L25}

\bibitem[{{Storey} \& {Zeippen}(2000)}]{2000MNRAS.312..813S}
{Storey}, P.~J., \& {Zeippen}, C.~J. 2000, \mnras, 312, 813, \dodoi{10.1046/j.1365-8711.2000.03184.x}

\bibitem[{{Suresh} {et~al.}(2015){Suresh}, {Bird}, {Vogelsberger}, {Genel}, {Torrey}, {Sijacki}, {Springel}, \& {Hernquist}}]{2015MNRAS.448..895S}
{Suresh}, J., {Bird}, S., {Vogelsberger}, M., {et~al.} 2015, \mnras, 448, 895, \dodoi{10.1093/mnras/stu2762}

\bibitem[{{Torrey} {et~al.}(2019){Torrey}, {Vogelsberger}, {Marinacci}, {Pakmor}, {Springel}, {Nelson}, {Naiman}, {Pillepich}, {Genel}, {Weinberger}, \& {Hernquist}}]{2019MNRAS.484.5587T}
{Torrey}, P., {Vogelsberger}, M., {Marinacci}, F., {et~al.} 2019, \mnras, 484, 5587, \dodoi{10.1093/mnras/stz243}

\bibitem[{{Tremonti} {et~al.}(2004){Tremonti}, {Heckman}, {Kauffmann}, {Brinchmann}, {Charlot}, {White}, {Seibert}, {Peng}, {Schlegel}, {Uomoto}, {Fukugita}, \& {Brinkmann}}]{2004ApJ...613..898T}
{Tremonti}, C.~A., {Heckman}, T.~M., {Kauffmann}, G., {et~al.} 2004, \apj, 613, 898, \dodoi{10.1086/423264}

\bibitem[{{Treu} {et~al.}(2022){Treu}, {Roberts-Borsani}, {Bradac}, {Brammer}, {Fontana}, {Henry}, {Mason}, {Morishita}, {Pentericci}, {Wang}, {Acebron}, {Bagley}, {Bergamini}, {Belfiori}, {Bonchi}, {Boyett}, {Boutsia}, {Calabr{\'o}}, {Caminha}, {Castellano}, {Dressler}, {Glazebrook}, {Grillo}, {Jacobs}, {Jones}, {Kelly}, {Leethochawalit}, {Malkan}, {Marchesini}, {Mascia}, {Mercurio}, {Merlin}, {Nanayakkara}, {Nonino}, {Paris}, {Poggianti}, {Rosati}, {Santini}, {Scarlata}, {Shipley}, {Strait}, {Trenti}, {Tubthong}, {Vanzella}, {Vulcani}, \& {Yang}}]{2022ApJ...935..110T}
{Treu}, T., {Roberts-Borsani}, G., {Bradac}, M., {et~al.} 2022, \apj, 935, 110, \dodoi{10.3847/1538-4357/ac8158}

\bibitem[{{Troncoso} {et~al.}(2014){Troncoso}, {Maiolino}, {Sommariva}, {Cresci}, {Mannucci}, {Marconi}, {Meneghetti}, {Grazian}, {Cimatti}, {Fontana}, {Nagao}, \& {Pentericci}}]{2014A&A...563A..58T}
{Troncoso}, P., {Maiolino}, R., {Sommariva}, V., {et~al.} 2014, \aap, 563, A58, \dodoi{10.1051/0004-6361/201322099}

\bibitem[{{Ucci} {et~al.}(2023){Ucci}, {Dayal}, {Hutter}, {Kobayashi}, {Gottl{\"o}ber}, {Yepes}, {Hunt}, {Legrand}, \& {Tortora}}]{2023MNRAS.518.3557U}
{Ucci}, G., {Dayal}, P., {Hutter}, A., {et~al.} 2023, \mnras, 518, 3557, \dodoi{10.1093/mnras/stac2654}

\bibitem[{{Venturi} {et~al.}(2024){Venturi}, {Carniani}, {Parlanti}, {Kohandel}, {Curti}, {Pallottini}, {Vallini}, {Arribas}, {Bunker}, {Cameron}, {Castellano}, {Ferrara}, {Fontana}, {Gallerani}, {Gelli}, {Maiolino}, {Ntormousi}, {Pacifici}, {Pentericci}, {Salvadori}, \& {Vanzella}}]{2024arXiv240303977V}
{Venturi}, G., {Carniani}, S., {Parlanti}, E., {et~al.} 2024, arXiv e-prints, arXiv:2403.03977, \dodoi{10.48550/arXiv.2403.03977}

\bibitem[{{Vogelsberger} {et~al.}(2020){Vogelsberger}, {Marinacci}, {Torrey}, \& {Puchwein}}]{2020NatRP...2...42V}
{Vogelsberger}, M., {Marinacci}, F., {Torrey}, P., \& {Puchwein}, E. 2020, Nature Reviews Physics, 2, 42, \dodoi{10.1038/s42254-019-0127-2}

\bibitem[{{Vogelsberger} {et~al.}(2014{\natexlab{a}}){Vogelsberger}, {Genel}, {Springel}, {Torrey}, {Sijacki}, {Xu}, {Snyder}, {Bird}, {Nelson}, \& {Hernquist}}]{2014Natur.509..177V}
{Vogelsberger}, M., {Genel}, S., {Springel}, V., {et~al.} 2014{\natexlab{a}}, \nat, 509, 177, \dodoi{10.1038/nature13316}

\bibitem[{{Vogelsberger} {et~al.}(2014{\natexlab{b}}){Vogelsberger}, {Genel}, {Springel}, {Torrey}, {Sijacki}, {Xu}, {Snyder}, {Nelson}, \& {Hernquist}}]{2014MNRAS.444.1518V}
---. 2014{\natexlab{b}}, \mnras, 444, 1518, \dodoi{10.1093/mnras/stu1536}

\bibitem[{{Wang} {et~al.}(2024){Wang}, {Cheng}, {Ge}, {Meng}, {Daddi}, {Yan}, {Ji}, {Jin}, {Jones}, {Malkan}, {Arrabal Haro}, {Brammer}, {Oguri}, {Hou}, \& {Zhang}}]{2024ApJ...967L..42W}
{Wang}, X., {Cheng}, C., {Ge}, J., {et~al.} 2024, \apjl, 967, L42, \dodoi{10.3847/2041-8213/ad4ced}

\bibitem[{{Wilkins} {et~al.}(2023){Wilkins}, {Vijayan}, {Lovell}, {Roper}, {Irodotou}, {Caruana}, {Seeyave}, {Kuusisto}, {Thomas}, \& {Parris}}]{2023MNRAS.519.3118W}
{Wilkins}, S.~M., {Vijayan}, A.~P., {Lovell}, C.~C., {et~al.} 2023, \mnras, 519, 3118, \dodoi{10.1093/mnras/stac3280}

\bibitem[{{York} {et~al.}(2000){York}, {Adelman}, {Anderson}, {Anderson}, {Annis}, {Bahcall}, {Bakken}, {Barkhouser}, {Bastian}, {Berman}, {Boroski}, {Bracker}, {Briegel}, {Briggs}, {Brinkmann}, {Brunner}, {Burles}, {Carey}, {Carr}, {Castander}, {Chen}, {Colestock}, {Connolly}, {Crocker}, {Csabai}, {Czarapata}, {Davis}, {Doi}, {Dombeck}, {Eisenstein}, {Ellman}, {Elms}, {Evans}, {Fan}, {Federwitz}, {Fiscelli}, {Friedman}, {Frieman}, {Fukugita}, {Gillespie}, {Gunn}, {Gurbani}, {de Haas}, {Haldeman}, {Harris}, {Hayes}, {Heckman}, {Hennessy}, {Hindsley}, {Holm}, {Holmgren}, {Huang}, {Hull}, {Husby}, {Ichikawa}, {Ichikawa}, {Ivezi{\'c}}, {Kent}, {Kim}, {Kinney}, {Klaene}, {Kleinman}, {Kleinman}, {Knapp}, {Korienek}, {Kron}, {Kunszt}, {Lamb}, {Lee}, {Leger}, {Limmongkol}, {Lindenmeyer}, {Long}, {Loomis}, {Loveday}, {Lucinio}, {Lupton}, {MacKinnon}, {Mannery}, {Mantsch}, {Margon}, {McGehee}, {McKay}, {Meiksin}, {Merelli}, {Monet}, {Munn}, {Narayanan}, {Nash}, {Neilsen}, {Neswold}, {Newberg}, {Nichol}, {Nicinski},
  {Nonino}, {Okada}, {Okamura}, {Ostriker}, {Owen}, {Pauls}, {Peoples}, {Peterson}, {Petravick}, {Pier}, {Pope}, {Pordes}, {Prosapio}, {Rechenmacher}, {Quinn}, {Richards}, {Richmond}, {Rivetta}, {Rockosi}, {Ruthmansdorfer}, {Sandford}, {Schlegel}, {Schneider}, {Sekiguchi}, {Sergey}, {Shimasaku}, {Siegmund}, {Smee}, {Smith}, {Snedden}, {Stone}, {Stoughton}, {Strauss}, {Stubbs}, {SubbaRao}, {Szalay}, {Szapudi}, {Szokoly}, {Thakar}, {Tremonti}, {Tucker}, {Uomoto}, {Vanden Berk}, {Vogeley}, {Waddell}, {Wang}, {Watanabe}, {Weinberg}, {Yanny}, {Yasuda}, \& {SDSS Collaboration}}]{2000AJ....120.1579Y}
{York}, D.~G., {Adelman}, J., {Anderson}, John~E., J., {et~al.} 2000, \aj, 120, 1579, \dodoi{10.1086/301513}

\bibitem[{{Zahid} {et~al.}(2012){Zahid}, {Bresolin}, {Kewley}, {Coil}, \& {Dav{\'e}}}]{2012ApJ...750..120Z}
{Zahid}, H.~J., {Bresolin}, F., {Kewley}, L.~J., {Coil}, A.~L., \& {Dav{\'e}}, R. 2012, \apj, 750, 120, \dodoi{10.1088/0004-637X/750/2/120}

\bibitem[{{Zahid} {et~al.}(2013){Zahid}, {Geller}, {Kewley}, {Hwang}, {Fabricant}, \& {Kurtz}}]{2013ApJ...771L..19Z}
{Zahid}, H.~J., {Geller}, M.~J., {Kewley}, L.~J., {et~al.} 2013, \apjl, 771, L19, \dodoi{10.1088/2041-8205/771/2/L19}

\bibitem[{{Zahid} {et~al.}(2011){Zahid}, {Kewley}, \& {Bresolin}}]{2011ApJ...730..137Z}
{Zahid}, H.~J., {Kewley}, L.~J., \& {Bresolin}, F. 2011, \apj, 730, 137, \dodoi{10.1088/0004-637X/730/2/137}

\bibitem[{{Zaritsky} {et~al.}(1994){Zaritsky}, {Kennicutt}, \& {Huchra}}]{1994ApJ...420...87Z}
{Zaritsky}, D., {Kennicutt}, Robert~C., J., \& {Huchra}, J.~P. 1994, \apj, 420, 87, \dodoi{10.1086/173544}

\end{thebibliography}
